\documentclass{aa}

\usepackage{graphicx}
\usepackage{txfonts}
\usepackage{xcolor}
\usepackage{natbib}
\usepackage{hyperref}
\begin{document}

   \title{Morphologies for DECaLS Galaxies through a combination of non-parametric indices and machine learning methods}

   \subtitle{A comprehensive catalog using the Galaxy Morphology Extractor (\texttt{galmex}) code}

   \author{V. M. Sampaio
          \inst{1,2,}
          \and
          Y. Jaffé\inst{1,2}
          \and
          C. Lima-Dias\inst{3}
          \and
          S. V\'eliz Astudillo\inst{3}
          \and 
          M. Mart\'inez-Mar\'in\inst{4,2}
          \and 
          H. M\'endez-Hern\'andez\inst{3}
          \and
          R. Herrera-Camus\inst{4,2}
          \and
          A. Monachesi\inst{3}
          }

   \institute{Instituto de Física, Universidad Técnica Federico Santa María, Av. España 1680, Valparaíso, Chile\\
              \email{vitorms999@gmail.com}
         \and
             Millennium Nucleus for Galaxies (MINGAL), Chile\\
         \and
             Departamento de Astronomía, Universidad de La Serena, Avda. Raúl Bitrán 1305, La Serena, Chile\\
         \and
             Departamento de Astronomía, Facultad Ciencias Físicas y Matemáticas, Universidad de Concepción, Av. Esteban Iturra s/n Barrio Universitario, Casilla 160, Concepción, Chile\\
             }

\date{Received 11/2025; accepted 02/26, in original form 03/2026}

  \abstract
   {Galaxy morphology encodes key information about formation and evolution. Large imaging surveys require automated, reproducible methods beyond visual inspection. Non--parametric indices provide an useful framework, but their performance must be quantitatively assessed.}
   {We present a homogeneous catalog of non--parametric morphological indices for DECaLS galaxies with effective radii larger than 2 arcsec. Our goal is to evaluate the reliability of indices in separating spirals and ellipticals, test their consistency with existing classification schemes, and establish their applicability for the upcoming surveys focused in the southern hemisphere.}
   {We developed \texttt{galmex}, a modular Python package for preprocessing images and measuring a variety of non--parametric indices. Using bona-fide spirals and ellipticals as control samples, we assessed the discriminatory power of each index, and compared them with CNN-based T-Types and Galaxy Zoo DECaLS labels. We use the indices as input for a Light Gradient Boosted Machine (LightGBM) to obtain probabilistic classifications.}
   {Concentration is the most reliable parameter from the Concentratiom + Asymmetry + Smoothness system (CAS), while asymmetry--based indices (A and S) are limited to detecting disturbed morphologies. MEGG indices (M20, Entropy, Gini, G2) provide stronger separation and trace a gradient with T--Type. By using a simple binary (0/1) label for ellipticals/spirals, classifiers trained on non--parametric indices achieve high accuracy and well--calibrated probabilities, dominated by entropy, concentration, and Gini.}
   {We release the first public catalog of CA$\rm [A_S]$S+MEGG indices for DECaLS, together with \texttt{galmex}. We combine the non-parametric indices with machine learning framework to derive spiral/elliptical separation for galaxies below $z \sim 0.15$ through a probabilistic approach.}

   \keywords{galaxies: general; galaxies: structure; galaxies: spiral; galaxies: elliptical and lenticular, cD}
   \titlerunning{Morphology in DECaLS}
   \authorrunning{Sampaio et al.(2026)}

   \maketitle

\section{Introduction}

Early galaxy classification schemes \citep[e.g.][]{1926ApJ....64..321H} established the distinction between ellipticals, spirals, and lenticulars, emphasizing that structural appearance is not merely descriptive but encodes a galaxy’s formation and evolutionary history. During formation, the angular momentum of progenitor molecular clouds plays a decisive role in determining the initial morphology of galaxies. Systems with high specific angular momentum preferentially settle into rotationally supported disks, while low-angular-momentum clouds are more prone to collapse into spheroid-dominated structures \citep[e.g.][]{1969ApJ...155..393P, 2015ApJ...812...29T}. However, morphology is not static. Over cosmic time, both internal processes and environmental interactions can restructure galaxies, altering their stellar distributions, kinematics, and star formation activity. These transformations can be gradual---through secular processes---or rapid, driven by violent interactions or gas removal events \citep[e.g.][]{1972ApJ...178..623T, 1991ApJ...370L..65B, 2004ARA&A..42..603K, 2013MNRAS.432..336W}.

It is now well known that morphology reflects the interplay between internal and environmental mechanisms. Internal drivers include bar-driven secular evolution \citep{2004ARA&A..42..603K, 2013MNRAS.432L..56S}, disk instabilities \citep{2009Natur.457..451D, 2016ASSL..418..355B}, and stellar or AGN feedback \citep{2008MNRAS.387.1431D, 2012ARA&A..50..455F}, which can redistribute angular momentum, trigger or quench star formation, and alter bulge-to-disk ratios. Environmental processes are particularly relevant in dense regions of the cosmic web, where galaxy--galaxy interactions, harassment, and ram-pressure stripping can significantly reshape systems \citep{1972ApJ...176....1G, 1980ApJ...237..692L, 1999MNRAS.308..947A, 1999MNRAS.302..771J, 2000ApJ...540..113B, 2005ApJ...622L...9S}. The morphology--density relation \citep{Dressler, 1997ApJ...490..577D} encapsulates these environmental trends, and drastic environmental-driven morphological transition are observed as, for example, in ``jellyfish'' galaxies \citep{2017ApJ...844...48P, 2018MNRAS.476.4753J, 2019MNRAS.485.1157B}.

The cumulative effects of these mechanisms suggest a broad evolutionary pathway in which many galaxies migrate from star-forming, disk-dominated systems to quiescent, spheroidal ones. However, the build of this bimodality and the connection between star formation and morphology can depend on redshift. In the local universe, star-forming spirals populate the ``blue cloud,'' while quiescent ellipticals dominate the ``red sequence,'' with transitional systems lying in the ``green valley'' \citep{Strateva1, 2004ApJ...600..681B, 2014MNRAS.440..889S}. Towards higher redshift, $z\sim~2$ star-forming galaxies have clumpy morphologies \citep{2011ApJ...731...65F}, with galactic winds that are mainly driven by outflows from prominent star-forming clumps \citep{2011ApJ...733..101G} and have not yet formed a stable disc (or any disc at all). On the other hand there's observational evidence of the relation of colour and morphology at high redshift \citep[e.g.][]{2005MNRAS.357..903C} and suggestion of disk galaxies at very high redshifts \citep[e.g.][]{2022ApJ...938L...2F}. This highlights how the investigation of galaxy structural transformation is complex, with the signatures of the underlying mechanisms can be subtle and hard to disentangle observationally.

Despite its importance, there is no universal method to classify galaxy morphology. Visual classification remains intuitive and effective at low redshift \citep{1987rsac.book.....S, 1994cag..book.....S, 2010ApJS..186..427N}, but is limited by subjectivity (especially at higher redshifts) and applicability to very large samples. Parametric approaches, such as S\'ersic profile fitting \citep{1963BAAA....6...41S, 1968adga.book.....S, 2002AJ....124..266P, 2002ApJS..142....1S, 2010AJ....139.2097P, 2011ApJS..196...11S}, although simple in form, are not directly applicable to all types of galaxies due to assumed symmetries. Degeneracies between fitted parameters (e.g. bulge-to-disk ratios, effective radii, Sérsic index) often produce multiple statistically acceptable but physically distinct solutions \citep{2004AJ....128..163L}. Substructures such as compact nuclei, bars, or spiral arms can further bias fits, while even bulges themselves are not uniformly well described by high Sérsic indices \citep{1999ApJ...523..566C}. Finally, these methods assume galaxies follow smooth, symmetric light profiles, an assumption that breaks down in irregular, clumpy, or merging systems, yielding degenerate structural parameters \citep{2011MNRAS.411..385A}. Non-parametric indices---including Concentration, Asymmetry, and Smoothness (CAS; \citealt{2003ApJS..147....1C})---provide a model-independent approach, enabling structural characterization across diverse morphologies. The shape asymmetry ($\rm A_S$, \citealt{2016MNRAS.456.3032P}) can also be relevant to define disturbed systems, thus forming the $\rm CA[A_S]S$ system. Beyond the $\rm CA[A_S]S$, the combination of M20 \citep{2004AJ....128..163L}, Shannon Entropy \citep[E,][]{2015ApJ...814...55F}, Gini index \citep[G][]{2004AJ....128..163L}, and gradient pattern asymmetry \citep[G2][]{rosa2018gradient}---the MEGG system---has demonstrated improved performance in separating early- and late-type galaxies in the $z\leq0.1$ universe \citep{2024MNRAS.528...82K}. Still, the measurement of non-parametric indexes is heavily dependent on image pre-processing steps (e.g. object detection, cleaning, and segmentation mask). More recently, Machine- and deep-learning methods now enable automated classification of millions of galaxies \citep[e.g.][]{2020A&C....3000334B, 2022MNRAS.509.3966W}, though their interpretability depends strongly on the adopted training sets and classification schema, thus requiring extra caution. 

In this first paper, we provide a homogeneous, publicly available, catalog of non-parametric morphological indices for all galaxies below $z\leq 0.15$ in the Dark Energy Camera Legacy Survey \citep[DECaLS,][]{2019AJ....157..168D} observed in the $r$ band. The measurements are produced with the newly developed Galaxy Morphology Extractor (\texttt{galmex}) package, that, differently from the available codes in the literature, it has a modular structure that allows fine-tuning of every image pre-processing steps, and metric definitions. This structure is particularly suitable for delivering reliable $\rm CA[A_S]S+MEGG$ indices with flexible options. Focusing on this catalog, we limit this first paper to the fundamental separation between spirals and ellipticals. Using Galaxy Zoo classifications as training labels, we employ Light Gradient Boosted Machines (LightGBM) to derive probabilistic classifications for all galaxies in DECaLS, calibrated directly in the non-parametric parameter space. The treatment of disturbed systems---including mergers, tidally perturbed, and ram-pressure-stripped galaxies---will be presented in forthcoming work (Sampaio et al. in prep), just as the extension of this method towards higher redshifts (Vélliz Astudillo et al. in prep.).

This paper is organized as follows. Section 2 describes our data selection from DECaLS, the definition of our labeled spiral and elliptical and spiral control samples, and the adopted morphological indicators. Section 3 introduces the \texttt{galmex} package and its preprocessing and measurement procedures. Section 4 evaluates the performance of the indices and their consistency with previous classifications. Section 5 applies these metrics to a Light Gradient Boosting Machine (LightGBM) to derive probabilistic classifications up to $z = 0.15$. Section 6 summarizes our conclusions. We assume a flat $\Lambda$CDM cosmology with $[\Omega_{M}, \Omega_{\Lambda}, H_{0}] = [0.27, 0.73, 72 , {\rm km , s^{-1} , Mpc^{-1}}]$ \citep{2016A&A...594A..13P}, and report magnitudes in the AB system.

\section{Data}
\label{sec:data}
To develop our galaxy classification technique,  we select galaxies from the DECaLS\footnote{Data products were retrieved from the Legacy Surveys Data Release~10, available at \url{https://www.legacysurvey.org}}, in the r band. The choice of the Legacy sample is motivated by the combination of large sky coverage, good depth and multi-wavelength coverage achieved by the survey in the southern hemisphere. Additionally, it also has a substantial overlap with upcoming 4MOST spectroscopic surveys (e.g. CHileAN Cluster Evolution Survey -- CHANCES, \citealt{2023Msngr.190...31H};  WHT Enhanced Area Velocity Explorer -- WEAVE, \citealt{2024MNRAS.530.2688J}), thus being fundamental a reliable morphological classification of systems in the southern hemisphere. 

Given that LS–DR10 reaches a median $5\sigma$ depth of 23.5 in the r-band, with nearly uniform image quality across the footprint, we impose a bright magnitude limit of $m_{\rm r}\leq21$. This places our galaxies more than 2 magnitudes above the nominal survey depth, ensuring high S/N per pixel in both the central regions and the outskirts. Furthermore, non-parametric indices are intrinsically pixel–based measurements, and their reliability deteriorates rapidly as the number of galaxy pixels decreases. Thus, by requiring effective radius\footnote{The effective radius is provided by the Legacy Survey database in the column \texttt{shape\_r}} greater than 2 arcsec we minimize the biases and increased scatter that arise when these indices are estimated for barely resolved, undersampled systems. Finally, to avoid galaxies dominated by the effect of the PSF, which can also deeply influence the non-parametric indices estimation \citep{2022MNRAS.509.3966W}, we only select galaxies with $K\geq20$, where K is defined as
\begin{equation}
    K = \left(\frac{4\times R_{\rm e}}{FWHM}\right)^2,
\end{equation}
where $R_{\rm e}$ and $FWHM$ are the effective radius and the point spread function full width at half maximum ($\sim 1.3$ arcsec for DECam in the r-band), respectively. This results in an initial sample of 6,716,178 galaxies.

\subsection{Observational limits of DECam}
\label{sec:obs_limits}
\begin{figure}
    \centering
    \includegraphics[width = 0.7\columnwidth]{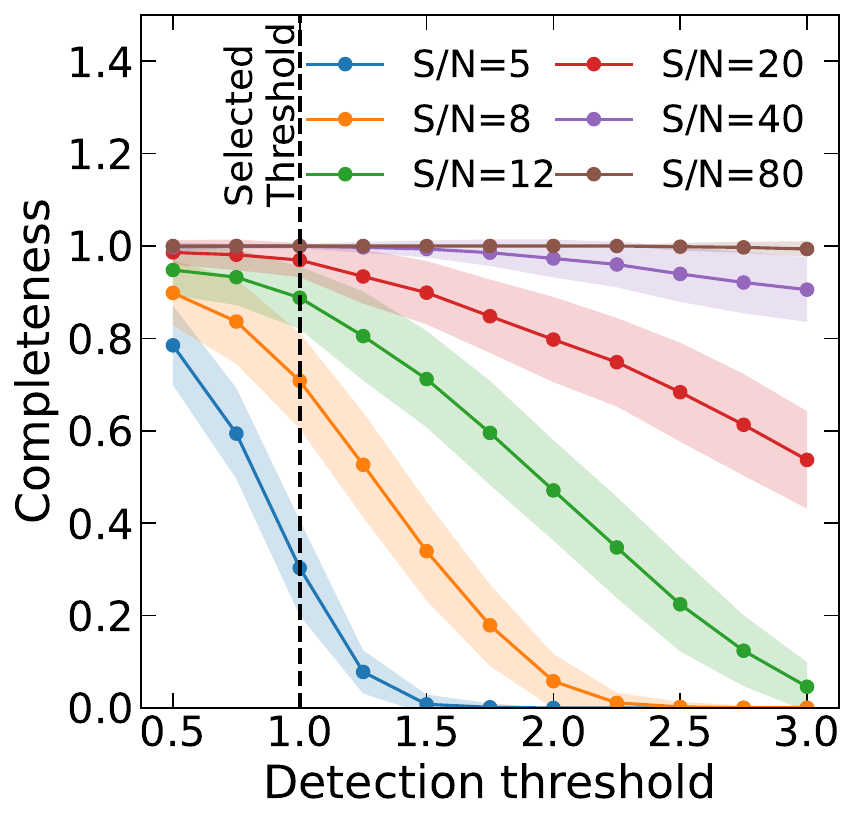}
    \caption{Detection completeness as a function of the \textsc{SExtractor} detection threshold $k$ (in units of the background rms) for simulated S\'ersic galaxies at different ${\rm S/N}$. Solid lines show the median completeness across $1000$ realizations per ${\rm S/N}$; shaded bands indicate the $1\sigma$ scatter. The vertical dashed line marks our adopted threshold $k=1$, which maintains $\gtrsim70\%$ completeness for ${\rm S/N}\ge8$ while limiting spurious detections.}
    \label{fig:SN_threshold}
\end{figure}

To investigate the completeness and the limiting surface brightness that we provide reliable classifications, we carried out controlled simulations of galaxies modeled with Sérsic profiles spanning a wide range of Sérsic indices ($n$), ellipticities, position angles, and redshifts. We first quantify how detection completeness depends on the \textsc{SExtractor} detection threshold. For each ${\rm S/N}\in\{5,\,8,\,12,\,20,\,40,\,80\}$ we simulate $1000$ S\'ersic galaxies with parameters drawn uniformly from $1\le n\le5$, $1\le R_{\rm eff}\le10$~arcsec, axis ratio $0.3\le q\le1$, and position angle $0^\circ\le\theta\le90^\circ$. Completeness is defined as the fraction of input sources recovered by the detection algorithm (\textsc{SExtractor}). As shown in Fig.~\ref{fig:SN_threshold}, increasing the threshold suppresses detections at low ${\rm S/N}$, while high-${\rm S/N}$ sources remain nearly unaffected. Guided by these curves, we adopt a threshold of $k=1$ (in units of the background rms), which preserves a completeness $\gtrsim95\%$ for ${\rm S/N}\ge20$ while limiting spurious detections from background fluctuations. Thus, we remove galaxies with S/N smaller than 20 from our sample (2\%).

\begin{figure*}
    \centering
    \includegraphics[width=0.8\textwidth]{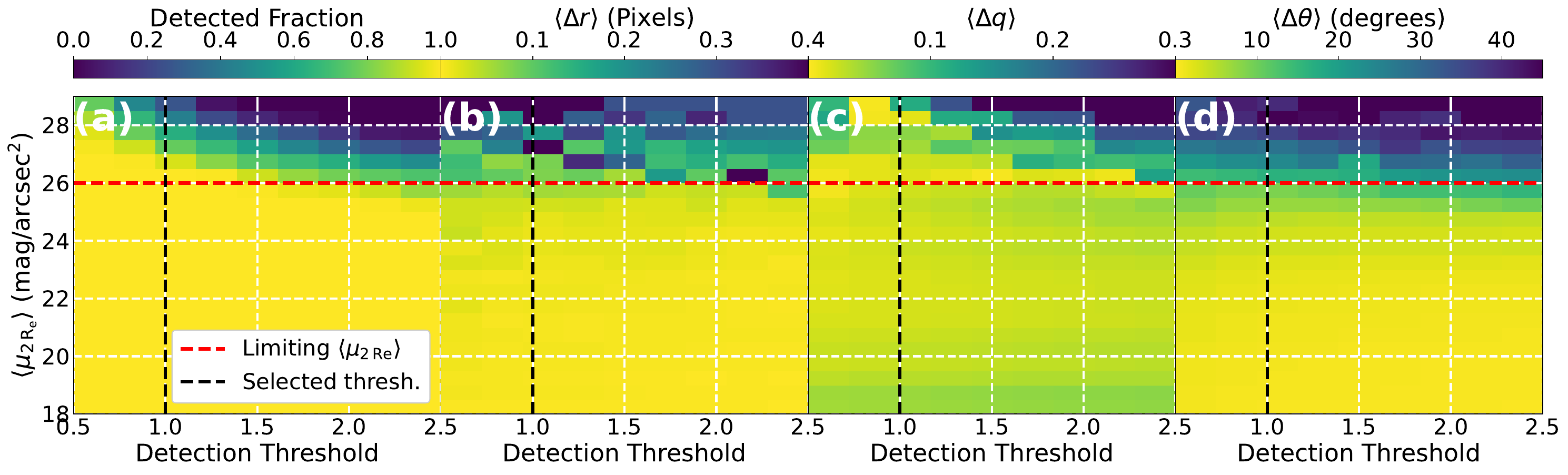}
    \caption{Panel (a): Detection completeness in the $\langle\mu_{\rm 2Re}\rangle$ \textit{vs.} detection threshold. Panels (b), (c), and (d) average difference between true and measured central position, eccentricity, and position angle, respectively, in the same grid as panel (a). We also highlight two different lines: 1) the black dashed line shows the detection threshold adopted in our pipeline; and 2) the red dashed line shows the conservative threshold in surface brightness, such that we can still recover reliable galaxy properties.}
    \label{fig:detection_grid}
\end{figure*}

Second, we investigate how the combination of detection threshold and object surface brightness can affect both detection completeness and shape parameters estimates (central coordinate -- $r$, eccentricity -- $e$ -- and position angle -- $\theta$). In Fig.~\ref{fig:detection_grid} we show the variations of such parameters in the mean surface brightness within $2R_{\rm e}$ ($\langle\mu_{\rm 2R_{\rm eff}}\rangle$, in mag\,arcsec$^{-2}$) \textit{vs.} the detection threshold. Each cell is colored with the average difference (across different Sérsic indexes) between true and measured values. The vertical black dashed line marks the threshold adopted in our pipeline. Furthermore, we also highlight that the main differences between true and measured properties occur for objects with $\langle\mu_{\rm 2R_{\rm eff}}\rangle$ fainter than 26 mag\,arcsec$^{-2}$. Therefore, we limit our analysis to objects brighter than $\langle\mu_{\rm 2R_{\rm eff}}\rangle$ = 26 mag\,arcsec$^{-2}$, which is highlighted by the horizontal red dashed line, and reduces our sample to 6,088,103 galaxies.

\subsection{Defining labeled subsamples}

Despite providing morphology for all galaxies, the morphological classification using non-parametric indexes relies on a labeled dataset to define the separations between different morphological types. In this subsection we describe the definition of spiral and elliptical subsamples, which will be used as the basis to derive morphological probabilities for our entire galaxy set. 

\subsubsection{Visual morphologies from Galaxy Zoo}

A natural first step in morphological analysis is the binary classification between spirals and ellipticals. In the context of DECaLS, the Galaxy Zoo-DECaLS (GZ DECaLS, hereon) project \citep{2022MNRAS.509.3966W} provides large-scale visual classifications. However, the classification scheme adopted in GZ DECaLS classify galaxies between "smooth" or "disk/feature", which is not a direct mapping onto "spiral" or "elliptical". Notably, the separation between "smooth" vs "disk/feature" is considerably subjective and not necessarily exclusive. For example, a disk-dominated system may be classified as ``smooth'' if the disk lacks clear features, while some bulge-dominated galaxies may still receive non-negligible ``disk/feature'' votes.

\begin{figure}
    \centering
    \includegraphics[width=0.7\linewidth]{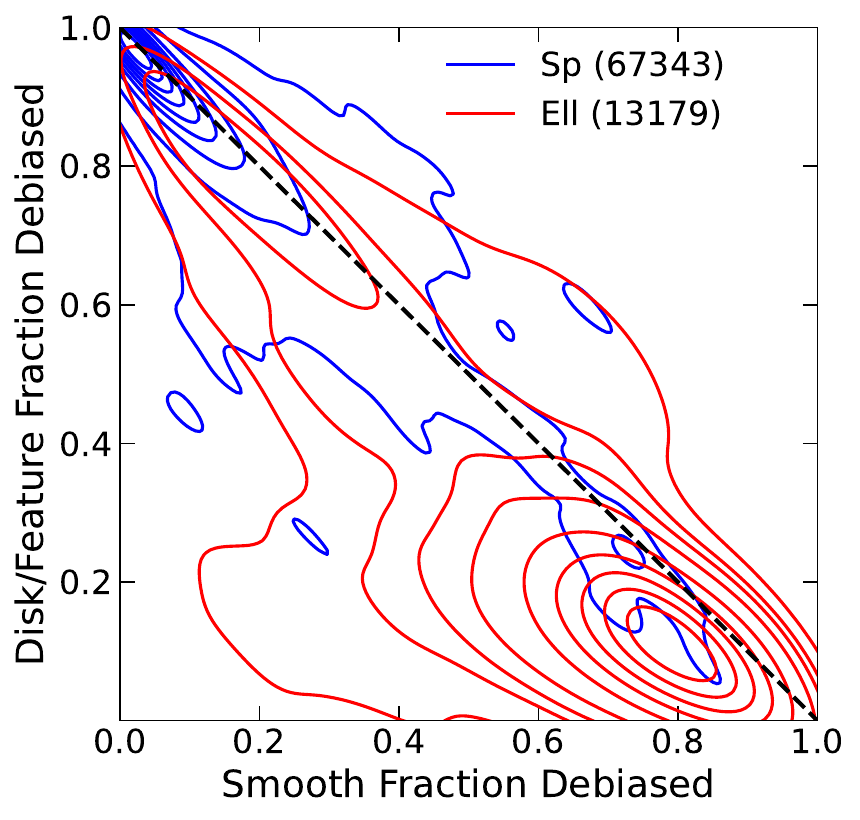}
    \caption{Distribution of GZ 1 selected spiral and elliptical sub-samples in the $f_{\rm smooth}$ versus $f_{\rm disk}$ (see text for the definition) diagram, according to GZ DECaLS results. The dashed black line shows the expected anti-correlation line.}
    \label{fig:contour_plot}
\end{figure}

We therefore turn to the original Galaxy Zoo 1 (GZ1, hereafter) project \citep{2008MNRAS.389.1179L}, which provides explicit spiral and elliptical classifications for SDSS galaxies. Hereafter, we define the spiral (simply "Sp" hereon) and elliptical ("Ell" hereafter) subsamples according to the GZ 1 project\footnote{The GZ 1 project provides direct classifcation of spirals and ellipticals, avoiding the need to adopt a threshold in the vote fraction.}, and that are also on the DECaLS footprint. Since the difference between SDSS and DECam pixels scales are somewhat small (0.396 vs. 0.261 px, respectively), and they have comparable PSFs in the r-band (1.18" for DECaLS vs. 1.32" for SDSS) we don't expect these labels to change between surveys. This is reinforced by Fig.~\ref{fig:contour_plot}, in which we show the distribution of the elliptical and spiral subsamples in the top-level classification scheme of GZ DECaLS. Namely, we define $f_{\rm smooth}$ (x-axis), and $f_{\rm disk}$ (y-axis) as the debiased fraction\footnote{In practice, galaxies are binned by absolute magnitude and physical size; within each bin and for each answer ("smooth" or "features/disk"), the vote-fraction distributions at each redshift are shifted to match those of the lowest-redshift slice ($0.02 < z < 0.03$), yielding the fraction expected if every galaxy were observed at $z\sim0.02$ and keeping the fraction above any chosen threshold constant with redshift.} of votes that the object is "smooth" or "disk/feature", respectively, in the GZ DECaLS. Both the Sp and Ell samples lie well within the anti-correlation line (black dashed line), even though ellipticals show a larger spread, highlighting that these are robust subsamples even though having their label defined in a different survey. 

We use Sp and Ell galaxies as control samples, and also as benchmarks for calibrating non-parametric morphological indices. In this first paper, we will focus on providing the morphology for galaxies within the redshift coverage of both GZ 1 and GZ DECaLS\footnote{In the original GZ DECaLS catalog, z is the spectroscopic redshift retrieved from the Nasa-Sloan Atlas catalog \citep[NSAtlas][]{2011AJ....142...31B}.}, namely systems below redshift 0.15. Imposing this redshift cut, we end up with a control sample of 80,516 galaxies\footnote{Although the number of spiral galaxies is about six times that of ellipticals (13,179 Ell and 67,343 Sp), we explain how we address this imbalance in Sec.~\ref{sec:def_training}}. For completeness, a detailed comparison between GZ1 and GZ DECaLS is presented in Appendix \ref{ap:GZ_comparison}. In a few words, our analysis shows that differences in the Galaxy Zoo classification schemes can significantly impact the purity of selected samples. Furthermore, the extension of our method to higher redshifts (up to 0.5) and the impact of redshift in the non-parametric indices will be discussed in a future paper (Vélliz Astudillo et al. in prep.). Yet, by artificially redshifting galaxies closer than $z<0.03$ to $z=0.15$, in steps of $0.03$, we find that the metrics do not vary by more than 10\%, thus ensuring consistency across the entire redshift range.

Finally, the redshift limit in the control sample implies that, for consistency, we must limit the galaxies that we classify to the same redshift range. Although spectroscopic redshift is only available for 7\% of our sample, we apply this cut using the photometric redshift, which is shown to be consistent with the spec z (see Appendix~\ref{ap:redshift}). A caveat of adopting the labels from GZ 1 is that it is limited in magnitude to 17.78 in the r-band, whereas the DECaLS is able to provide deeper observations. Therefore, we adopt the upcomming CHANCES low-z sub survey conservative magnitude limit of 18.5 (Méndez-Hernández et al. in prep.) for our sample, resulting in a final sample of 1,744,454 galaxies (from which 80,516 are labeled between "spiral" or "elliptical").

\subsubsection{Automated classifications from deep learning}

Beyond direct visual classifications, we also incorporate automated morphological estimates obtained with convolutional neural networks (CNNs), in order to compare the indices performance with both visual inspection from GZ DECaLS, and as a function of T-Type (Sec.~\ref{sec:comparison_previous}). To this end, we adopt the catalog of \citet{2018MNRAS.476.3661D}, that train CNNs on Galaxy Zoo~2 questions to predict a continuous T-Type for SDSS galaxies, encompassing both our Sp and Ell subsamples. T-Type is estimated as a continuous numerical proxy for the classical Hubble sequence, through the equation:
\begin{equation}
    \text{T-Type} = -4.6P(\text{Ell}) -2.4P(\text{S0}) + 2.5P(\text{Sab}) + 6.1P(\text{Scd}),
\end{equation}
where $P({\rm X})$ denotes the CNN attributed probability of a galaxy being classified as a given morphology, with X representing Elliptical (Ell), lenticular (S0), A-B spiral (Sab), and C-D spiral (Scd). This provides a quantitative mapping onto the classical Hubble sequence, ranging from ellipticals ($\text{T-Type} \approx -3$) through lenticulars ($\text{T-Type} \approx -0$) and spirals ($\text{T-Type} \approx 1-5$). We highlight that we do not use the T-Type as a label in any step of our method, being included in the catalog only for connecting non-parametric indices to previous machine-learning classifications of galaxy morphology.

\section{Non-parametric morphological estimation}

We chose a non-parametric method to characterize the structure of galaxies, given that they do not rely on any assumption about the light profile of the observed galaxies, have direct physical interpretation and extensive use in the literature to connect structural parameters and galaxy evolution related mechanisms \citep{1996MNRAS.279L..47A, 2000ApJ...529..886C,2008ApJ...672..177L, 2008MNRAS.386..909C}.  However, a fundamental step when measuring non-parametric indexes is the need for image preprocessing. Here we present our own python package (Sec.~\ref{sec:galmex}) to perform image processing and metrics measurements. The choice of creating our own code is to ensure transparency and the need for fine-tuning, which is not found in non-modular existing codes with the same purpose \citep[e.g.][]{2015ApJ...814...55F,2019MNRAS.483.4140R}.

\subsection{The \texttt{galmex} package}
\label{sec:galmex}
The Galaxy Morphology Extractor\footnote{A full tutorial and description are available in \href{https://galmex.readthedocs.io/en/latest/}{"read the docs"} or \href{https://github.com/vitorms99/galmex}{github}.} (\texttt{galmex}) is a user-friendly Python package designed to reliably estimate non-parametric morphological indices from imaging surveys. The code is designed with a modular architecture, allowing each stage (preprocessing, segmentation, measurement, output) to be accessed independently. Users can therefore customize the workflow, integrate new routines, or apply only a subset of the available tools. In addition to a command-line interface (CLI) optimized for large-scale processing, \texttt{galmex} also includes a graphical user interface (GUI) for more interactive analysis and visualization. This design makes the package suitable both for bulk catalog production and for detailed inspection of individual galaxies. Next we detail the pre-processing steps adopted prior to measuring the indexes: 

\begin{enumerate}
    \item Cutout creation -- For each target we read right ascension, declination, and a prior Petrosian angular scale from the input catalog, and then request the stamp in the r-band from the Legacy Survey (DR10) cutout service. The linear size of the cutout in pixels is set as the the reported effective radius multiplied by a factor of 20 (10 effective radius around the galaxy);
    
    \item Background subtraction -- We estimate and remove the sky using a frame-based statistic around the image edges, since our cutouts are made with size given as a function of the effective radius of the galaxy ($10\times R_{eff}$). Specifically, we select a border containing a fixed fraction of the image area and compute background statistics on those pixels with sigma-clipping enabled to suppress contamination from secondary sources near the image border. In practice we set the frame width by an image-area fraction of 0.2, enable sigma-clipping, and reject pixels above a $2.5\sigma$ threshold; the resulting background model is subtracted from the science image to produce a background-subtracted frame for all subsequent steps;

    \item Object detection -- Sources are identified on the background-subtracted image with the analogue of Source Extractor \citep{1996A&AS..117..393B}, transcribed to python -- SEP  \citep[SExtractor-in-Python,][]{2016zndo....159035B} -- using a matched-filter option. We adopt a per-pixel detection threshold of $1.0\sigma$ relative to the measured background noise, require a minimum footprint of 10 connected pixels, and deblend with 32 thresholds at a contrast parameter of 0.005; we pass SEP the measured background standard deviation so its internal thresholding is on the correct noise scale. SEP returns a normalized catalog (centroid x, y; ellipse a, b; position angle $\theta$ (in radians); npix; mag) and a first segmentation map. The primary galaxy is selected as the label at the cutout center; if the center falls on background, an error is raised highlighting that no object was detected at the image center.

    \item Cleaning (removal of secondaries) -- To mitigate contamination from stars and neighboring galaxies, we generate a cleaned image using an isophotal "painting" procedure that respects the target’s geometry. Starting from the detection segmentation, all labels other than the main object are treated as contaminants; their pixels are replaced via elliptical-isophote interpolation oriented by the galaxy’s position angle $\theta$. Operationally, the algorithm scans concentric elliptical annuli and replaces masked pixels with interpolated values from adjacent pixels along the same isophote, which preserves the target’s radial structure while suppressing flux from secondaries. This yields a “galaxy-only” image used for all light-profile quantities that follow;

    \item Characteristic radii estimation -- We compute Petrosian profiles on the cleaned image using both circular and elliptical annuli, anchored to the SEP-measured center (x, y), axes (a, b), and position angle $\theta$. The Petrosian radius ($R_{P}$) follows the standard $\eta(R) = 0.2$ threshold with an optimized search: a guided (bisection-style) evaluation of the curve with cubic interpolation and taking into account neighboring points ($\text{crossing point} \pm 3$). After $R_{P}$ is determined, we derive the circular and elliptical half-light radii by integrating the growth curve to the 50\% level, restricting the search to $2\times R_{\rm P}$, with a 1-pixel step. We also report a Kron-style radius computed within the same outer bound. This procedure is executed twice—first with circular annuli and then with elliptical annuli—so different analyses can use the most appropriate geometry.

\end{enumerate}

An example of the pre-processing procedure is shown in Fig.~\ref{ap:image_preprocessing}.

\subsection{Robustness of \texttt{galmex} applied to DECam images}

\begin{figure*}
    \centering
    \includegraphics[width=0.8\textwidth]{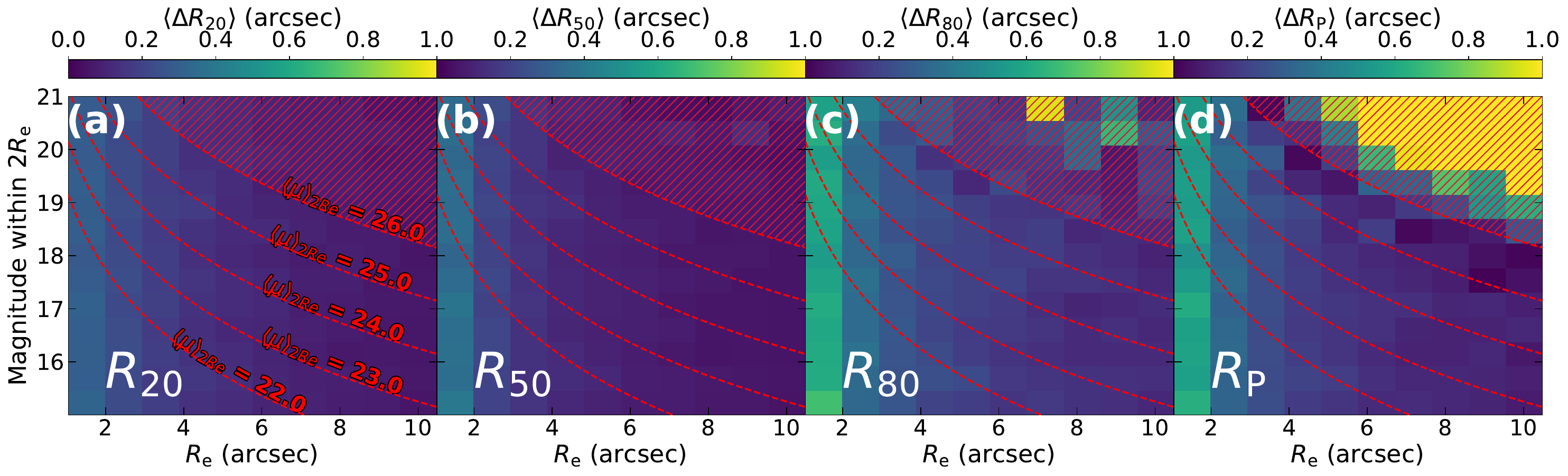}
    \caption{Recovery of characteristic radii across size–flux space. Each panel shows the map of the average absolute difference (in arcsec) between the measured and reference values of a given radius -- $R_{20}$ (a), $R_{50}$ (b), $R_{80}$ (c), $R_{\rm P}$ (d) in the apparent magnitude \textit{vs.} $R_{\rm e}$. The dashed red lines define the approximate average surface brightness when assuming a circular (q = 1) Sérsic profile. The hatched region above the $\langle \mu_{\rm 2Re}\rangle$ denotes the adopted threshold in this work. Galaxies with surface brightness smaller than 26 mag\,s$^{-2}$ can yield unreliable shape parameters and characteristic radii. In particular, the hatched region overlaps significantly with the region in which the error in $R_{\rm P}$ exceeds 1 arcsec ($\sim 4$ pixels). This effect is more visible in $R_{\rm P}$ due to it being the most outter radii in comparison to the others, thus being more prone to background contamination.}
    \label{fig:radii_grid}
\end{figure*}

In this section we test how well \texttt{galmex} recover galaxy properties using the DECam-like simulated Sérsic profiles described in Sec.~\ref{sec:obs_limits}. In particular we focus on the radii encompassing 20, 50, and 80\% of total flux, due to its tracing of the growth curve and direct relation to the Concentration index, and the Petrosian radius\footnote{Notably, \texttt{galmex} rely only on the shape properties estimated during object detection step to reliably calculate the Petrosian radius, whereas \texttt{statmorph} \citep{2019MNRAS.483.4140R} estimate may depend also on the segmentation mask used as input.}, which is extensively used in the literature to define the segmentation mask -- i.e. the region that will be taken into account in metrics computation. In Fig.~\ref{fig:radii_grid}, we show the average difference between true and measured $R_{20}$ (panel a), $R_{50}$ (panel b), $R_{80}$ (panel c), and $R_{\rm P}$ (panel d), in the apparent magnitude \textit{vs.} effective radius grid. For these computations, we use elliptical apertures, and discuss in Appendix~\ref{ap:rp} how the use of circular apertures to calculate characteristic radii can introduce significant bias in the analysis. Notably, the combination of apparent magnitude and effective radius define an average surface brightness\footnote{The surface brightness depends on the eccentricity of the object, but we adopt a simple case of circular Sérsic profile ($q = 1$).}, which is shown by the red dashed lines. The red hatched region denotes the region fainter than our adopted limit in average surface brightness (26 mag\,arcsec$^{-2}$). Noteworthy, this is the region where we find the larger offsets (particularly in panel d), again reinforcing that our adopted thresholds ensure that we are providing reliable metrics for all the objects. In particular, Fig~\ref{fig:radii_grid} reveals that we recover the characteristic radii with average differences smaller than 0.6 arcsec in most of the cases, which corresponds for a difference of 2.3 pixels in the DECam resolution ($\sim 0.262$ pixels/arcsec).

Finally, a key step in the computation of non-parametric indices is the definition of the segmentation mask. To ensure consistency across galaxies of different magnitudes and redshifts, we compared the mean pixel intensity in the r-band as a function of radius, written as a function of the Petrosian radius ($k \times R_{\rm p}$). By scaling the mask with $R_{\rm p}$, we guarantee a relative aperture size that adapts to the galaxy’s intrinsic light profile, providing a homogeneous basis for comparison. Following \cite{2024MNRAS.528...82K, 2004AJ....128..163L}, we define the conservative threshold of $k=1$. We highlight that, differently from \texttt{statmorph}, we use the same segmentation mask for all the metrics, which also ensure a more direct interpretability of their performance in separating ellipticals and spirals. For completeness, we show in Appendix~\ref{ap:segmentation} how segmentation affects our results, in particular on how segmentation affect the separation bewteen spirals and ellipticals in the non-parametric indices diagrams.

\begin{figure*}
    \centering
    \includegraphics[width = 0.9\textwidth]{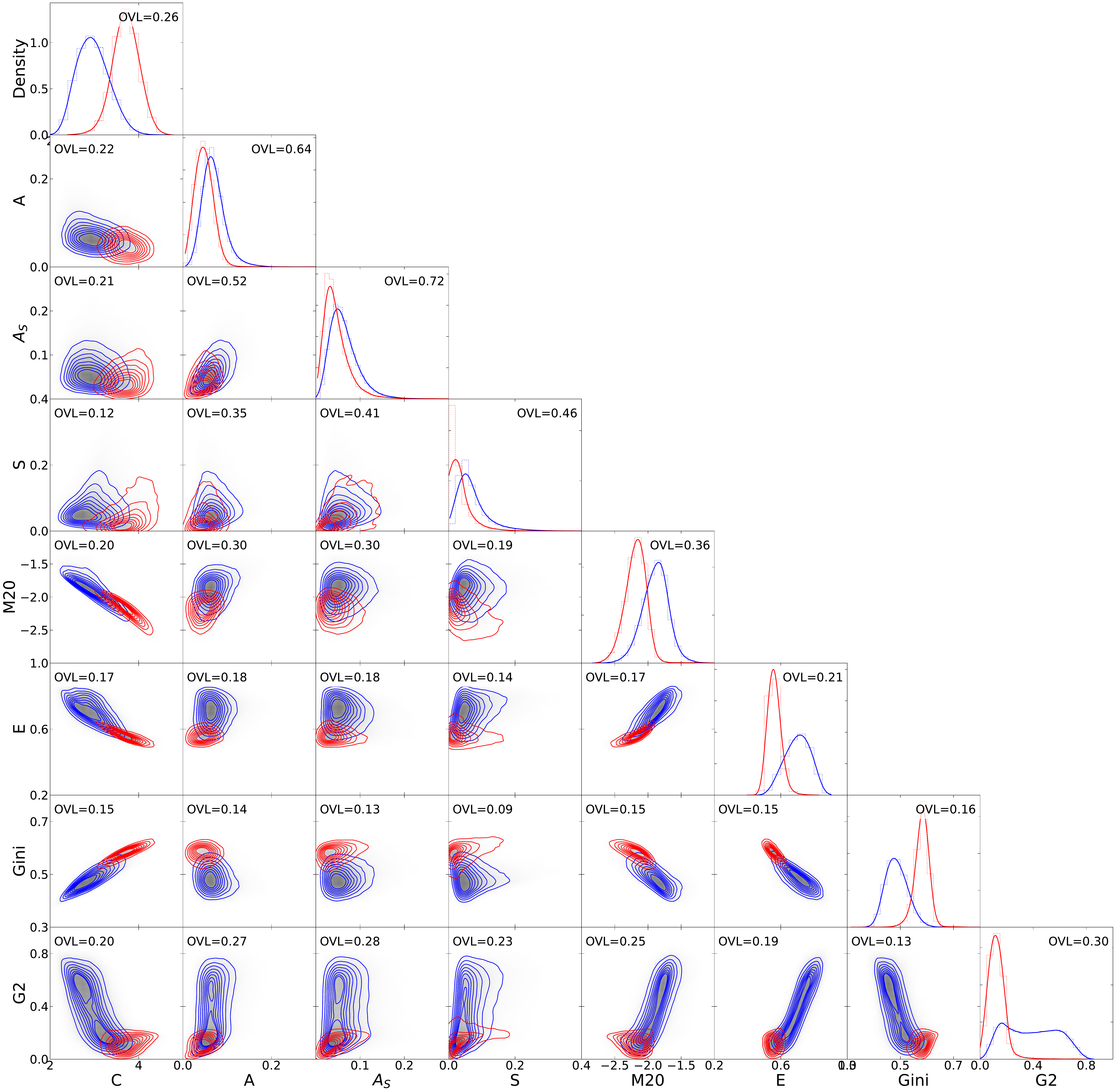}
    \caption{Distribution of Spiral (blue curves) and Elliptical (red curve) galaxies in 2D diagrams combing the different non-parametric indices. In each panel, we also include the overlap between the spiral and elliptical distributions, which is calculated using equations \ref{eq:OVL_1D} and \ref{eq:OVL_2D} for histograms and 2D diagrams, respectively.}
    \label{fig:2d_hist}
\end{figure*}

\section{Results}

\subsection{Non-parametric morphological properties of galaxies}

Figure~\ref{fig:2d_hist} shows the one and two-dimensional distributions of spiral and elliptical control samples across the CAS and MEGG parameter spaces. The contours highlight the normalized density distributions for each class (15, 25, 50, 60, 70, 80, 90, and 95\%), enabling a quantitative comparison of their separation. Overall, the C[$\rm A_S$]AS parameters retain their classical behavior. Concentration shows the strongest discriminatory power, with ellipticals occupying systematically higher values than spirals, consistent with their centrally concentrated light profiles. Asymmetry ($\rm A$ and $\rm A_S$) and Smoothness are more effective at rejecting extreme outliers (e.g.\ mergers), but their distributions overlap significantly between spirals and ellipticals, limiting their power as stand-alone classifiers. This behavior has been reported in previous works \citep[e.g.][]{2024MNRAS.528...82K}, and is confirmed here with the larger DECaLS samples.

The MEGG parameters provide complementary information. The Gini index and entropy exhibit clear trends, with ellipticals clustering at high $\rm G$ and low $\rm E$, while spirals show the opposite behavior. The M20 parameter retains sensitivity to bright off-center regions, helping to separate star-forming disks from smooth spheroids, although with substantial overlap. The E index stands out as the most effective single discriminator: spirals and ellipticals are distributed with minimal overlap. This corroborates previous results that the MEGG system provide robust morphological separation in both local and intermediate-redshift samples \citep{2020A&C....3000334B, 2024MNRAS.528...82K, 2025MNRAS.539.2765K}.

To move beyond a purely visual comparison, we quantified the degree of overlap between the spiral and elliptical distributions using the overlap coefficient (OVL). For a single index $X$, we computed normalized histograms on shared bin edges for each class and defined the 1D overlap as
\begin{equation}
\label{eq:OVL_1D}
\mathrm{OVL}_{\mathrm{1D}}(X) = \sum_{k=1}^{K} \min \big[ p_k(X),\, q_k(X) \big],
\end{equation}
where $p_k$ and $q_k$ are the spiral and elliptical probabilities in bin $k$. Values close to unity indicate nearly indistinguishable distributions, while values near zero indicate strong separation. For a indices-pair $(X,Y)$, we applied an empirical probability–integral transform to each axis, mapping both classes onto the unit square $(0,1)^2$, and then computed a two-dimensional histogram intersection,
\begin{equation}
\label{eq:OVL_2D}
\mathrm{OVL}_{\mathrm{2D}}(X,Y) = \sum_{i,j} \min \big[ P_{ij}(U,V),\, Q_{ij}(U,V) \big],
\end{equation}
with $P_{ij}$ and $Q_{ij}$ being the spiral and elliptical probabilities in bin $(i,j)$. This normalization ensures that OVL values are comparable across different index pairs.

Quantitatively, the most effective single indices are Concentration, Entropy, and Gini, with $\mathrm{OVL}_{\mathrm{1D}} \simeq 0.18-0.21$, followed by M20 and G2 with $\mathrm{OVL}_{\mathrm{1D}} \simeq 0.26-0.27$. Asymmetry, Shape Asymmetry, and Smoothness show substantially larger overlaps ($\gtrsim 0.5$), confirming that they are better suited to identifying disturbed morphologies than to separating spirals from ellipticals. Among all 2D projections, the best separation is found for the involving the Gini index, showcasing that this flux-inequality measure is reliable when separating late- and early-type galaxies.

While empirical linear divisions in each 2D plane to separate spirals and ellipticals thresholds can be drawn, the overlap between the distributions, particularly in $\rm A, A_{S}$ and S, suggests that no single cut provides a reliable classification. Instead, the joint use of CA$\rm [A_S]$S+MEGG indices in a probabilistic framework (Sect.~\ref{sec:lightgbm}) provides a more robust approach to assigning morphological classes. In summary, the CA[$\rm A_S$]S parameters reproduce the expected trends but with considerable overlap, while the MEGG indices---especially E and G---deliver superior discriminatory power. 

\begin{figure*}
    \centering
    \includegraphics[width=0.9\linewidth]{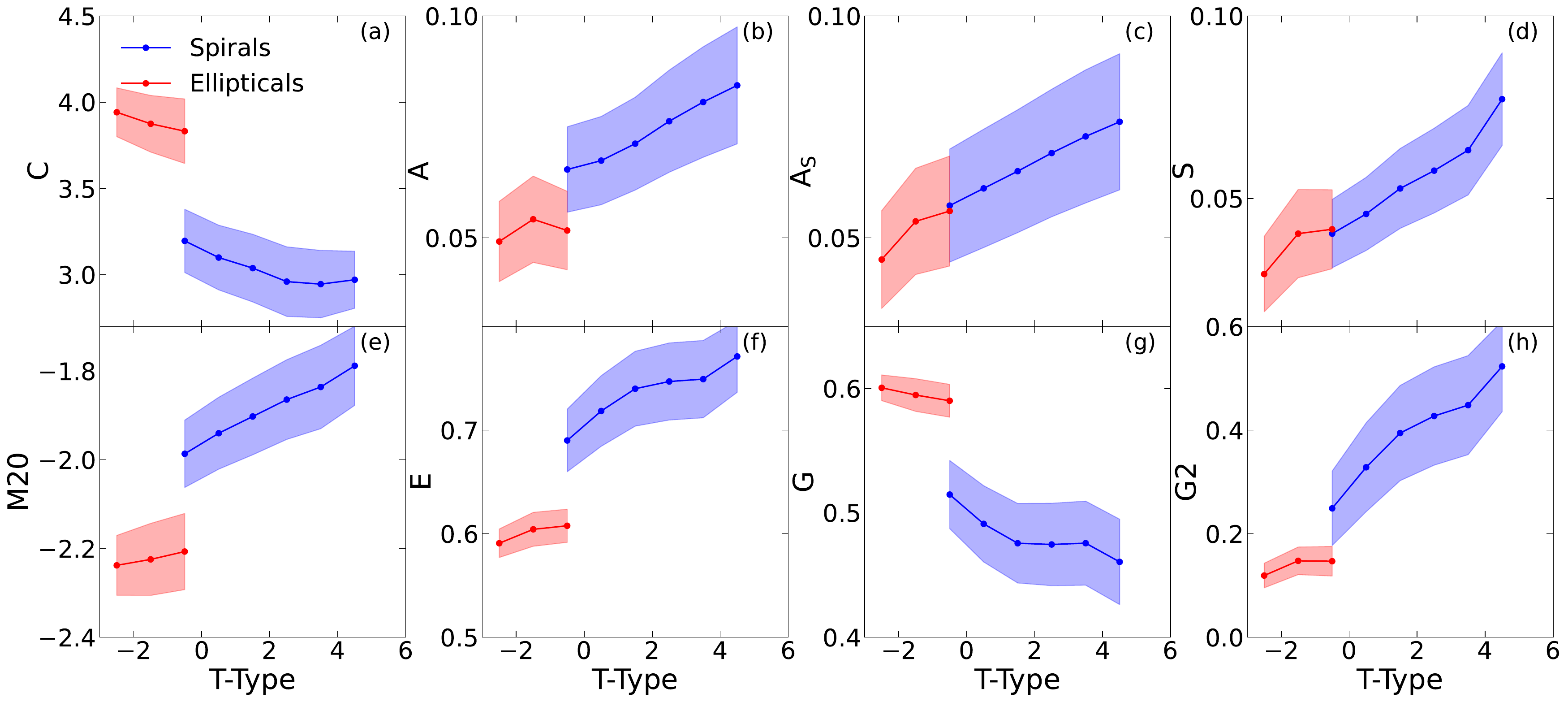}
    \caption{Metrics variations with respect to CNN based T-Type for the Sp (blue) and Ell (red) subsamples. We highlight that, although both the Sp/Ell classification and T-Type are based on SDSS data (Galaxy Zoo 1 and 2, respectively), our results show consistency even when using DECam observations.}
    \label{fig:ttype_variation}
\end{figure*}

\subsection{Comparison to previous classifications}
\label{sec:comparison_previous}
In this subsection we compare the CA$\rm [A_S]$S+MEGG indices with two independent morphological classification schemes in order to place them on a common scale and test their consistency. First, we investigate their variation as a function of the CNN based T-Type from \citet{2018MNRAS.476.3661D}. This allows us to assess whether the indices trace the expected early--to--late morphological sequence in a monotonic way. Second, we examine how the same indices vary across the GZ-DECaLS top-level separation ($f_{\rm smooth}$ vs. $f_{\rm disk}$). 

\subsubsection{C[$\rm A_S$]AS+MEGG vs. CNN based T-Type}

Figure~\ref{fig:ttype_variation} shows the median values and $1\sigma$ scatter of the C[$\rm A_S$]AS and MEGG indices as a function of CNN-based T-Type. For robustness, medians and scatters are computed only for T-Type bins containing at least 1\% of the corresponding control subsample (Sp or Ell). As a first check, we confirm that the GZ 1 control samples are fully consistent with this scheme: spiral galaxies lie dominantly at $\text{T-Type}>0$, while ellipticals occupy $\text{T-Type}<0$. 

The CAS indices show the expected broad separation between early- and late-type morphologies. Concentration shows a discontinuity separation between early- and late-type morphologies, varying from $\langle {\rm C} \rangle \sim 4.0$ at $\text{T-Type}=-3$ to $\sim 3.0$ at $\text{T-Type}=5$, clearly distinguishing Ell from Sp. This discontinuity may indicate that the T-Type is not as continuous as expected, which may follow from one or a combination of the following reasons: 1) \citet{2018MNRAS.476.3661D} use different CNN models for the $\text{T-Type} \sim 0$ region; 2) the T-Type estimation carries bias from the training dataset; and 3) the equation used to map T-Type continuously is somewhat arbitrary, and not necessarily reflect the continuous transition expected from negative to positive T-Type values. In contrast, A, $\rm A_S$, and S remain nearly constant across $\text{T-Type}<0$, but increase slightly toward later types. Their variation is modest ($\Delta A, \Delta A_S, \Delta S \lesssim 0.08$), consistent with their limited discriminatory power for separating Sp from Ell.

\begin{figure*}
    \centering
    \includegraphics[width=0.8\linewidth]{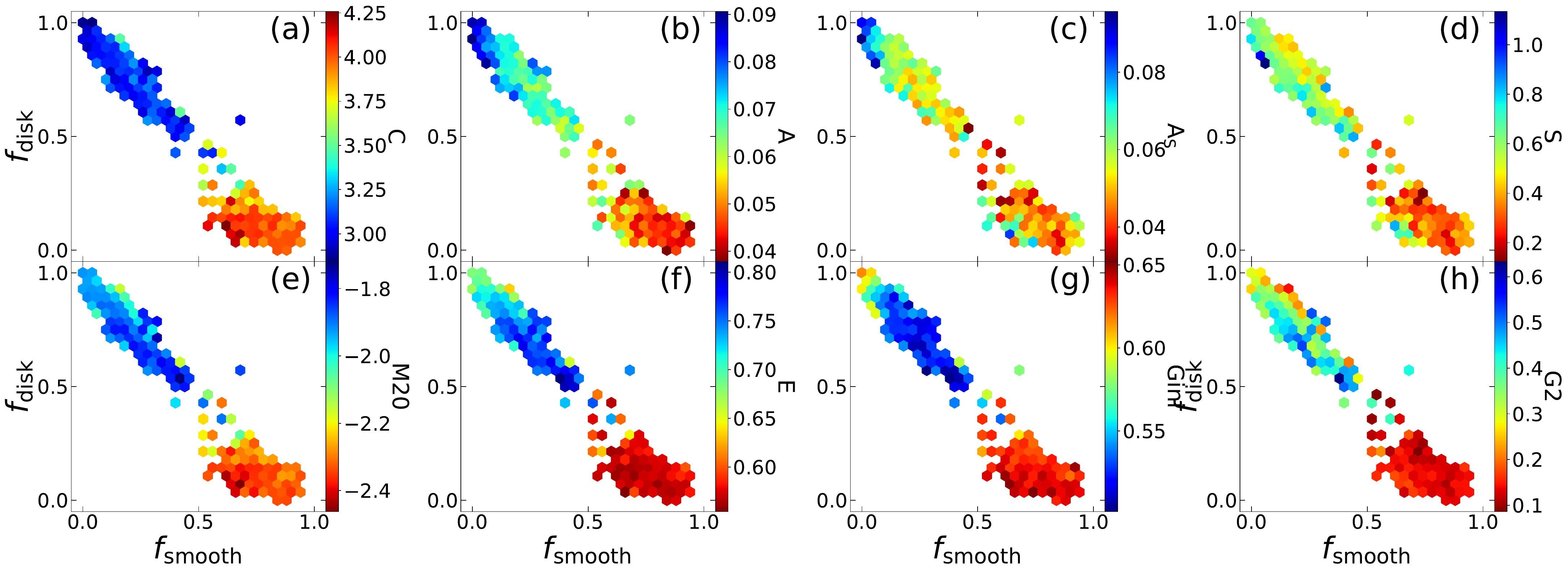}
    \caption{Metrics variation in the Smooth Debiased vs. Disk/Feature Debiased diagram. In this case we merge the spiral and elliptical subsamples in order to get a full picture of the metrics variation across this diagram.}
    \label{fig:frac_variation}
\end{figure*}

The MEGG indices exhibit both clear early–late separation and strong internal trends within the spiral sequence. M20 increases from $\langle {\rm M20} \rangle \simeq -2.3$ at $\text{T-Type}=-3$ to $\simeq -1.8$ at $\text{T-Type}=5$, possibly due to the increasing prominence of bright off-center regions in late-type spirals. E and Gini display steep, opposite variations: ellipticals cluster at high Gini ($\gtrsim 0.6$) and low entropy ($\lesssim 0.6$), while spirals reach $\langle {\rm G} \rangle \simeq 0.45$ and $\langle {\rm E} \rangle \simeq 0.8$ at $\text{T-Type}\sim 5$. The G2 index provides the sharpest discrimination: it remains near zero for ellipticals, increases steadily through early spirals, and reaches $\langle {\rm G2} \rangle \gtrsim 0.45$ for the latest types. This steep gradient at $\text{T-Type}>0$ demonstrates that M20, E, and G not only separates ellipticals from spirals (with a confidence of more than 3-sigma), but also effectively resolves substructure within the spiral sequence.

\subsubsection{Non parametric indices vs. Visual Classification}

Fig~\ref{fig:frac_variation} presents the variation of C$\rm [A_S]$AS and MEGG indices across the GZ-DECaLS $f_{\rm disk}$ vs. $f_{\rm smooth}$ plane. Particularly for Fig.~\ref{fig:frac_variation}, we merge the spiral and elliptical subsamples rather than analyzing them separately, in order to provide a complete view of the parameters variation. We restrict the hexbin maps to bins containing at least ten galaxies, and we scale the colorbars in a consistent way such that regions dominated by ellipticals appear in redder tones.

Overall, the indices vary across this diagram in good agreement with the Galaxy Zoo classifications. Concentration C increases steadily toward the smooth-dominated corner, while E decreases and Gini increases, reproducing the contrast between bulge-dominated and disk-dominated systems. A, $\rm A_S$, and S peak in the high $f_{\rm disk}$ regime, consistent with the visual impression of clumpier and more irregular morphologies. M20 also increases in this region, reflecting the prominence of bright off-center structures in spiral galaxies. Finally, G2 shows a marked gradient from smooth to disk-dominated systems, again underscoring its effectiveness as a discriminator.

These trends demonstrate that non-parametric indices are broadly consistent with human visual assessments from Galaxy Zoo, capturing the same underlying morphological differences directly from the pixel data. In other words, CA$\rm [A_S]$S+MEGG indices to some extent mimic what classifiers perceived by eye. This motivates the next step of our analysis, where we employ these indices as input features for a machine-learning framework (Sect.~\ref{sec:lightgbm}) to assign probabilistic classifications across the full DECaLS sample.

\begin{figure*}
    \centering
    \includegraphics[width=0.8\linewidth]{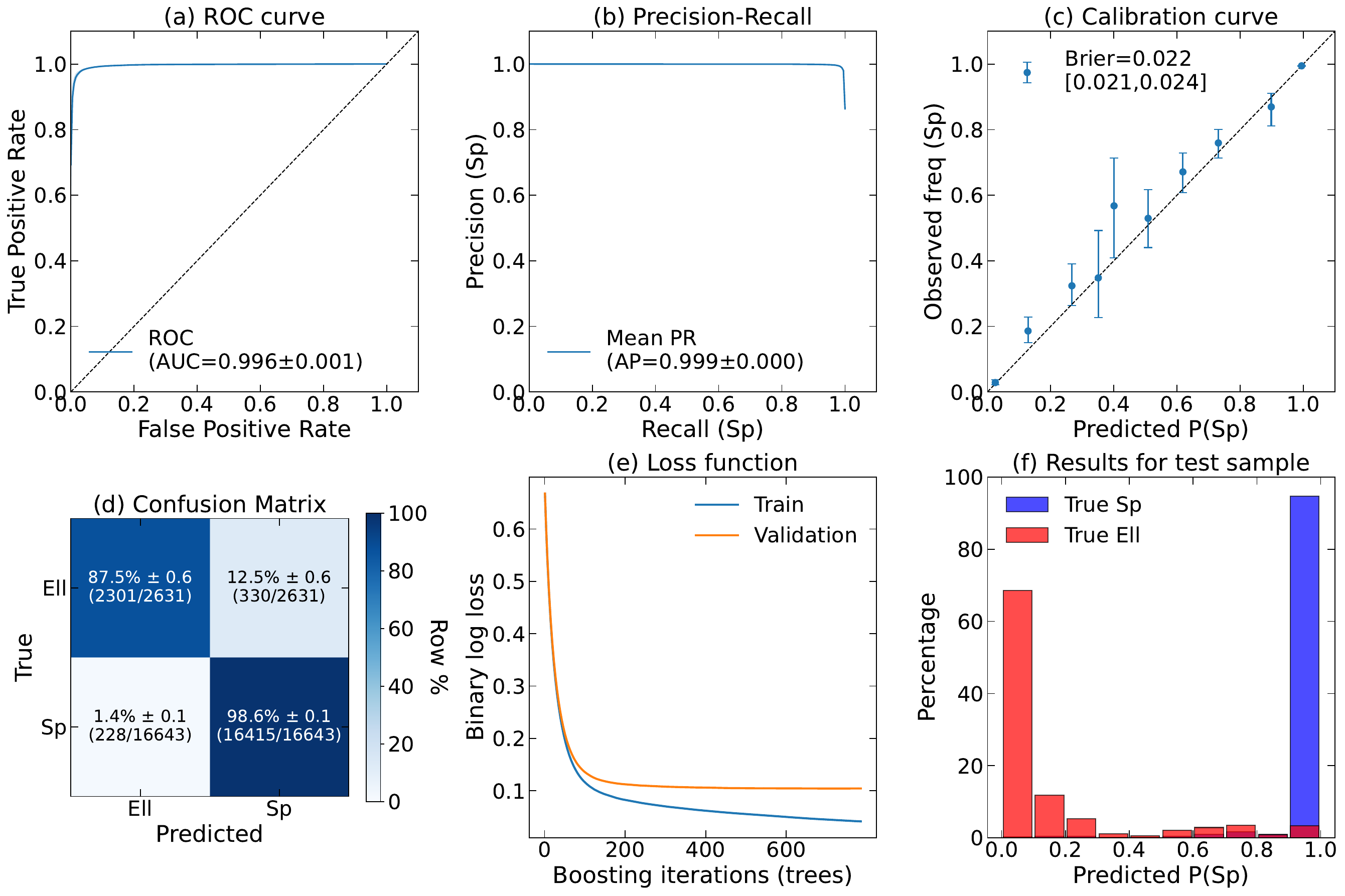}
    \caption{Results of LightGBM using the non-parametric indexes as input, and trained in the GZ 1 selected spirals and ellipticals. Panel (a) and (b) shows the ROC curve and the Precision-recall curve, respectively. Within these two panels, we also add the area under the curve (AUC) and the mean AP. Panel (c) shows the calibration curve, highlighting that our method aligns well with the expected 1 to 1 line, thus ensuring that our method is able to provide calibrated probabilities. Panel (d) shows the row-normalized confusion matrix. Panel (e) shows the log-loss function. See the text for a description on why there is a difference between the train and validation. Panel (f) show the predicted spiral probability for galaxies in our test subsample. Notably, our method shows high accuracy.}
    \label{fig:lightgbm_results}
\end{figure*}

\section{$\rm CA[A_S]S+MEGG$ indices as input for Machine Learning classification}

\label{sec:lightgbm}

\begin{figure}
    \centering
    \includegraphics[width=\columnwidth]{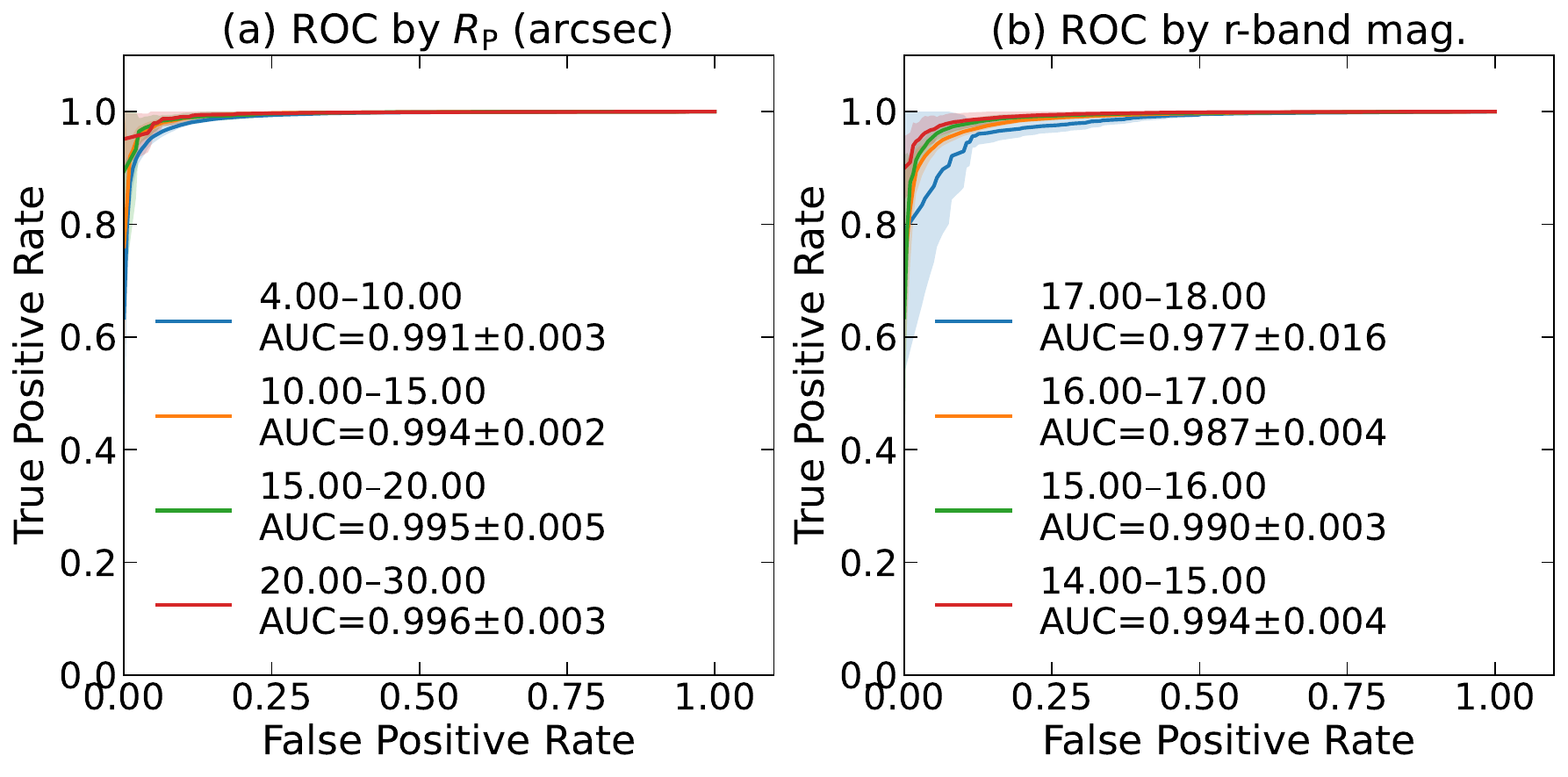}
    \caption{Sanity check of classifier performance as a function of observational regime. Receiver operating characteristic (ROC) curves for the spiral-vs.-smooth classifier evaluated in bins of (a) apparent size, using the Petrosian radius (arcsec), and (b) $r$-band Petrosian magnitude.  Curves show the mean ROC across cross-validation folds, with shaded regions indicating the $\pm 1\sigma$ scatter between folds; the corresponding AUC values (mean $\pm$ standard deviation) are listed in the legend for each bin. 
    Performance remains high across the full range, with the expected mild degradation toward the smallest/faintest galaxies where morphology measurements are noisier and resolution is lower.}

    \label{fig:roc_variation}
    
\end{figure}

To move beyond qualitative trends and improve the accuracy of separating spirals and ellipticals, we combine the measured non-parametric indices with the visual classifications from GZ 1 to train a supervised machine-learning model. This approach use the discriminatory power of the $\rm CA[A_S]S+MEGG$ parameter space, while adopting the decision boundaries from reliable visual labels, and enabling the derivation of probabilistic morphological classifications. By doing so, we transform the indices from descriptive diagnostics into quantitative predictors, allowing us to assign each galaxy a probability of being spiral or elliptical in a homogeneous way.

\begin{figure}
    \centering
    \includegraphics[width=\columnwidth]{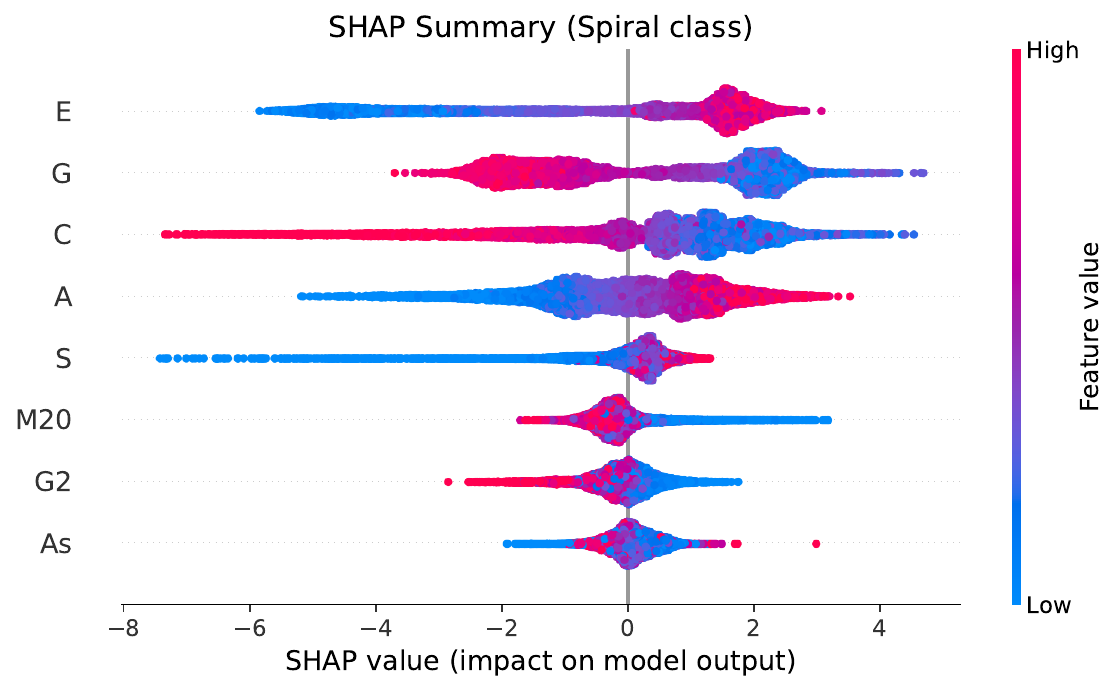}
    \caption{SHAP summary plot for the LightGBM model predicting Spiral galaxy classification. Each point represents the SHAP value of a single feature for one galaxy, showing its impact on the model output. The horizontal axis indicates the contribution (positive or negative) to the prediction, while the vertical axis lists the most important features ranked by overall impact. The color gradient encodes the feature value from low (blue) to high (red), highlighting how different ranges of feature values drive the prediction toward or away from the Spiral class.}
    \label{fig:lightgbm_shap}
    
\end{figure}

\subsection{Defining a training set}
\label{sec:def_training}
We use our spiral and elliptical subsamples as training set for the machine learning. We use the GZ1 label (elliptical vs.\ spiral) as the target $y\in\{0,1\}$, with spiral as the positive class (1). The combined sample is then divided into pool (60\%), calibration (15\%) and test sample (25\%). Because spirals largely outnumber ellipticals in our sample, we address class imbalance in two ways. First, all splits preserve the class ratio in the pool, calibration/validation, and test sets. Second, we apply SMOTE\footnote{SMOTE generates synthetic samples through nearest-neighbor interpolation in the minority class feature space, which helps avoid the overfitting associated with simple duplication. Nonetheless, as with any resampling technique, it can introduce bias if classes are highly overlapping in the full space. Since (most) of the metrics show very clear distinction between Ell and Sp (Fig.\ref{fig:2d_hist}), it is a safe procedure. We also computed the results without using the SMOTE, for which we find similar results. Namely, the overall accuracy decreases by ~0.5\%, which follows from a decrease in the accuracy specifically for Ell (also small ~2\%), while the spirals remain "untouched".} (Synthetic Minority Over-sampling Technique, \citet{2011arXiv1106.1813C}) only within the training sets of the cross-validation and in the training portion of the pool set: synthetic minority examples are generated by interpolating between nearest neighbors of the minority class in the C$\rm [A_S]$AS+MEGG feature space. No over-sampling is applied to calibration/validation or test sets, ensuring unbiased performance estimates and well-calibrated probabilities. 

\subsection{Results using Light Gradient Boosting Machine}

To assess the discriminative power of the full set of non--parametric morphological indices, we employed the Light Gradient Boosting Machine (LightGBM; \citealt{ke2017lightgbm}). LightGBM is a decision--tree--based ensemble algorithm that implements gradient boosting in a highly efficient manner. In contrast to classical classifiers that rely on a linear or logistic boundary in the feature space, gradient boosting iteratively builds an ensemble of weak learners (decision trees), where each subsequent tree corrects the residual errors of the previous ensemble. LightGBM improves on standard implementations of gradient boosting by using a leaf--wise tree growth strategy and histogram--based binning of features, allowing for faster training, lower memory usage, and the ability to handle large, imbalanced datasets. These properties make LightGBM particularly suitable for our morphological classification problem, where the input feature space is moderately high--dimensional and the class distribution between spirals and ellipticals is not balanced. Furthermore, the algorithm provides well--calibrated probabilistic outputs and interpretable measures of feature importance, both of which are essential for a robust scientific interpretation.

Figure~\ref{fig:lightgbm_results} summarizes the performance of the LightGBM classifier. We detail each panel, from leftmost top row, to rightmost bottom row, in the following:
\begin{enumerate}
    \item The ROC curve, which quantifies the trade--off between true positive rate and false positive rate for varying classification thresholds. The resulting Area Under the Curve (AUC = $0.995 \pm 0.001$) indicates near--perfect separability between spiral and elliptical galaxies;

    \item The Precision--Recall (PR) curve, focusing on the performance for the spiral class. The extremely high average precision ([AP] = $0.999 \pm 0.000$) further confirms that the classifier maintains excellent purity across the full range of recall values;

    \item The probability calibration curve. This diagnostic compares the raw model outputs (predicted probability of being a spiral) against the empirical fraction of true spirals in corresponding probability bins. If the classifier is perfectly calibrated, points will fall along the one–to–one diagonal: for example, among all galaxies assigned a spiral probability of $70\%$, about $70\%$ should actually be spirals. The plotted blue points represent the mean observed frequencies in probability bins, with errorbars denoting the 95\% confidence interval. The close alignment with the diagonal line indicates that the LightGBM predictions are almost perfectly calibrated across the full probability range. In the same panel, we show the Brier score \citep{1950MWRv...78....1B}, which provides a quantitative summary of calibration and refinement. It measures the mean squared error between predicted probabilities and the true binary outcomes, taking values between 0 (perfect) and 1 (worst possible). Our measured Brier score of $0.020^{+0.002}_{-0.002}$ is extremely low, meaning that the probabilities are not only discriminative but also reliable. This complements the ROC and PR curves: a model can achieve high AUC or [AP] while still producing poorly calibrated probabilities, but in this case LightGBM achieves both high discrimination and excellent calibration;

    \item The confusion matrix expressed in row--normalized percentages. LightGBM correctly identifies $98.6\% \pm 0.3$ of spiral galaxies and $87.5\% \pm 0.6$ of ellipticals. Misclassifications are rare, amounting to only $\sim 1.4\pm 0.3 \% $ of spirals classified as ellipticals and $\sim 12.5 \pm 0.9 \%$ of ellipticals classified as spirals. The latter can follow from the presence of S0s within the elliptical label in GZ1;

    \item The learning (loss) curves of the LightGBM classifier, showing the binary cross–entropy (log loss) as a function of boosting iterations (trees). We plot loss on the raw training fold (no SMOTE) and on an independent validation fold. The validation curve drops rapidly and then flattens without an upturn, indicating no overfitting. The training curve remains below the validation curve, as expected from the generalization gap\footnote{In supervised learning the loss evaluated on the data used for fitting is systematically lower than the loss on unseen data. This difference is the generalization gap. A nonzero separation between train and validation curves is therefore expected and, within bounds, an evidence of a model that fits the data while still generalizing.}. The training loss remains strictly above zero because we optimize probabilistic log loss under regularization and early stopping; pushing log loss to zero would require assigning probabilities of exactly 0/1 to every training object—a behavior typical of overfitting and inconsistent with the probabilistic approach we adopt;
    
    \item  The distribution of predicted spiral probabilities for the true spiral and elliptical systems in the test sample. The strong bimodality, with spirals peaking near unity and ellipticals near zero, highlights the high confidence of the model predictions. Only a negligible fraction of objects occupy the intermediate regime, reinforcing the robustness of the classification.

\end{enumerate} 

The very high performance (AUC $\simeq 0.99$, AP $\simeq 1$) largely reflects the fact that our target label is intentionally simple (``spiral'' vs.\ ``early-type'' as defined by GZ1) and that the adopted non-parametric morphology vector is designed to separate these two regimes efficiently. We understand that it would be better to have a good separation between ellipticals, lenticulars and spirals, however separating ellipticals and lenticulars is a longstanding problem in the literature, and for the current exercise, the labels from GZ1 are the most robust we can use. To address the concern that the result could be driven by a restricted subset of large, high-S/N systems, we performed a sanity check by measuring ROC performance in bins of apparent size and brightness (Fig.~\ref{fig:roc_variation}). Using out-of-fold calibrated probabilities, the classifier remains highly performant across the full range probed: for $R_{\mathrm{P}}$ (arcsec) the AUC varies only mildly from $0.991\pm0.003$ (4--10 arcsec) up to $0.996\pm0.003$ (20--30 arcsec), and as a function of $r$ magnitude it decreases smoothly from $0.994\pm0.004$ (14--15) to $0.977\pm0.016$ (17--18). This controlled degradation toward the faintest/smallest bins—where morphology is intrinsically harder due to lower resolution and surface-brightness sensitivity—supports the interpretation that the high global AUC/AP is not an artifact of a single easy regime, but rather that the separation remains robust over most of the parameter space while behaving as expected where the task becomes observationally more challenging.

Moreover, since we feed the LightGBM with eight different indexes, it is important also to investigate which are contributing the most to define the desired probability. Thus, we applied SHAP (SHapley Additive exPlanations; \citealt{NIPS2017_8a20a862}) values to the LightGBM model, shown in Figure~\ref{fig:lightgbm_shap}. Each point in the summary plot corresponds to a galaxy, with its horizontal position encoding the SHAP value (i.e. the marginal contribution of that feature to the probability of being classified as a spiral), and the color denoting the normalized feature value. Negative SHAP values (to the left) lower the spiral probability, while positive values (to the right) increase it. For instance, galaxies with low G (blue points) tend to shift the classification toward spiral, reflecting the clumpy light distribution of disks. Similarly, high entropy increases spiral likelihood, while low values support elliptical classifications. Overall, the SHAP values not only corroborate the feature importance ranking but also provide physical interpretability by linking specific morphological traits to the classifier’s decision process. 

Yet, we highlight one particular caveat of the adopted procedure. Because GZ1 provides only a binary spiral vs.\ elliptical label for bright SDSS galaxies, the ``elliptical'' class inevitably contains a non-negligible fraction of lenticular (S0) systems: in single-band imaging, S0s share the smooth, centrally concentrated appearance of ellipticals, yet they are physically disk galaxies, often with weak spiral structure and subtle lenses/bars. This mixing is important for calibration: the model is trained (and isotonic-calibrated) to reproduce GZ1's operational definition of ``elliptical'', so the resulting $P(\mathrm{Sp})$ should be interpreted as the probability of being spiral versus a mixed early-type (E$+$S0) class, rather than a pure E vs.\ Sp separation. Not by chance, the accuracy in the confusion matrix for ellipticals is smaller than the one for spirals, as we expect most of S0's to be included within the ellipticals subset. Nevertheless, S0's can also be misclassified as a disk galaxy, particularly with an edge-on line-of-sight.

In summary, the LightGBM model is able to provide a high accuracy for spiral probability through the use of structural features of galaxies, quantified through the non-parametric indexes. In this regard, the most important set to define the spiral probability are G, C, and E, all of which show a great separation between GZ 1 selected spirals and ellipticals. We incorporate the P(spiral) for all the galaxies in our sample in the provided catalog, for which the columns and respective descriptions can be found in Appendix~\ref{ap:example_table}.

\section{Conclusions and summary}

In this work we provide the first homogeneous catalog of non–parametric morphological indices for galaxies in the Dark Energy Camera Legacy Survey (DECaLS, part of the Legacy survey, data release 10), limited to systems with effective radius larger than 2 arcsec and brighter than 18.5 in the r-band apparent magnitude. Using our newly developed Python package \texttt{galmex}, we measure the full CA[$\rm A_S$]S+MEGG set of non–parametric indices in a uniform way for more than one million DECaLS galaxies, and derive probabilistic spiral/elliptical classifications for about 1.7 million objects at $z \leq 0.15$. The modular, transparent design of \texttt{galmex} ensures that every preprocessing and measurement step can be inspected, reproduced, and adapted, turning the catalog and the code into a long–lived resource for the community.  

Compared to previous morphology catalogs based on visual inspection or parametric profile fitting, our work delivers: (i) a deeper and wider–area dataset in the southern hemisphere, fully processed with a single, well–tested pipeline; (ii) a consistent set of C[$\rm A_S$]AS+MEGG indices measured with segmentation and Petrosian apertures tuned on realistic DECam simulations; and (iii) calibrated probabilistic classifications directly in the non–parametric parameter space. This combination provides a more homogeneous and physically interpretable view of galaxy structure than either visual labels alone or purely Sérsic–based decompositions.

Our main conclusions about the reliability and use of non–parametric indices for morphology are:

\begin{enumerate}
    \item C[$\rm A_S$]AS and MEGG indices -- Using bona fide samples of spirals and ellipticals defined from Galaxy Zoo~1, we confirm that concentration is the most reliable CAS parameter for separating early and late types, whereas asymmetry–based indices ($A$, $A_S$, and $S$) exhibit substantial overlap between the two classes and are therefore best suited to separating strongly disturbed systems rather than to performing a clean spiral–elliptical split (Fig.~\ref{fig:2d_hist}). In contrast, all indices in the MEGG system ($M_{20}$, entropy, Gini, and $G_2$) provide strong and consistent separation, highlighting their robustness as tracers of bulge– versus disk–dominated morphologies.

    \item Connection with T--Type and visual classification -- When compared with CNN--based T--Types (Fig.~\ref{fig:ttype_variation}), the indices not only recover the global spiral–elliptical division, but also trace a continuous gradient along the Hubble sequence. The trend is particularly steep for $M_{20}$, $E$, Gini, and $G_2$, which respond to substructure and clumpiness in spiral galaxies. Moreover, the indices reproduce the distributions obtained from Galaxy Zoo visual classifications (Fig.~\ref{fig:frac_variation}), demonstrating that non–parametric indices capture, to first order, the same morphological traits perceived by human classifiers.

    \item Machine–learning classification -- Using their discriminatory power, we provide the indices as input features to a binary LightGBM classifier (Figs.~\ref{fig:lightgbm_results} and \ref{fig:lightgbm_shap}), focused in discriminating between spirals and early-type systems (E+S0s). The model achieves high accuracy ($97\%$) and produces well–calibrated probabilities of a galaxy being spiral, with entropy, concentration, and Gini consistently emerging as the most influential features. 
\end{enumerate}

An important caveat that emerges from our analysis is that the reliability of control samples depends strongly on the adopted visual classification scheme. In particular, the top--level separation in Galaxy Zoo DECaLS into ``smooth'' versus ``disk/feature'' categories is not equivalent to the classical early-- versus late--type division. We show that the distributions of non--parametric indices for these DECaLS classes differ significantly from those of ellipticals and spirals selected from Galaxy Zoo~1, with the largest discrepancies appearing when comparing ``smooth'' to ellipticals. This mismatch reflects the subjectivity of the ``smooth'' category, which can include both bulge--dominated disks and genuine ellipticals, and leads to systematically different metric distributions. Moreover, training a machine--learning classifier on the DECaLS smooth/disk labels results in degraded performance compared to using the GZ1 spiral/elliptical subsamples, directly affecting both the reliability and purity of the resulting classifications. These biases are further compounded by the dependence of vote fractions on redshift, luminosity, and Petrosian radius, which imprint observational effects onto the labels themselves. Together, these results highlight that the choice of training set and classification scheme is not a neutral decision: it can propagate systematic biases into automated classifications, underscoring the need for careful sample definition when bridging visual projects and machine--learning frameworks.

In summary, this work establishes a transparent and reproducible framework for morphological classification in wide--field imaging surveys. The combination of a publicly available catalog, a modular software package, and a calibrated machine--learning classifier provides the community with an extremely versatile toolset to study galaxy evolution. Because the catalog covers the full DECaLS footprint and overlaps with major spectroscopic programs in the southern hemisphere (e.g. 4MOST/CHANCES and WEAVE), it enables a broad range of new science: from mapping morphology as a function of environment, mass, and star–formation activity, to selecting rare disturbed systems, such as mergers and jellyfish galaxies, in a uniform way. The natural next step is to move beyond the simple spiral--elliptical dichotomy and explicitly incorporate disturbed and transitioning systems, as well as to extend the methodology toward higher redshifts. This will allow us to probe more directly the dynamical processes that drive morphological transformation, providing a more complete picture of galaxy evolution across environments and cosmic time.

\begin{acknowledgements}
The full acknowledgements are available in Appendix
\end{acknowledgements}

\bibliographystyle{aa} 
\bibliography{bib} 

@ARTICLE{1926ApJ....64..321H,
       author = {{Hubble}, E.~P.},
        title = "{Extragalactic nebulae.}",
      journal = {\apj},
         year = 1926,
        month = dec,
       volume = {64},
        pages = {321-369},
          doi = {10.1086/143018},
       adsurl = {https://ui.adsabs.harvard.edu/abs/1926ApJ....64..321H}
}

@book{VanRossum2009,
  author    = {Van Rossum, Guido and Drake, Fred L.},
  title     = {Python 3 Reference Manual},
  year      = {2009},
  publisher = {CreateSpace},
  address   = {Scotts Valley, CA}
}

@article{Harris2020,
  author  = {Harris, Charles R. and Millman, K. Jarrod and van der Walt, St{\'e}fan J. and
             Gommers, Ralf and Virtanen, Pauli and Cournapeau, David and Wieser, Eric and
             Taylor, Julian and Berg, Sebastian and Smith, Nathaniel J. and others},
  title   = {Array programming with {NumPy}},
  journal = {Nature},
  year    = {2020},
  volume  = {585},
  pages   = {357--362}
}

@article{Virtanen2020,
  author  = {Virtanen, Pauli and Gommers, Ralf and Oliphant, Travis E. and Haberland, Matt and
             Reddy, Tyler and Cournapeau, David and Burovski, Evgeni and Peterson, Pearu and
             Weckesser, Warren and Bright, Jonathan and others},
  title   = {{SciPy} 1.0: Fundamental Algorithms for Scientific Computing in {Python}},
  journal = {Nature Methods},
  year    = {2020},
  volume  = {17},
  pages   = {261--272}
}

@article{Astropy2013,
  author  = {{Astropy Collaboration} and Robitaille, Thomas P. and Tollerud, Erik J. and Greenfield, Perry and
             Droettboom, Michael and Bray, Erik and Aldcroft, Thomas and Davis, Matt and Ginsburg, Adam and
             Price-Whelan, Adrian M. and others},
  title   = {Astropy: A community {Python} package for astronomy},
  journal = {Astronomy \& Astrophysics},
  year    = {2013},
  volume  = {558},
  pages   = {A33}
}

@article{Astropy2018,
  author  = {{Astropy Collaboration} and Price-Whelan, Adrian M. and Sip{\H o}cz, Brigitta M. and G{\"u}nther, H. M. and
             Lim, Pey Lian and Crawford, Sam M. and Conseil, Simon and Shupe, David L. and Craig, Matthew W. and
             Dencheva, Nadia and others},
  title   = {The {Astropy} Project: Building an Open-science Project and Status of the v2.0 Core Package},
  journal = {The Astronomical Journal},
  year    = {2018},
  volume  = {156},
  pages   = {123}
}

@article{Astropy2022,
  author  = {{Astropy Collaboration} and Price-Whelan, Adrian M. and Lim, Pey Lian and Earl, Nicholas and Starkman, Nathaniel and
             Bradley, Larry and Shupe, David L. and Patil, Aarya A. and Corrales, Lia and Brasseur, C. E. and others},
  title   = {The {Astropy} Project: Sustaining and Growing a Community-oriented Open-source Project and the {Astropy} v5.0 Core Package},
  journal = {The Astrophysical Journal},
  year    = {2022},
  volume  = {935},
  pages   = {167}
}

@inproceedings{McKinney2010,
  author    = {McKinney, Wes},
  title     = {Data Structures for Statistical Computing in {Python}},
  booktitle = {Proceedings of the 9th Python in Science Conference},
  year      = {2010},
  editor    = {van der Walt, St{\'e}fan and Millman, Jarrod},
  pages     = {56--61}
}

@article{Reback2022,
  author  = {Reback, Jeff and Jbrockmendel and McKinney, Wes and Van den Bossche, Joris and Augspurger, Tom and
             Cloud, Phillip and Hawkins, Simon and Roeschke, Matthew and Sinclair, Mark and Klein, Adam and others},
  title   = {pandas-dev/pandas: {Pandas} 1.4.0},
  journal = {Zenodo},
  year    = {2022},
  doi     = {10.5281/zenodo.3509134}
}

@article{Hunter2007,
  author  = {Hunter, John D.},
  title   = {Matplotlib: A 2D Graphics Environment},
  journal = {Computing in Science \& Engineering},
  year    = {2007},
  volume  = {9},
  number  = {3},
  pages   = {90--95}
}

@ARTICLE{2002AJ....124..266P,
       author = {{Peng}, Chien Y. and {Ho}, Luis C. and {Impey}, Chris D. and {Rix}, Hans-Walter},
        title = "{Detailed Structural Decomposition of Galaxy Images}",
      journal = {\aj},
         year = 2002,
        month = jul,
       volume = {124},
       number = {1},
        pages = {266-293},
          doi = {10.1086/340952},
archivePrefix = {arXiv},
       eprint = {astro-ph/0204182},
 primaryClass = {astro-ph},
       adsurl = {https://ui.adsabs.harvard.edu/abs/2002AJ....124..266P}
}

@ARTICLE{2010AJ....139.2097P,
       author = {{Peng}, Chien Y. and {Ho}, Luis C. and {Impey}, Chris D. and {Rix}, Hans-Walter},
        title = "{Detailed Decomposition of Galaxy Images. II. Beyond Axisymmetric Models}",
      journal = {\aj},
         year = 2010,
        month = jun,
       volume = {139},
       number = {6},
        pages = {2097-2129},
          doi = {10.1088/0004-6256/139/6/2097},
archivePrefix = {arXiv},
       eprint = {0912.0731},
 primaryClass = {astro-ph.CO},
       adsurl = {https://ui.adsabs.harvard.edu/abs/2010AJ....139.2097P}
}

@ARTICLE{2011ApJS..196...11S,
       author = {{Simard}, Luc and {Mendel}, J. Trevor and {Patton}, David R. and {Ellison}, Sara L. and {McConnachie}, Alan W.},
        title = "{A Catalog of Bulge+disk Decompositions and Updated Photometry for 1.12 Million Galaxies in the Sloan Digital Sky Survey}",
      journal = {\apjs},
         year = 2011,
        month = sep,
       volume = {196},
       number = {1},
          eid = {11},
        pages = {11},
          doi = {10.1088/0067-0049/196/1/11},
archivePrefix = {arXiv},
       eprint = {1107.1518},
 primaryClass = {astro-ph.CO},
       adsurl = {https://ui.adsabs.harvard.edu/abs/2011ApJS..196...11S}
}

@ARTICLE{2002ApJS..142....1S,
       author = {{Simard}, Luc and {Willmer}, Christopher N.~A. and {Vogt}, Nicole P. and {Sarajedini}, Vicki L. and {Phillips}, Andrew C. and {Weiner}, Benjamin J. and {Koo}, David C. and {Im}, Myungshin and {Illingworth}, Garth D. and {Faber}, S.~M.},
        title = "{The DEEP Groth Strip Survey. II. Hubble Space Telescope Structural Parameters of Galaxies in the Groth Strip}",
      journal = {\apjs},
         year = 2002,
        month = sep,
       volume = {142},
       number = {1},
        pages = {1-33},
          doi = {10.1086/341399},
archivePrefix = {arXiv},
       eprint = {astro-ph/0205025},
 primaryClass = {astro-ph},
       adsurl = {https://ui.adsabs.harvard.edu/abs/2002ApJS..142....1S}
}

@ARTICLE{1969ApJ...155..393P,
       author = {{Peebles}, P.~J.~E.},
        title = "{Origin of the Angular Momentum of Galaxies}",
      journal = {\apj},
         year = 1969,
        month = feb,
       volume = {155},
        pages = {393},
          doi = {10.1086/149876},
       adsurl = {https://ui.adsabs.harvard.edu/abs/1969ApJ...155..393P}
}

@ARTICLE{2015ApJ...812...29T,
       author = {{Teklu}, Adelheid F. and {Remus}, Rhea-Silvia and {Dolag}, Klaus and {Beck}, Alexander M. and {Burkert}, Andreas and {Schmidt}, Andreas S. and {Schulze}, Felix and {Steinborn}, Lisa K.},
        title = "{Connecting Angular Momentum and Galactic Dynamics: The Complex Interplay between Spin, Mass, and Morphology}",
      journal = {\apj},
         year = 2015,
        month = oct,
       volume = {812},
       number = {1},
          eid = {29},
        pages = {29},
          doi = {10.1088/0004-637X/812/1/29},
archivePrefix = {arXiv},
       eprint = {1503.03501},
 primaryClass = {astro-ph.GA},
       adsurl = {https://ui.adsabs.harvard.edu/abs/2015ApJ...812...29T}
}

@ARTICLE{2013MNRAS.432L..56S,
       author = {{Sanchez-Janssen}, R. and {Gadotti}, D.~A.},
        title = "{Evidence for secular evolution of disc structural parameters in massive  barred galaxies.}",
      journal = {\mnras},
         year = 2013,
        month = may,
       volume = {432},
        pages = {L56-L60},
          doi = {10.1093/mnrasl/slt037},
archivePrefix = {arXiv},
       eprint = {1208.3682},
 primaryClass = {astro-ph.CO},
       adsurl = {https://ui.adsabs.harvard.edu/abs/2013MNRAS.432L..56S}
}

@INPROCEEDINGS{2016ASSL..418..355B,
       author = {{Bournaud}, Fr{\'e}d{\'e}ric},
        title = "{Bulge Growth Through Disc Instabilities in High-Redshift Galaxies}",
    booktitle = {Galactic Bulges},
         year = 2016,
       editor = {{Laurikainen}, Eija and {Peletier}, Reynier and {Gadotti}, Dimitri},
       series = {Astrophysics and Space Science Library},
       volume = {418},
        month = jan,
        pages = {355},
          doi = {10.1007/978-3-319-19378-6_13},
archivePrefix = {arXiv},
       eprint = {1503.07660},
 primaryClass = {astro-ph.GA},
       adsurl = {https://ui.adsabs.harvard.edu/abs/2016ASSL..418..355B}
}

@ARTICLE{2009Natur.457..451D,
       author = {{Dekel}, A. and {Birnboim}, Y. and {Engel}, G. and {Freundlich}, J. and {Goerdt}, T. and {Mumcuoglu}, M. and {Neistein}, E. and {Pichon}, C. and {Teyssier}, R. and {Zinger}, E.},
        title = "{Cold streams in early massive hot haloes as the main mode of galaxy formation}",
      journal = {\nat},
         year = 2009,
        month = jan,
       volume = {457},
       number = {7228},
        pages = {451-454},
          doi = {10.1038/nature07648},
archivePrefix = {arXiv},
       eprint = {0808.0553},
 primaryClass = {astro-ph},
       adsurl = {https://ui.adsabs.harvard.edu/abs/2009Natur.457..451D}
}

@ARTICLE{2008MNRAS.387.1431D,
       author = {{Dalla Vecchia}, Claudio and {Schaye}, Joop},
        title = "{Simulating galactic outflows with kinetic supernova feedback}",
      journal = {\mnras},
         year = 2008,
        month = jul,
       volume = {387},
       number = {4},
        pages = {1431-1444},
          doi = {10.1111/j.1365-2966.2008.13322.x},
archivePrefix = {arXiv},
       eprint = {0801.2770},
 primaryClass = {astro-ph},
       adsurl = {https://ui.adsabs.harvard.edu/abs/2008MNRAS.387.1431D}
}

@ARTICLE{2012ARA&A..50..455F,
       author = {{Fabian}, A.~C.},
        title = "{Observational Evidence of Active Galactic Nuclei Feedback}",
      journal = {\araa},
         year = 2012,
        month = sep,
       volume = {50},
        pages = {455-489},
          doi = {10.1146/annurev-astro-081811-125521},
archivePrefix = {arXiv},
       eprint = {1204.4114},
 primaryClass = {astro-ph.CO},
       adsurl = {https://ui.adsabs.harvard.edu/abs/2012ARA&A..50..455F}
}

@ARTICLE{1980ApJ...237..692L,
       author = {{Larson}, R.~B. and {Tinsley}, B.~M. and {Caldwell}, C.~N.},
        title = "{The evolution of disk galaxies and the origin of S0 galaxies}",
      journal = {\apj},
         year = 1980,
        month = may,
       volume = {237},
        pages = {692-707},
          doi = {10.1086/157917},
       adsurl = {https://ui.adsabs.harvard.edu/abs/1980ApJ...237..692L}
}

@ARTICLE{2000ApJ...540..113B,
       author = {{Balogh}, Michael L. and {Navarro}, Julio F. and {Morris}, Simon L.},
        title = "{The Origin of Star Formation Gradients in Rich Galaxy Clusters}",
      journal = {\apj},
         year = 2000,
        month = sep,
       volume = {540},
       number = {1},
        pages = {113-121},
          doi = {10.1086/309323},
archivePrefix = {arXiv},
       eprint = {astro-ph/0004078},
 primaryClass = {astro-ph},
       adsurl = {https://ui.adsabs.harvard.edu/abs/2000ApJ...540..113B}
}

@ARTICLE{1999MNRAS.302..771J,
       author = {{Johnston}, Kathryn V. and {Sigurdsson}, Steinn and {Hernquist}, Lars},
        title = "{Measuring mass-loss rates from Galactic satellites}",
      journal = {\mnras},
         year = 1999,
        month = feb,
       volume = {302},
       number = {4},
        pages = {771-789},
          doi = {10.1046/j.1365-8711.1999.02200.x},
archivePrefix = {arXiv},
       eprint = {astro-ph/9805291},
 primaryClass = {astro-ph},
       adsurl = {https://ui.adsabs.harvard.edu/abs/1999MNRAS.302..771J}
}

@ARTICLE{1972ApJ...176....1G,
       author = {{Gunn}, James E. and {Gott}, J. Richard, III},
        title = "{On the Infall of Matter Into Clusters of Galaxies and Some Effects on Their Evolution}",
      journal = {\apj},
         year = 1972,
        month = aug,
       volume = {176},
        pages = {1},
          doi = {10.1086/151605},
       adsurl = {https://ui.adsabs.harvard.edu/abs/1972ApJ...176....1G}
}

@ARTICLE{1999MNRAS.308..947A,
       author = {{Abadi}, Mario G. and {Moore}, Ben and {Bower}, Richard G.},
        title = "{Ram pressure stripping of spiral galaxies in clusters}",
      journal = {\mnras},
         year = 1999,
        month = oct,
       volume = {308},
       number = {4},
        pages = {947-954},
          doi = {10.1046/j.1365-8711.1999.02715.x},
archivePrefix = {arXiv},
       eprint = {astro-ph/9903436},
 primaryClass = {astro-ph},
       adsurl = {https://ui.adsabs.harvard.edu/abs/1999MNRAS.308..947A}
}

@ARTICLE{1972ApJ...178..623T,
       author = {{Toomre}, Alar and {Toomre}, Juri},
        title = "{Galactic Bridges and Tails}",
      journal = {\apj},
         year = 1972,
        month = dec,
       volume = {178},
        pages = {623-666},
          doi = {10.1086/151823},
       adsurl = {https://ui.adsabs.harvard.edu/abs/1972ApJ...178..623T}
}

@ARTICLE{1991ApJ...370L..65B,
       author = {{Barnes}, Joshua E. and {Hernquist}, Lars E.},
        title = "{Fueling Starburst Galaxies with Gas-rich Mergers}",
      journal = {\apjl},
         year = 1991,
        month = apr,
       volume = {370},
        pages = {L65},
          doi = {10.1086/185978},
       adsurl = {https://ui.adsabs.harvard.edu/abs/1991ApJ...370L..65B}
}

@ARTICLE{2013MNRAS.432..336W,
       author = {{Wetzel}, Andrew R. and {Tinker}, Jeremy L. and {Conroy}, Charlie and {van den Bosch}, Frank C.},
        title = "{Galaxy evolution in groups and clusters: satellite star formation histories and quenching time-scales in a hierarchical Universe}",
      journal = {\mnras},
         year = 2013,
        month = jun,
       volume = {432},
       number = {1},
        pages = {336-358},
          doi = {10.1093/mnras/stt469},
archivePrefix = {arXiv},
       eprint = {1206.3571},
 primaryClass = {astro-ph.CO},
       adsurl = {https://ui.adsabs.harvard.edu/abs/2013MNRAS.432..336W}
}

@ARTICLE{2004ARA&A..42..603K,
       author = {{Kormendy}, John and {Kennicutt}, Jr., Robert C.},
        title = "{Secular Evolution and the Formation of Pseudobulges in Disk Galaxies}",
      journal = {\araa},
         year = 2004,
        month = sep,
       volume = {42},
       number = {1},
        pages = {603-683},
          doi = {10.1146/annurev.astro.42.053102.134024},
archivePrefix = {arXiv},
       eprint = {astro-ph/0407343},
 primaryClass = {astro-ph},
       adsurl = {https://ui.adsabs.harvard.edu/abs/2004ARA&A..42..603K}
}

@ARTICLE{2011ApJ...731...65F,
       author = {{F{\"o}rster Schreiber}, N.~M. and {Shapley}, A.~E. and {Erb}, D.~K. and {Genzel}, R. and {Steidel}, C.~C. and {Bouch{\'e}}, N. and {Cresci}, G. and {Davies}, R.},
        title = "{Constraints on the Assembly and Dynamics of Galaxies. I. Detailed Rest-frame Optical Morphologies on Kiloparsec Scale of z \raisebox{-0.5ex}\textasciitilde 2 Star-forming Galaxies}",
      journal = {\apj},
         year = 2011,
        month = apr,
       volume = {731},
       number = {1},
          eid = {65},
        pages = {65},
          doi = {10.1088/0004-637X/731/1/65},
archivePrefix = {arXiv},
       eprint = {1011.1507},
 primaryClass = {astro-ph.CO},
       adsurl = {https://ui.adsabs.harvard.edu/abs/2011ApJ...731...65F}
}

@ARTICLE{2011ApJ...733..101G,
       author = {{Genzel}, R. and {Newman}, S. and {Jones}, T. and {F{\"o}rster Schreiber}, N.~M. and {Shapiro}, K. and {Genel}, S. and {Lilly}, S.~J. and {Renzini}, A. and {Tacconi}, L.~J. and {Bouch{\'e}}, N. and {Burkert}, A. and {Cresci}, G. and {Buschkamp}, P. and {Carollo}, C.~M. and {Ceverino}, D. and {Davies}, R. and {Dekel}, A. and {Eisenhauer}, F. and {Hicks}, E. and {Kurk}, J. and {Lutz}, D. and {Mancini}, C. and {Naab}, T. and {Peng}, Y. and {Sternberg}, A. and {Vergani}, D. and {Zamorani}, G.},
        title = "{The Sins Survey of z \raisebox{-0.5ex}\textasciitilde 2 Galaxy Kinematics: Properties of the Giant Star-forming Clumps}",
      journal = {\apj},
         year = 2011,
        month = jun,
       volume = {733},
       number = {2},
          eid = {101},
        pages = {101},
          doi = {10.1088/0004-637X/733/2/101},
archivePrefix = {arXiv},
       eprint = {1011.5360},
 primaryClass = {astro-ph.CO},
       adsurl = {https://ui.adsabs.harvard.edu/abs/2011ApJ...733..101G}
}

@ARTICLE{2005MNRAS.357..903C,
       author = {{Cassata}, P. and {Cimatti}, A. and {Franceschini}, A. and {Daddi}, E. and {Pignatelli}, E. and {Fasano}, G. and {Rodighiero}, G. and {Pozzetti}, L. and {Mignoli}, M. and {Renzini}, A.},
        title = "{The evolution of the galaxy B-band rest-frame morphology to z\raisebox{-0.5ex}\textasciitilde 2: new clues from the K20/GOODS sample}",
      journal = {\mnras},
         year = 2005,
        month = mar,
       volume = {357},
       number = {3},
        pages = {903-917},
          doi = {10.1111/j.1365-2966.2005.08657.x},
archivePrefix = {arXiv},
       eprint = {astro-ph/0411768},
 primaryClass = {astro-ph},
       adsurl = {https://ui.adsabs.harvard.edu/abs/2005MNRAS.357..903C}
}

@ARTICLE{2011AJ....142...31B,
       author = {{Blanton}, Michael R. and {Kazin}, Eyal and {Muna}, Demitri and {Weaver}, Benjamin A. and {Price-Whelan}, Adrian},
        title = "{Improved Background Subtraction for the Sloan Digital Sky Survey Images}",
      journal = {\aj},
         year = 2011,
        month = jul,
       volume = {142},
       number = {1},
          eid = {31},
        pages = {31},
          doi = {10.1088/0004-6256/142/1/31},
archivePrefix = {arXiv},
       eprint = {1105.1960},
 primaryClass = {astro-ph.IM},
       adsurl = {https://ui.adsabs.harvard.edu/abs/2011AJ....142...31B}
}

@ARTICLE{2019MNRAS.483.4140R,
       author = {{Rodriguez-Gomez}, Vicente and {Snyder}, Gregory F. and {Lotz}, Jennifer M. and {Nelson}, Dylan and {Pillepich}, Annalisa and {Springel}, Volker and {Genel}, Shy and {Weinberger}, Rainer and {Tacchella}, Sandro and {Pakmor}, R{\"u}diger and {Torrey}, Paul and {Marinacci}, Federico and {Vogelsberger}, Mark and {Hernquist}, Lars and {Thilker}, David A.},
        title = "{The optical morphologies of galaxies in the IllustrisTNG simulation: a comparison to Pan-STARRS observations}",
      journal = {\mnras},
         year = 2019,
        month = mar,
       volume = {483},
       number = {3},
        pages = {4140-4159},
          doi = {10.1093/mnras/sty3345},
archivePrefix = {arXiv},
       eprint = {1809.08239},
 primaryClass = {astro-ph.GA},
       adsurl = {https://ui.adsabs.harvard.edu/abs/2019MNRAS.483.4140R}
}

@ARTICLE{1996A&AS..117..393B,
       author = {{Bertin}, E. and {Arnouts}, S.},
        title = "{SExtractor: Software for source extraction.}",
      journal = {\aaps},
         year = 1996,
        month = jun,
       volume = {117},
        pages = {393-404},
          doi = {10.1051/aas:1996164},
       adsurl = {https://ui.adsabs.harvard.edu/abs/1996A&AS..117..393B}
}

@ARTICLE{2008MNRAS.386..909C,
       author = {{Conselice}, Christopher J. and {Rajgor}, Sheena and {Myers}, Robert},
        title = "{The structures of distant galaxies - I. Galaxy structures and the merger rate to z \raisebox{-0.5ex}\textasciitilde 3 in the Hubble Ultra-Deep Field}",
      journal = {\mnras},
         year = 2008,
        month = may,
       volume = {386},
       number = {2},
        pages = {909-927},
          doi = {10.1111/j.1365-2966.2008.13069.x},
archivePrefix = {arXiv},
       eprint = {0711.2333},
 primaryClass = {astro-ph},
       adsurl = {https://ui.adsabs.harvard.edu/abs/2008MNRAS.386..909C}
}

@ARTICLE{2008ApJ...672..177L,
       author = {{Lotz}, Jennifer M. and {Davis}, M. and {Faber}, S.~M. and {Guhathakurta}, P. and {Gwyn}, S. and {Huang}, J. and {Koo}, D.~C. and {Le Floc'h}, E. and {Lin}, Lihwai and {Newman}, J. and {Noeske}, K. and {Papovich}, C. and {Willmer}, C.~N.~A. and {Coil}, A. and {Conselice}, C.~J. and {Cooper}, M. and {Hopkins}, A.~M. and {Metevier}, A. and {Primack}, J. and {Rieke}, G. and {Weiner}, B.~J.},
        title = "{The Evolution of Galaxy Mergers and Morphology at z < 1.2 in the Extended Groth Strip}",
      journal = {\apj},
         year = 2008,
        month = jan,
       volume = {672},
       number = {1},
        pages = {177-197},
          doi = {10.1086/523659},
archivePrefix = {arXiv},
       eprint = {astro-ph/0602088},
 primaryClass = {astro-ph},
       adsurl = {https://ui.adsabs.harvard.edu/abs/2008ApJ...672..177L}
}

@software{2016zndo....159035B,
       author = {{Barbary}, Kyle and {Boone}, Kyle and {McCully}, Curtis and {Craig}, Matt and {Deil}, Christoph and {Rose}, Benjamin},
        title = "{kbarbary/sep: v1.0.0}",
         year = 2016,
        month = sep,
          eid = {10.5281/zenodo.159035},
          doi = {10.5281/zenodo.159035},
      version = {v1.0.0},
    publisher = {Zenodo},
       adsurl = {https://ui.adsabs.harvard.edu/abs/2016zndo....159035B}
}

@ARTICLE{1996MNRAS.279L..47A,
       author = {{Abraham}, R.~G. and {Tanvir}, N.~R. and {Santiago}, B.~X. and {Ellis}, R.~S. and {Glazebrook}, K. and {van den Bergh}, S.},
        title = "{Galaxy morphology to I=25 mag in the Hubble Deep Field}",
      journal = {\mnras},
         year = 1996,
        month = apr,
       volume = {279},
       number = {3},
        pages = {L47-L52},
          doi = {10.1093/mnras/279.3.L47},
archivePrefix = {arXiv},
       eprint = {astro-ph/9602044},
 primaryClass = {astro-ph},
       adsurl = {https://ui.adsabs.harvard.edu/abs/1996MNRAS.279L..47A}
}

@ARTICLE{2024MNRAS.530.2688J,
       author = {{Jin}, Shoko and {Trager}, Scott C. and {Dalton}, Gavin B. and {Aguerri}, J. Alfonso L. and {Drew}, J.~E. and {Falc{\'o}n-Barroso}, Jes{\'u}s and {G{\"a}nsicke}, Boris T. and {Hill}, Vanessa and {Iovino}, Angela and {Pieri}, Matthew M. and {Poggianti}, Bianca M. and {Smith}, D.~J.~B. and {Vallenari}, Antonella and {Abrams}, Don Carlos and {Aguado}, David S. and {Antoja}, Teresa and {Arag{\'o}n-Salamanca}, Alfonso and {Ascasibar}, Yago and {Babusiaux}, Carine and {Balcells}, Marc and {Barrena}, R. and {Battaglia}, Giuseppina and {Belokurov}, Vasily and {Bensby}, Thomas and {Bonifacio}, Piercarlo and {Bragaglia}, Angela and {Carrasco}, Esperanza and {Carrera}, Ricardo and {Cornwell}, Daniel J. and {Dom{\'\i}nguez-Palmero}, Lilian and {Duncan}, Kenneth J. and {Famaey}, Benoit and {Fari{\~n}a}, Cecilia and {Gonzalez}, Oscar A. and {Guest}, Steve and {Hatch}, Nina A. and {Hess}, Kelley M. and {Hoskin}, Matthew J. and {Irwin}, Mike and {Knapen}, Johan H. and {Koposov}, Sergey E. and {Kuchner}, Ulrike and {Laigle}, Clotilde and {Lewis}, Jim and {Longhetti}, Marcella and {Lucatello}, Sara and {M{\'e}ndez-Abreu}, Jairo and {Mercurio}, Amata and {Molaeinezhad}, Alireza and {Mongui{\'o}}, Maria and {Morrison}, Sean and {Murphy}, David N.~A. and {Peralta de Arriba}, Luis and {P{\'e}rez}, Isabel and {P{\'e}rez-R{\`a}fols}, Ignasi and {Pic{\'o}}, Sergio and {Raddi}, Roberto and {Romero-G{\'o}mez}, Merc{\`e} and {Royer}, Fr{\'e}d{\'e}ric and {Siebert}, Arnaud and {Seabroke}, George M. and {Som}, Debopam and {Terrett}, David and {Thomas}, Guillaume and {Wesson}, Roger and {Worley}, C. Clare and {Alfaro}, Emilio J. and {Allende Prieto}, Carlos and {Alonso-Santiago}, Javier and {Amos}, Nicholas J. and {Ashley}, Richard P. and {Balaguer-N{\'u}{\~n}ez}, Lola and {Balbinot}, Eduardo and {Bellazzini}, Michele and {Benn}, Chris R. and {Berlanas}, Sara R. and {Bernard}, Edouard J. and {Best}, Philip and {Bettoni}, Daniela and {Bianco}, Andrea and {Bishop}, Georgia and {Blomqvist}, Michael and {Boeche}, Corrado and {Bolzonella}, Micol and {Bonoli}, Silvia and {Bosma}, Albert and {Britavskiy}, Nikolay and {Busarello}, Gianni and {Caffau}, Elisabetta and {Cantat-Gaudin}, Tristan and {Castro-Ginard}, Alfred and {Couto}, Guilherme and {Carbajo-Hijarrubia}, Juan and {Carter}, David and {Casamiquela}, Laia and {Conrado}, Ana M. and {Corcho-Caballero}, Pablo and {Costantin}, Luca and {Deason}, Alis and {de Burgos}, Abel and {De Grandi}, Sabrina and {Di Matteo}, Paola and {Dom{\'\i}nguez-G{\'o}mez}, Jes{\'u}s and {Dorda}, Ricardo and {Drake}, Alyssa and {Dutta}, Rajeshwari and {Erkal}, Denis and {Feltzing}, Sofia and {Ferr{\'e}-Mateu}, Anna and {Feuillet}, Diane and {Figueras}, Francesca and {Fossati}, Matteo and {Franciosini}, Elena and {Frasca}, Antonio and {Fumagalli}, Michele and {Gallazzi}, Anna and {Garc{\'\i}a-Benito}, Rub{\'e}n and {Gentile Fusillo}, Nicola and {Gebran}, Marwan and {Gilbert}, James and {Gledhill}, T.~M. and {Gonz{\'a}lez Delgado}, Rosa M. and {Greimel}, Robert and {Guarcello}, Mario Giuseppe and {Guerra}, Jose and {Gullieuszik}, Marco and {Haines}, Christopher P. and {Hardcastle}, Martin J. and {Harris}, Amy and {Haywood}, Misha and {Helmi}, Amina and {Hernandez}, Nauzet and {Herrero}, Artemio and {Hughes}, Sarah and {Ir{\v{s}}i{\v{c}}}, Vid and {Jablonka}, Pascale and {Jarvis}, Matt J. and {Jordi}, Carme and {Kondapally}, Rohit and {Kordopatis}, Georges and {Krogager}, Jens-Kristian and {La Barbera}, Francesco and {Lam}, Man I. and {Larsen}, S{\o}ren S. and {Lemasle}, Bertrand and {Lewis}, Ian J. and {Lhom{\'e}}, Emilie and {Lind}, Karin and {Lodi}, Marcello and {Longobardi}, Alessia and {Lonoce}, Ilaria and {Magrini}, Laura and {Ma{\'\i}z Apell{\'a}niz}, Jes{\'u}s and {Marchal}, Olivier and {Marco}, Amparo and {Martin}, Nicolas F. and {Matsuno}, Tadafumi and {Maurogordato}, Sophie and {Merluzzi}, Paola and {Miralda-Escud{\'e}}, Jordi and {Molinari}, Emilio and {Monari}, Giacomo and {Morelli}, Lorenzo and {Mottram}, Christopher J. and {Naylor}, Tim and {Negueruela}, Ignacio and {O{\~n}orbe}, Jose and {Pancino}, Elena and {Peirani}, S{\'e}bastien and {Peletier}, Reynier F. and {Pozzetti}, Lucia and {Rainer}, Monica and {Ramos}, Pau and {Read}, Shaun C. and {Rossi}, Elena Maria and {R{\"o}ttgering}, Huub J.~A. and {Rubi{\~n}o-Mart{\'\i}n}, Jose Alberto and {Sabater}, Jose and {San Juan}, Jos{\'e} and {Sanna}, Nicoletta and {Schallig}, Ellen and {Schiavon}, Ricardo P. and {Schultheis}, Mathias and {Serra}, Paolo and {Shimwell}, Timothy W. and {Sim{\'o}n-D{\'\i}az}, Sergio and {Smith}, Russell J. and {Sordo}, Rosanna and {Sorini}, Daniele and {Soubiran}, Caroline and {Starkenburg}, Else and {Steele}, Iain A. and {Stott}, John and {Stuik}, Remko and {Tolstoy}, Eline and {Tortora}, Crescenzo and {Tsantaki}, Maria and {Van der Swaelmen}, Mathieu and {van Weeren}, Reinout J. and {Vergani}, Daniela},
        title = "{The wide-field, multiplexed, spectroscopic facility WEAVE: Survey design, overview, and simulated implementation}",
      journal = {\mnras},
         year = 2024,
        month = may,
       volume = {530},
       number = {3},
        pages = {2688-2730},
          doi = {10.1093/mnras/stad557},
archivePrefix = {arXiv},
       eprint = {2212.03981},
 primaryClass = {astro-ph.IM},
       adsurl = {https://ui.adsabs.harvard.edu/abs/2024MNRAS.530.2688J}
}

@ARTICLE{2022ApJ...938L...2F,
       author = {{Ferreira}, Leonardo and {Adams}, Nathan and {Conselice}, Christopher J. and {Sazonova}, Elizaveta and {Austin}, Duncan and {Caruana}, Joseph and {Ferrari}, Fabricio and {Verma}, Aprajita and {Trussler}, James and {Broadhurst}, Tom and {Diego}, Jose and {Frye}, Brenda L. and {Pascale}, Massimo and {Wilkins}, Stephen M. and {Windhorst}, Rogier A. and {Zitrin}, Adi},
        title = "{Panic! at the Disks: First Rest-frame Optical Observations of Galaxy Structure at z > 3 with JWST in the SMACS 0723 Field}",
      journal = {\apjl},
         year = 2022,
        month = oct,
       volume = {938},
       number = {1},
          eid = {L2},
        pages = {L2},
          doi = {10.3847/2041-8213/ac947c},
archivePrefix = {arXiv},
       eprint = {2207.09428},
 primaryClass = {astro-ph.GA},
       adsurl = {https://ui.adsabs.harvard.edu/abs/2022ApJ...938L...2F}
}

@ARTICLE{1999ApJ...523..566C,
       author = {{Carollo}, C. Marcella},
        title = "{The Centers of Early- to Intermediate-Type Spiral Galaxies: A Structural Analysis}",
      journal = {\apj},
         year = 1999,
        month = oct,
       volume = {523},
       number = {2},
        pages = {566-574},
          doi = {10.1086/307753},
       adsurl = {https://ui.adsabs.harvard.edu/abs/1999ApJ...523..566C}
}

@ARTICLE{2011MNRAS.411..385A,
       author = {{Andrae}, Ren{\'e} and {Jahnke}, Knud and {Melchior}, Peter},
        title = "{Parametrizing arbitrary galaxy morphologies: potentials and pitfalls}",
      journal = {\mnras},
         year = 2011,
        month = feb,
       volume = {411},
       number = {1},
        pages = {385-401},
          doi = {10.1111/j.1365-2966.2010.17690.x},
archivePrefix = {arXiv},
       eprint = {1009.2508},
 primaryClass = {astro-ph.CO},
       adsurl = {https://ui.adsabs.harvard.edu/abs/2011MNRAS.411..385A}
}

@ARTICLE{2005ApJ...622L...9S,
       author = {{Springel}, Volker and {Hernquist}, Lars},
        title = "{Formation of a Spiral Galaxy in a Major Merger}",
      journal = {\apjl},
         year = 2005,
        month = mar,
       volume = {622},
       number = {1},
        pages = {L9-L12},
          doi = {10.1086/429486},
archivePrefix = {arXiv},
       eprint = {astro-ph/0411379},
 primaryClass = {astro-ph},
       adsurl = {https://ui.adsabs.harvard.edu/abs/2005ApJ...622L...9S}
}

@ARTICLE{Dressler,
       author = {{Dressler}, A.},
        title = "{Galaxy morphology in rich clusters: implications for the formation and evolution of galaxies.}",
      journal = {\apj},
         year = 1980,
        month = mar,
       volume = {236},
        pages = {351-365},
          doi = {10.1086/157753},
       adsurl = {https://ui.adsabs.harvard.edu/abs/1980ApJ...236..351D}
}

@ARTICLE{1997ApJ...490..577D,
       author = {{Dressler}, Alan and {Oemler}, Augustus, Jr. and {Couch}, Warrick J. and {Smail}, Ian and {Ellis}, Richard S. and {Barger}, Amy and {Butcher}, Harvey and {Poggianti}, Bianca M. and {Sharples}, Ray M.},
        title = "{Evolution since z = 0.5 of the Morphology-Density Relation for Clusters of Galaxies}",
      journal = {\apj},
         year = 1997,
        month = dec,
       volume = {490},
       number = {2},
        pages = {577-591},
          doi = {10.1086/304890},
archivePrefix = {arXiv},
       eprint = {astro-ph/9707232},
 primaryClass = {astro-ph},
       adsurl = {https://ui.adsabs.harvard.edu/abs/1997ApJ...490..577D}
}

@ARTICLE{2017ApJ...844...48P,
       author = {{Poggianti}, Bianca M. and {Moretti}, Alessia and {Gullieuszik}, Marco and {Fritz}, Jacopo and {Jaff{\'e}}, Yara and {Bettoni}, Daniela and {Fasano}, Giovanni and {Bellhouse}, Callum and {Hau}, George and {Vulcani}, Benedetta and {Biviano}, Andrea and {Omizzolo}, Alessandro and {Paccagnella}, Angela and {D'Onofrio}, Mauro and {Cava}, Antonio and {Sheen}, Y. -K. and {Couch}, Warrick and {Owers}, Matt},
        title = "{GASP. I. Gas Stripping Phenomena in Galaxies with MUSE}",
      journal = {\apj},
         year = 2017,
        month = jul,
       volume = {844},
       number = {1},
          eid = {48},
        pages = {48},
          doi = {10.3847/1538-4357/aa78ed},
archivePrefix = {arXiv},
       eprint = {1704.05086},
 primaryClass = {astro-ph.GA},
       adsurl = {https://ui.adsabs.harvard.edu/abs/2017ApJ...844...48P}
}

@ARTICLE{2018MNRAS.476.4753J,
       author = {{Jaff{\'e}}, Yara L. and {Poggianti}, Bianca M. and {Moretti}, Alessia and {Gullieuszik}, Marco and {Smith}, Rory and {Vulcani}, Benedetta and {Fasano}, Giovanni and {Fritz}, Jacopo and {Tonnesen}, Stephanie and {Bettoni}, Daniela and {Hau}, George and {Biviano}, Andrea and {Bellhouse}, Callum and {McGee}, Sean},
        title = "{GASP. IX. Jellyfish galaxies in phase-space: an orbital study of intense ram-pressure stripping in clusters}",
      journal = {\mnras},
         year = 2018,
        month = jun,
       volume = {476},
       number = {4},
        pages = {4753-4764},
          doi = {10.1093/mnras/sty500},
archivePrefix = {arXiv},
       eprint = {1802.07297},
 primaryClass = {astro-ph.GA},
       adsurl = {https://ui.adsabs.harvard.edu/abs/2018MNRAS.476.4753J}
}

@ARTICLE{2019MNRAS.485.1157B,
       author = {{Bellhouse}, Callum and {Jaff{\'e}}, Y.~L. and {McGee}, S.~L. and {Poggianti}, B.~M. and {Smith}, R. and {Tonnesen}, S. and {Fritz}, J. and {Hau}, G.~K.~T. and {Gullieuszik}, M. and {Vulcani}, B. and {Fasano}, G. and {Moretti}, A. and {George}, K. and {Bettoni}, D. and {D'Onofrio}, M. and {Omizzolo}, A. and {Sheen}, Y. -K.},
        title = "{GASP. XV. A MUSE view of extreme ram-pressure stripping along the line of sight: physical properties of the jellyfish galaxy JO201}",
      journal = {\mnras},
         year = 2019,
        month = may,
       volume = {485},
       number = {1},
        pages = {1157-1170},
          doi = {10.1093/mnras/stz460},
archivePrefix = {arXiv},
       eprint = {1902.04486},
 primaryClass = {astro-ph.GA},
       adsurl = {https://ui.adsabs.harvard.edu/abs/2019MNRAS.485.1157B}
}

@ARTICLE{Strateva1,
       author = {{Strateva}, Iskra and {Ivezi{\'c}}, {\v{Z}}eljko and
         {Knapp}, Gillian R. and {Narayanan}, Vijay K. and
         {Strauss}, Michael A. and {Gunn}, James E. and {Lupton}, Robert H. and
         {Schlegel}, David and {Bahcall}, Neta A. and {Brinkmann}, Jon and
         {Brunner}, Robert J. and {Budav{\'a}ri}, Tam{\'a}s and
         {Csabai}, Istv{\'a}n and {Castander}, Francisco Javier and
         {Doi}, Mamoru and {Fukugita}, Masataka and {Gy{\H{o}}ry}, Zsuzsanna and
         {Hamabe}, Masaru and {Hennessy}, Greg and {Ichikawa}, Takashi and
         {Kunszt}, Peter Z. and {Lamb}, Don Q. and {McKay}, Timothy A. and
         {Okamura}, Sadanori and {Racusin}, Judith and {Sekiguchi}, Maki and
         {Schneider}, Donald P. and {Shimasaku}, Kazuhiro and {York}, Donald},
        title = "{Color Separation of Galaxy Types in the Sloan Digital Sky Survey Imaging Data}",
      journal = {\aj},
         year = 2001,
        month = oct,
       volume = {122},
       number = {4},
        pages = {1861-1874},
          doi = {10.1086/323301},
archivePrefix = {arXiv},
       eprint = {astro-ph/0107201},
 primaryClass = {astro-ph},
       adsurl = {https://ui.adsabs.harvard.edu/abs/2001AJ....122.1861S}
}

@ARTICLE{2004ApJ...600..681B,
       author = {{Baldry}, Ivan K. and {Glazebrook}, Karl and {Brinkmann}, Jon and {Ivezi{\'c}}, {\v{Z}}eljko and {Lupton}, Robert H. and {Nichol}, Robert C. and {Szalay}, Alexander S.},
        title = "{Quantifying the Bimodal Color-Magnitude Distribution of Galaxies}",
      journal = {\apj},
         year = 2004,
        month = jan,
       volume = {600},
       number = {2},
        pages = {681-694},
          doi = {10.1086/380092},
archivePrefix = {arXiv},
       eprint = {astro-ph/0309710},
 primaryClass = {astro-ph},
       adsurl = {https://ui.adsabs.harvard.edu/abs/2004ApJ...600..681B}
}

@ARTICLE{2014MNRAS.440..889S,
       author = {{Schawinski}, Kevin and {Urry}, C. Megan and {Simmons}, Brooke D. and {Fortson}, Lucy and {Kaviraj}, Sugata and {Keel}, William C. and {Lintott}, Chris J. and {Masters}, Karen L. and {Nichol}, Robert C. and {Sarzi}, Marc and {Skibba}, Ramin and {Treister}, Ezequiel and {Willett}, Kyle W. and {Wong}, O. Ivy and {Yi}, Sukyoung K.},
        title = "{The green valley is a red herring: Galaxy Zoo reveals two evolutionary pathways towards quenching of star formation in early- and late-type galaxies}",
      journal = {\mnras},
         year = 2014,
        month = may,
       volume = {440},
       number = {1},
        pages = {889-907},
          doi = {10.1093/mnras/stu327},
archivePrefix = {arXiv},
       eprint = {1402.4814},
 primaryClass = {astro-ph.GA},
       adsurl = {https://ui.adsabs.harvard.edu/abs/2014MNRAS.440..889S}
}

@BOOK{1994cag..book.....S,
       author = {{Sandage}, Allan and {Bedke}, John},
        title = "{The Carnegie atlas of galaxies}",
         year = 1994,
       volume = {638},
       adsurl = {https://ui.adsabs.harvard.edu/abs/1994cag..book.....S}
}

@BOOK{1987rsac.book.....S,
       author = {{Sandage}, Allan and {Tammann}, G.~A.},
        title = "{A Revised Shapley-Ames Catalog of Bright Galaxies}",
         year = 1987,
       adsurl = {https://ui.adsabs.harvard.edu/abs/1987rsac.book.....S}
}

@ARTICLE{2010ApJS..186..427N,
       author = {{Nair}, Preethi B. and {Abraham}, Roberto G.},
        title = "{A Catalog of Detailed Visual Morphological Classifications for 14,034 Galaxies in the Sloan Digital Sky Survey}",
      journal = {\apjs},
         year = 2010,
        month = feb,
       volume = {186},
       number = {2},
        pages = {427-456},
          doi = {10.1088/0067-0049/186/2/427},
archivePrefix = {arXiv},
       eprint = {1001.2401},
 primaryClass = {astro-ph.CO},
       adsurl = {https://ui.adsabs.harvard.edu/abs/2010ApJS..186..427N}
}

@ARTICLE{1963BAAA....6...41S,
       author = {{S{\'e}rsic}, J.~L.},
        title = "{Influence of the atmospheric and instrumental dispersion on the brightness distribution in a galaxy}",
      journal = {Boletin de la Asociacion Argentina de Astronomia La Plata Argentina},
         year = 1963,
        month = feb,
       volume = {6},
        pages = {41-43},
       adsurl = {https://ui.adsabs.harvard.edu/abs/1963BAAA....6...41S}
}

@BOOK{1968adga.book.....S,
       author = {{Sersic}, Jose Luis},
        title = "{Atlas de Galaxias Australes}",
         year = 1968,
       adsurl = {https://ui.adsabs.harvard.edu/abs/1968adga.book.....S}
}

@ARTICLE{2003ApJS..147....1C,
       author = {{Conselice}, Christopher J.},
        title = "{The Relationship between Stellar Light Distributions of Galaxies and Their Formation Histories}",
      journal = {\apjs},
         year = 2003,
        month = jul,
       volume = {147},
       number = {1},
        pages = {1-28},
          doi = {10.1086/375001},
archivePrefix = {arXiv},
       eprint = {astro-ph/0303065},
 primaryClass = {astro-ph},
       adsurl = {https://ui.adsabs.harvard.edu/abs/2003ApJS..147....1C}
}

@ARTICLE{2004AJ....128..163L,
       author = {{Lotz}, Jennifer M. and {Primack}, Joel and {Madau}, Piero},
        title = "{A New Nonparametric Approach to Galaxy Morphological Classification}",
      journal = {\aj},
         year = 2004,
        month = jul,
       volume = {128},
       number = {1},
        pages = {163-182},
          doi = {10.1086/421849},
archivePrefix = {arXiv},
       eprint = {astro-ph/0311352},
 primaryClass = {astro-ph},
       adsurl = {https://ui.adsabs.harvard.edu/abs/2004AJ....128..163L}
}

@ARTICLE{2015ApJ...814...55F,
       author = {{Ferrari}, F. and {de Carvalho}, R.~R. and {Trevisan}, M.},
        title = "{Morfometryka{\textemdash}A New Way of Establishing Morphological Classification of Galaxies}",
      journal = {\apj},
         year = 2015,
        month = nov,
       volume = {814},
       number = {1},
          eid = {55},
        pages = {55},
          doi = {10.1088/0004-637X/814/1/55},
archivePrefix = {arXiv},
       eprint = {1509.05430},
 primaryClass = {astro-ph.GA},
       adsurl = {https://ui.adsabs.harvard.edu/abs/2015ApJ...814...55F}
}

@article{rosa2018gradient,
  title={Gradient pattern analysis applied to galaxy morphology},
  author={Rosa, RR and De Carvalho, RR and Sautter, RA and Barchi, PH and Stalder, DH and Moura, TC and Rembold, SB and Morell, DRF and Ferreira, NC},
  journal={Monthly Notices of the Royal Astronomical Society: Letters},
  volume={477},
  number={1},
  pages={L101--L105},
  year={2018},
  publisher={Oxford University Press}
}

@ARTICLE{2020A&C....3000334B,
       author = {{Barchi}, P.~H. and {de Carvalho}, R.~R. and {Rosa}, R.~R. and {Sautter}, R.~A. and {Soares-Santos}, M. and {Marques}, B.~A.~D. and {Clua}, E. and {Gon{\c{c}}alves}, T.~S. and {de S{\'a}-Freitas}, C. and {Moura}, T.~C.},
        title = "{Machine and Deep Learning applied to galaxy morphology - A comparative study}",
      journal = {Astronomy and Computing},
         year = 2020,
        month = jan,
       volume = {30},
          eid = {100334},
        pages = {100334},
          doi = {10.1016/j.ascom.2019.100334},
archivePrefix = {arXiv},
       eprint = {1901.07047},
 primaryClass = {astro-ph.IM},
       adsurl = {https://ui.adsabs.harvard.edu/abs/2020A&C....3000334B}
}

@ARTICLE{2019AJ....157..168D,
       author = {{Dey}, Arjun and {Schlegel}, David J. and {Lang}, Dustin and {Blum}, Robert and {Burleigh}, Kaylan and {Fan}, Xiaohui and {Findlay}, Joseph R. and {Finkbeiner}, Doug and {Herrera}, David and {Juneau}, St{\'e}phanie and {Landriau}, Martin and {Levi}, Michael and {McGreer}, Ian and {Meisner}, Aaron and {Myers}, Adam D. and {Moustakas}, John and {Nugent}, Peter and {Patej}, Anna and {Schlafly}, Edward F. and {Walker}, Alistair R. and {Valdes}, Francisco and {Weaver}, Benjamin A. and {Y{\`e}che}, Christophe and {Zou}, Hu and {Zhou}, Xu and {Abareshi}, Behzad and {Abbott}, T.~M.~C. and {Abolfathi}, Bela and {Aguilera}, C. and {Alam}, Shadab and {Allen}, Lori and {Alvarez}, A. and {Annis}, James and {Ansarinejad}, Behzad and {Aubert}, Marie and {Beechert}, Jacqueline and {Bell}, Eric F. and {BenZvi}, Segev Y. and {Beutler}, Florian and {Bielby}, Richard M. and {Bolton}, Adam S. and {Brice{\~n}o}, C{\'e}sar and {Buckley-Geer}, Elizabeth J. and {Butler}, Karen and {Calamida}, Annalisa and {Carlberg}, Raymond G. and {Carter}, Paul and {Casas}, Ricard and {Castander}, Francisco J. and {Choi}, Yumi and {Comparat}, Johan and {Cukanovaite}, Elena and {Delubac}, Timoth{\'e}e and {DeVries}, Kaitlin and {Dey}, Sharmila and {Dhungana}, Govinda and {Dickinson}, Mark and {Ding}, Zhejie and {Donaldson}, John B. and {Duan}, Yutong and {Duckworth}, Christopher J. and {Eftekharzadeh}, Sarah and {Eisenstein}, Daniel J. and {Etourneau}, Thomas and {Fagrelius}, Parker A. and {Farihi}, Jay and {Fitzpatrick}, Mike and {Font-Ribera}, Andreu and {Fulmer}, Leah and {G{\"a}nsicke}, Boris T. and {Gaztanaga}, Enrique and {George}, Koshy and {Gerdes}, David W. and {Gontcho}, Satya Gontcho A. and {Gorgoni}, Claudio and {Green}, Gregory and {Guy}, Julien and {Harmer}, Diane and {Hernandez}, M. and {Honscheid}, Klaus and {Huang}, Lijuan Wendy and {James}, David J. and {Jannuzi}, Buell T. and {Jiang}, Linhua and {Joyce}, Richard and {Karcher}, Armin and {Karkar}, Sonia and {Kehoe}, Robert and {Kneib}, Jean-Paul and {Kueter-Young}, Andrea and {Lan}, Ting-Wen and {Lauer}, Tod R. and {Le Guillou}, Laurent and {Le Van Suu}, Auguste and {Lee}, Jae Hyeon and {Lesser}, Michael and {Perreault Levasseur}, Laurence and {Li}, Ting S. and {Mann}, Justin L. and {Marshall}, Robert and {Mart{\'\i}nez-V{\'a}zquez}, C.~E. and {Martini}, Paul and {du Mas des Bourboux}, H{\'e}lion and {McManus}, Sean and {Meier}, Tobias Gabriel and {M{\'e}nard}, Brice and {Metcalfe}, Nigel and {Mu{\~n}oz-Guti{\'e}rrez}, Andrea and {Najita}, Joan and {Napier}, Kevin and {Narayan}, Gautham and {Newman}, Jeffrey A. and {Nie}, Jundan and {Nord}, Brian and {Norman}, Dara J. and {Olsen}, Knut A.~G. and {Paat}, Anthony and {Palanque-Delabrouille}, Nathalie and {Peng}, Xiyan and {Poppett}, Claire L. and {Poremba}, Megan R. and {Prakash}, Abhishek and {Rabinowitz}, David and {Raichoor}, Anand and {Rezaie}, Mehdi and {Robertson}, A.~N. and {Roe}, Natalie A. and {Ross}, Ashley J. and {Ross}, Nicholas P. and {Rudnick}, Gregory and {Safonova}, Sasha and {Saha}, Abhijit and {S{\'a}nchez}, F. Javier and {Savary}, Elodie and {Schweiker}, Heidi and {Scott}, Adam and {Seo}, Hee-Jong and {Shan}, Huanyuan and {Silva}, David R. and {Slepian}, Zachary and {Soto}, Christian and {Sprayberry}, David and {Staten}, Ryan and {Stillman}, Coley M. and {Stupak}, Robert J. and {Summers}, David L. and {Sien Tie}, Suk and {Tirado}, H. and {Vargas-Maga{\~n}a}, Mariana and {Vivas}, A. Katherina and {Wechsler}, Risa H. and {Williams}, Doug and {Yang}, Jinyi and {Yang}, Qian and {Yapici}, Tolga and {Zaritsky}, Dennis and {Zenteno}, A. and {Zhang}, Kai and {Zhang}, Tianmeng and {Zhou}, Rongpu and {Zhou}, Zhimin},
        title = "{Overview of the DESI Legacy Imaging Surveys}",
      journal = {\aj},
         year = 2019,
        month = may,
       volume = {157},
       number = {5},
          eid = {168},
        pages = {168},
          doi = {10.3847/1538-3881/ab089d},
archivePrefix = {arXiv},
       eprint = {1804.08657},
 primaryClass = {astro-ph.IM},
       adsurl = {https://ui.adsabs.harvard.edu/abs/2019AJ....157..168D}
}

@ARTICLE{2023Msngr.190...31H,
       author = {{Haines}, C. and {Jaff{\'e}}, Y. and {Tejos}, N. and {Monachesi}, A. and {Pompei}, E. and {Finoguenov}, A. and {Sif{\'o}n}, C. and {Lopez}, S. and {Manjunatha}, A.~B. and {Bilton}, L. and {Comparat}, J. and {Cuellar}, R. and {D'Ago}, G. and {Demarco}, R. and {Lima-Dias}, C. and {L{\"o}sch}, E. and {Merluzzi}, P. and {Smith Castelli}, A. and {Sodre}, L. and {Vinicius}, E. and {CHANCES Team}},
        title = "{CHANCES: A CHileAN Cluster galaxy Evolution Survey}",
      journal = {The Messenger},
         year = 2023,
        month = mar,
       volume = {190},
        pages = {31-33},
          doi = {10.18727/0722-6691/5308},
       adsurl = {https://ui.adsabs.harvard.edu/abs/2023Msngr.190...31H}
}

@ARTICLE{2016A&A...594A..13P,
       author = {{Planck Collaboration} and {Ade}, P.~A.~R. and {Aghanim}, N. and {Arnaud}, M. and {Ashdown}, M. and {Aumont}, J. and {Baccigalupi}, C. and {Banday}, A.~J. and {Barreiro}, R.~B. and {Bartlett}, J.~G. and {Bartolo}, N. and {Battaner}, E. and {Battye}, R. and {Benabed}, K. and {Beno{\^\i}t}, A. and {Benoit-L{\'e}vy}, A. and {Bernard}, J. -P. and {Bersanelli}, M. and {Bielewicz}, P. and {Bock}, J.~J. and {Bonaldi}, A. and {Bonavera}, L. and {Bond}, J.~R. and {Borrill}, J. and {Bouchet}, F.~R. and {Boulanger}, F. and {Bucher}, M. and {Burigana}, C. and {Butler}, R.~C. and {Calabrese}, E. and {Cardoso}, J. -F. and {Catalano}, A. and {Challinor}, A. and {Chamballu}, A. and {Chary}, R. -R. and {Chiang}, H.~C. and {Chluba}, J. and {Christensen}, P.~R. and {Church}, S. and {Clements}, D.~L. and {Colombi}, S. and {Colombo}, L.~P.~L. and {Combet}, C. and {Coulais}, A. and {Crill}, B.~P. and {Curto}, A. and {Cuttaia}, F. and {Danese}, L. and {Davies}, R.~D. and {Davis}, R.~J. and {de Bernardis}, P. and {de Rosa}, A. and {de Zotti}, G. and {Delabrouille}, J. and {D{\'e}sert}, F. -X. and {Di Valentino}, E. and {Dickinson}, C. and {Diego}, J.~M. and {Dolag}, K. and {Dole}, H. and {Donzelli}, S. and {Dor{\'e}}, O. and {Douspis}, M. and {Ducout}, A. and {Dunkley}, J. and {Dupac}, X. and {Efstathiou}, G. and {Elsner}, F. and {En{\ss}lin}, T.~A. and {Eriksen}, H.~K. and {Farhang}, M. and {Fergusson}, J. and {Finelli}, F. and {Forni}, O. and {Frailis}, M. and {Fraisse}, A.~A. and {Franceschi}, E. and {Frejsel}, A. and {Galeotta}, S. and {Galli}, S. and {Ganga}, K. and {Gauthier}, C. and {Gerbino}, M. and {Ghosh}, T. and {Giard}, M. and {Giraud-H{\'e}raud}, Y. and {Giusarma}, E. and {Gjerl{\o}w}, E. and {Gonz{\'a}lez-Nuevo}, J. and {G{\'o}rski}, K.~M. and {Gratton}, S. and {Gregorio}, A. and {Gruppuso}, A. and {Gudmundsson}, J.~E. and {Hamann}, J. and {Hansen}, F.~K. and {Hanson}, D. and {Harrison}, D.~L. and {Helou}, G. and {Henrot-Versill{\'e}}, S. and {Hern{\'a}ndez-Monteagudo}, C. and {Herranz}, D. and {Hildebrandt}, S.~R. and {Hivon}, E. and {Hobson}, M. and {Holmes}, W.~A. and {Hornstrup}, A. and {Hovest}, W. and {Huang}, Z. and {Huffenberger}, K.~M. and {Hurier}, G. and {Jaffe}, A.~H. and {Jaffe}, T.~R. and {Jones}, W.~C. and {Juvela}, M. and {Keih{\"a}nen}, E. and {Keskitalo}, R. and {Kisner}, T.~S. and {Kneissl}, R. and {Knoche}, J. and {Knox}, L. and {Kunz}, M. and {Kurki-Suonio}, H. and {Lagache}, G. and {L{\"a}hteenm{\"a}ki}, A. and {Lamarre}, J. -M. and {Lasenby}, A. and {Lattanzi}, M. and {Lawrence}, C.~R. and {Leahy}, J.~P. and {Leonardi}, R. and {Lesgourgues}, J. and {Levrier}, F. and {Lewis}, A. and {Liguori}, M. and {Lilje}, P.~B. and {Linden-V{\o}rnle}, M. and {L{\'o}pez-Caniego}, M. and {Lubin}, P.~M. and {Mac{\'\i}as-P{\'e}rez}, J.~F. and {Maggio}, G. and {Maino}, D. and {Mandolesi}, N. and {Mangilli}, A. and {Marchini}, A. and {Maris}, M. and {Martin}, P.~G. and {Martinelli}, M. and {Mart{\'\i}nez-Gonz{\'a}lez}, E. and {Masi}, S. and {Matarrese}, S. and {McGehee}, P. and {Meinhold}, P.~R. and {Melchiorri}, A. and {Melin}, J. -B. and {Mendes}, L. and {Mennella}, A. and {Migliaccio}, M. and {Millea}, M. and {Mitra}, S. and {Miville-Desch{\^e}nes}, M. -A. and {Moneti}, A. and {Montier}, L. and {Morgante}, G. and {Mortlock}, D. and {Moss}, A. and {Munshi}, D. and {Murphy}, J.~A. and {Naselsky}, P. and {Nati}, F. and {Natoli}, P. and {Netterfield}, C.~B. and {N{\o}rgaard-Nielsen}, H.~U. and {Noviello}, F. and {Novikov}, D. and {Novikov}, I. and {Oxborrow}, C.~A. and {Paci}, F. and {Pagano}, L. and {Pajot}, F. and {Paladini}, R. and {Paoletti}, D. and {Partridge}, B. and {Pasian}, F. and {Patanchon}, G. and {Pearson}, T.~J. and {Perdereau}, O. and {Perotto}, L. and {Perrotta}, F. and {Pettorino}, V. and {Piacentini}, F. and {Piat}, M. and {Pierpaoli}, E. and {Pietrobon}, D. and {Plaszczynski}, S. and {Pointecouteau}, E. and {Polenta}, G. and {Popa}, L. and {Pratt}, G.~W. and {Pr{\'e}zeau}, G. and {Prunet}, S. and {Puget}, J. -L. and {Rachen}, J.~P. and {Reach}, W.~T. and {Rebolo}, R. and {Reinecke}, M. and {Remazeilles}, M. and {Renault}, C. and {Renzi}, A. and {Ristorcelli}, I. and {Rocha}, G. and {Rosset}, C. and {Rossetti}, M. and {Roudier}, G. and {Rouill{\'e} d'Orfeuil}, B. and {Rowan-Robinson}, M. and {Rubi{\~n}o-Mart{\'\i}n}, J.~A. and {Rusholme}, B. and {Said}, N. and {Salvatelli}, V. and {Salvati}, L. and {Sandri}, M. and {Santos}, D. and {Savelainen}, M. and {Savini}, G. and {Scott}, D. and {Seiffert}, M.~D. and {Serra}, P. and {Shellard}, E.~P.~S. and {Spencer}, L.~D. and {Spinelli}, M. and {Stolyarov}, V. and {Stompor}, R. and {Sudiwala}, R. and {Sunyaev}, R. and {Sutton}, D. and {Suur-Uski}, A. -S. and {Sygnet}, J. -F. and {Tauber}, J.~A. and {Terenzi}, L. and {Toffolatti}, L. and {Tomasi}, M. and {Tristram}, M. and {Trombetti}, T. and {Tucci}, M. and {Tuovinen}, J. and {T{\"u}rler}, M. and {Umana}, G. and {Valenziano}, L. and {Valiviita}, J. and {Van Tent}, F. and {Vielva}, P. and {Villa}, F. and {Wade}, L.~A. and {Wandelt}, B.~D. and {Wehus}, I.~K. and {White}, M. and {White}, S.~D.~M. and {Wilkinson}, A. and {Yvon}, D. and {Zacchei}, A. and {Zonca}, A.},
        title = "{Planck 2015 results. XIII. Cosmological parameters}",
      journal = {\aap},
         year = 2016,
        month = sep,
       volume = {594},
          eid = {A13},
        pages = {A13},
          doi = {10.1051/0004-6361/201525830},
archivePrefix = {arXiv},
       eprint = {1502.01589},
 primaryClass = {astro-ph.CO},
       adsurl = {https://ui.adsabs.harvard.edu/abs/2016A&A...594A..13P}
}

@ARTICLE{2022MNRAS.509.3966W,
       author = {{Walmsley}, Mike and {Lintott}, Chris and {G{\'e}ron}, Tobias and {Kruk}, Sandor and {Krawczyk}, Coleman and {Willett}, Kyle W. and {Bamford}, Steven and {Kelvin}, Lee S. and {Fortson}, Lucy and {Gal}, Yarin and {Keel}, William and {Masters}, Karen L. and {Mehta}, Vihang and {Simmons}, Brooke D. and {Smethurst}, Rebecca and {Smith}, Lewis and {Baeten}, Elisabeth M. and {Macmillan}, Christine},
        title = "{Galaxy Zoo DECaLS: Detailed visual morphology measurements from volunteers and deep learning for 314 000 galaxies}",
      journal = {\mnras},
         year = 2022,
        month = jan,
       volume = {509},
       number = {3},
        pages = {3966-3988},
          doi = {10.1093/mnras/stab2093},
archivePrefix = {arXiv},
       eprint = {2102.08414},
 primaryClass = {astro-ph.GA},
       adsurl = {https://ui.adsabs.harvard.edu/abs/2022MNRAS.509.3966W}
}

@ARTICLE{2008MNRAS.389.1179L,
       author = {{Lintott}, Chris J. and {Schawinski}, Kevin and {Slosar}, An{\v{z}}e and {Land}, Kate and {Bamford}, Steven and {Thomas}, Daniel and {Raddick}, M. Jordan and {Nichol}, Robert C. and {Szalay}, Alex and {Andreescu}, Dan and {Murray}, Phil and {Vandenberg}, Jan},
        title = "{Galaxy Zoo: morphologies derived from visual inspection of galaxies from the Sloan Digital Sky Survey}",
      journal = {\mnras},
         year = 2008,
        month = sep,
       volume = {389},
       number = {3},
        pages = {1179-1189},
          doi = {10.1111/j.1365-2966.2008.13689.x},
archivePrefix = {arXiv},
       eprint = {0804.4483},
 primaryClass = {astro-ph},
       adsurl = {https://ui.adsabs.harvard.edu/abs/2008MNRAS.389.1179L}
}

@ARTICLE{2000AJ....119.2645B,
       author = {{Bershady}, Matthew A. and {Jangren}, Anna and {Conselice}, Christopher J.},
        title = "{Structural and Photometric Classification of Galaxies. I. Calibration Based on a Nearby Galaxy Sample}",
      journal = {\aj},
         year = 2000,
        month = jun,
       volume = {119},
       number = {6},
        pages = {2645-2663},
          doi = {10.1086/301386},
archivePrefix = {arXiv},
       eprint = {astro-ph/0002262},
 primaryClass = {astro-ph},
       adsurl = {https://ui.adsabs.harvard.edu/abs/2000AJ....119.2645B}
}

@ARTICLE{2016MNRAS.456.3032P,
       author = {{Pawlik}, M.~M. and {Wild}, V. and {Walcher}, C.~J. and {Johansson}, P.~H. and {Villforth}, C. and {Rowlands}, K. and {Mendez-Abreu}, J. and {Hewlett}, T.},
        title = "{Shape asymmetry: a morphological indicator for automatic detection of galaxies in the post-coalescence merger stages}",
      journal = {\mnras},
         year = 2016,
        month = mar,
       volume = {456},
       number = {3},
        pages = {3032-3052},
          doi = {10.1093/mnras/stv2878},
archivePrefix = {arXiv},
       eprint = {1512.02000},
 primaryClass = {astro-ph.GA},
       adsurl = {https://ui.adsabs.harvard.edu/abs/2016MNRAS.456.3032P}
}

@ARTICLE{2000ApJ...529..886C,
       author = {{Conselice}, Christopher J. and {Bershady}, Matthew A. and {Jangren}, Anna},
        title = "{The Asymmetry of Galaxies: Physical Morphology for Nearby and High-Redshift Galaxies}",
      journal = {\apj},
         year = 2000,
        month = feb,
       volume = {529},
       number = {2},
        pages = {886-910},
          doi = {10.1086/308300},
archivePrefix = {arXiv},
       eprint = {astro-ph/9907399},
 primaryClass = {astro-ph},
       adsurl = {https://ui.adsabs.harvard.edu/abs/2000ApJ...529..886C}
}

@ARTICLE{2024MNRAS.528...82K,
       author = {{Kolesnikov}, I. and {Sampaio}, V.~M. and {de Carvalho}, R.~R. and {Conselice}, C. and {Rembold}, S.~B. and {Mendes}, C.~L. and {Rosa}, R.~R.},
        title = "{Unveiling galaxy morphology through an unsupervised-supervised hybrid approach}",
      journal = {\mnras},
         year = 2024,
        month = feb,
       volume = {528},
       number = {1},
        pages = {82-107},
          doi = {10.1093/mnras/stad3934},
archivePrefix = {arXiv},
       eprint = {2401.08906},
 primaryClass = {astro-ph.IM},
       adsurl = {https://ui.adsabs.harvard.edu/abs/2024MNRAS.528...82K}
}

@ARTICLE{2025MNRAS.539.2765K,
       author = {{Kolesnikov}, I. and {Sampaio}, V.~M. and {de Carvalho}, R.~R. and {Conselice}, C.},
        title = "{Galaxy morphology in CANDELS: addressing evolutionary changes across 0.2 {\ensuremath{\leqslant}} z {\ensuremath{\leqslant}} 2.4 with hybrid classification approach}",
      journal = {\mnras},
         year = 2025,
        month = may,
       volume = {539},
       number = {3},
        pages = {2765-2779},
          doi = {10.1093/mnras/staf625},
archivePrefix = {arXiv},
       eprint = {2412.03778},
 primaryClass = {astro-ph.GA},
       adsurl = {https://ui.adsabs.harvard.edu/abs/2025MNRAS.539.2765K}
}

@ARTICLE{2018MNRAS.476.3661D,
       author = {{Dom{\'\i}nguez S{\'a}nchez}, H. and {Huertas-Company}, M. and {Bernardi}, M. and {Tuccillo}, D. and {Fischer}, J.~L.},
        title = "{Improving galaxy morphologies for SDSS with Deep Learning}",
      journal = {\mnras},
         year = 2018,
        month = feb,
       volume = {476},
       number = {3},
        pages = {3661-3676},
          doi = {10.1093/mnras/sty338},
archivePrefix = {arXiv},
       eprint = {1711.05744},
 primaryClass = {astro-ph.GA},
       adsurl = {https://ui.adsabs.harvard.edu/abs/2018MNRAS.476.3661D}
}

@ARTICLE{2011arXiv1106.1813C,
       author = {{Chawla}, N.~V. and {Bowyer}, K.~W. and {Hall}, L.~O. and {Kegelmeyer}, W.~P.},
        title = "{SMOTE: Synthetic Minority Over-sampling Technique}",
      journal = {arXiv e-prints},
         year = 2011,
        month = jun,
          eid = {arXiv:1106.1813},
        pages = {arXiv:1106.1813},
          doi = {10.48550/arXiv.1106.1813},
archivePrefix = {arXiv},
       eprint = {1106.1813},
 primaryClass = {cs.AI},
       adsurl = {https://ui.adsabs.harvard.edu/abs/2011arXiv1106.1813C}
}

@article{ke2017lightgbm,
  title={Lightgbm: A highly efficient gradient boosting decision tree},
  author={Ke, Guolin and Meng, Qi and Finley, Thomas and Wang, Taifeng and Chen, Wei and Ma, Weidong and Ye, Qiwei and Liu, Tie-Yan},
  journal={Advances in neural information processing systems},
  volume={30},
  year={2017}
}

@ARTICLE{1950MWRv...78....1B,
       author = {{Brier}, Glenn W.},
        title = "{Verification of Forecasts Expressed in Terms of Probability}",
      journal = {Monthly Weather Review},
         year = 1950,
        month = jan,
       volume = {78},
       number = {1},
        pages = {1},
          doi = {10.1175/1520-0493(1950)078<0001:VOFEIT>2.0.CO;2},
       adsurl = {https://ui.adsabs.harvard.edu/abs/1950MWRv...78....1B}
}

@inproceedings{NIPS2017_8a20a862,
 author = {Lundberg, Scott M and Lee, Su-In},
 booktitle = {Advances in Neural Information Processing Systems},
 editor = {I. Guyon and U. Von Luxburg and S. Bengio and H. Wallach and R. Fergus and S. Vishwanathan and R. Garnett},
 pages = {},
 publisher = {Curran Associates, Inc.},
 title = {A Unified Approach to Interpreting Model Predictions},
 url = {https://proceedings.neurips.cc/paper_files/paper/2017/file/8a20a8621978632d76c43dfd28b67767-Paper.pdf},
 volume = {30},
 year = {2017}
}
\begin{appendix}

\section{Acknowledgements}
We thank the referee for the suggestions that helped improving this paper. This research made use of the Python programming language \citep{VanRossum2009} and the packages NumPy \citep{Harris2020}, SciPy \citep{Virtanen2020}, Astropy \citep{Astropy2013,Astropy2018,Astropy2022}, pandas \citep{McKinney2010,Reback2022}, and Matplotlib \citep{Hunter2007}. This work has been supported by the Agencia Nacional de Investigación y Desarrollo (ANID) through the Millennium Science Initiative Program NCN2024\_112 (VMS, YLJ, HME); the BASAL project FB210003 (YLJ, HME, AM); the FONDECYT Regular projects 1241426 and 1230441 (YLJ) and 1251882 (AM); and the FONDECYT project 3250511 (CLD). VMS acknowledges additional support from ESO through grant ORP026/2021, and CLD from the ESO Comité Mixto through grant ORP037/2022. AM further acknowledges funding from the HORIZON-MSCA-2021-SE-01 Research and Innovation Programme under the Marie Sklodowska-Curie grant agreement No. 101086388. VMS thanks RRdC and IK for the fruitful discussions.

\section{Data Availability}

The data underlying this article can be requested to the corresponding author. We also make the separated elliptical and spiral catalogs available \href{https://www.dropbox.com/scl/fi/p48gv1xh6vwvdwo1v4jep/galmex_run_ellipticals_gz1_xmatch_gzdecals.csv?rlkey=q2rvi63qfas9cnrwo1ht2xznh&st=c14tipsz&dl=0}{here} and \href{https://www.dropbox.com/scl/fi/pwtjkm1lmwhcaiiyceayh/galmex_run_spirals_gz1_xmatch_gzdecals.csv?rlkey=ch9rmpzgy50l7yw1api98g3i6&st=43geqmlk&dl=0}{here}, respectively, and the full $\sim 1.7$ million galaxies catalog \href{https://www.dropbox.com/scl/fi/w748q3vx0pqr5nirvv0gz/galmex_run_ls_zphot0.15_magr18.5_mu26_shaper2_rband.csv?rlkey=v6esqil87iilgth6z1dvm8xq8&st=yfd6hx3u&dl=0}{here}. A 
\textit{readme} file can be found \href{https://www.dropbox.com/scl/fi/gbpz731u17wr6tu3ylk43/galmex_run_readme.txt?rlkey=b1htfzckfw9lyac5gm9dvug0m&st=zmfse4eh&dl=0}{here}. It will also be provided in the Strasbourg astronomical Data Center (CDS), and accessible from the Astrophysics Data System (ADS). All the codes used to generate results and plots of this paper are available at \url{https://github.com/vitorms99}.

\section{Comparison between Galaxy Zoo 1 and Galaxy Zoo DECALS}
\label{ap:GZ_comparison}

\begin{figure}
    \centering
    \includegraphics[width=0.8\columnwidth]{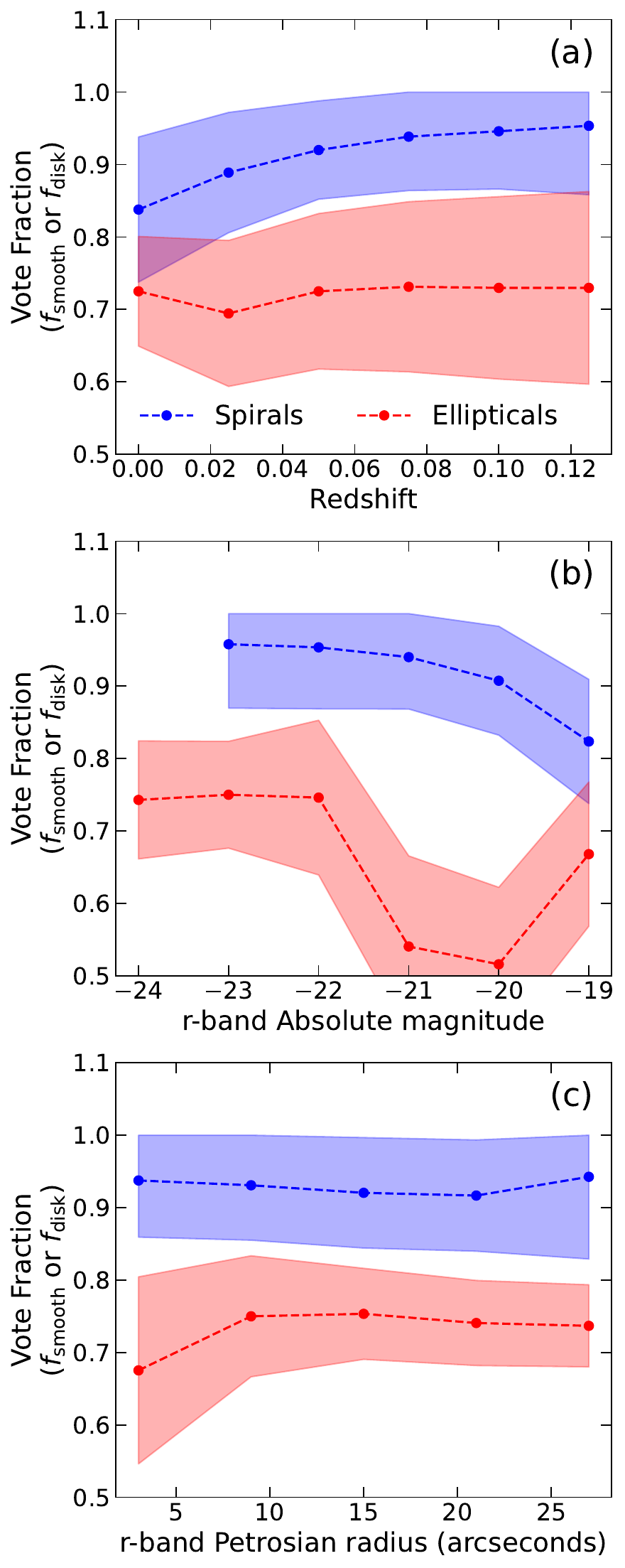}
    \caption{Variation of $f_{\rm smooth}$ for ellipticals, and $f_{\rm disk}$ for spirals, as a function of (from top to bottom) redshift, absolute magnitude in the r-band and Petrosian radius (in arcseconds).}
    \label{fig:gz_comparison}
\end{figure}

In this Appendix, we present a comparison between the Galaxy Zoo 1 and Galaxy Zoo DECaLS. In Fig.~\ref{fig:gz_comparison}, we show the variation of $f_{\rm smooth}$ for ellipticals, and $f_{\rm disk}$ for spirals, as a function of redshift, absolute magnitude in the r-band and Petrosian radius. First, $f_{\rm smooth}$ is always smaller than $f_{\rm disk}$. Irrespective of considered panel, the Galaxy Zoo 1 ellipticals is classified as "smooth" by roughly 70\% of the voters. This may indicate a direct influence of the adopted scheme in Galaxy Zoo DECaLS, in which the top-level question ("smooth" or "disk/feature") is considerably subjective, and the concept of an "smooth" is somewhat vague. Thus, even in elliptical galaxies (according to Galaxy Zoo 1), the vote fraction does not reach high percentages ($\geq 80\%$). This has relevant implications to CNN models that use the Galaxy Zoo DECaLS as training samples. 

\begin{figure*}
    \centering
    \includegraphics[width=0.8\textwidth]{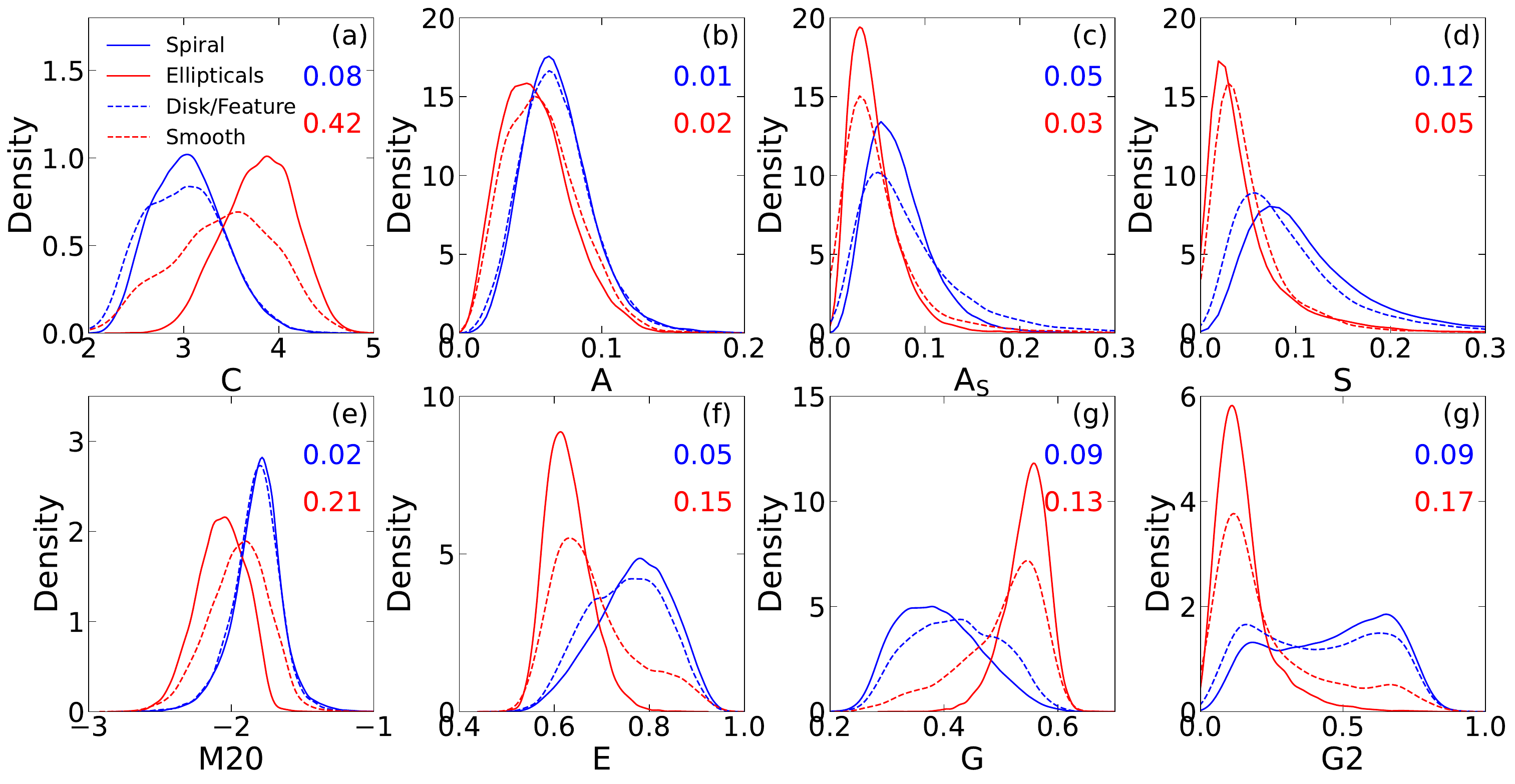}
    \caption{Non parametric indices distribution for the smooth (red dashed), disk/feature (blue dashed), elliptical (red solid), and spiral (blue solid) subsamples. In each panel we also show the energy distance value for the comparison between smooth and elliptical distributions (in red), and between disk/feature and spiral distributions (in blue). Notably, even though adopting a considerable restrictive threshold for the smooth and disk/feature subsamples, there are significant differences in the metrics distribution.}
    \label{fig:metrics_gz_comparison}
\end{figure*}

To investigate the variations in the metrics when using subsamples selected directly from the Galaxy Zoo DECaLS, we select galaxies "smooth" and "disk/feature" subsamples as it follows:
\begin{itemize}
    \item Smooth: ($f_{\rm smooth} \geq 0.7$) and ($f_{\rm disk} \leq 0.3$);
    \item Disk/Feature: ($f_{\rm smooth} \leq 0.3$) and ($f_{\rm disk} \geq 0.7$).
\end{itemize}
In Fig.~\ref{fig:metrics_gz_comparison}, we show the CA$\rm [A_S]$S+MEGG distributions for the "smooth" and "disk/feature" samples (dashed lines), besides the Galaxy Zoo 1 Spiral and Elliptical samples (solid lines). Quantitatively, we compare the smooth with the elliptical, and the disk/feature with the disk distributions using the energy distance parameter. The energy distance between two probability distributions $P$ and $Q$ is defined as
\begin{equation}
D_E(P,Q) \;=\; 2\,\mathbb{E}\bigl[\lVert X - Y \rVert \bigr]
 \;-\; \mathbb{E}\bigl[\lVert X - X' \rVert \bigr]
 \;-\; \mathbb{E}\bigl[\lVert Y - Y' \rVert \bigr],
\end{equation}
where $X,X' \sim P$ and $Y,Y' \sim Q$ are independent random variables, and $\mathbb{E}$ denotes the expectation of each comparison. This metric is non–negative and equals zero if and only if $P=Q$, making it a useful tool for quantifying differences. Notably, the larger differences are found in the comparison between smooth and elliptical subsamples, reinforcing that classifying galaxies as "smooth" or "disk/feature" is not equivalent to the first order separation between ellipticals and spirals. Moreover, panels (a), (f), and (g) show the results for the metrics pointed as the most relevant for the LightGBM method, with the difference in C (second in feature importance) being the largest among all non-parametric indexes. 

\begin{figure*}
    \centering
    \includegraphics[width=0.8\textwidth]{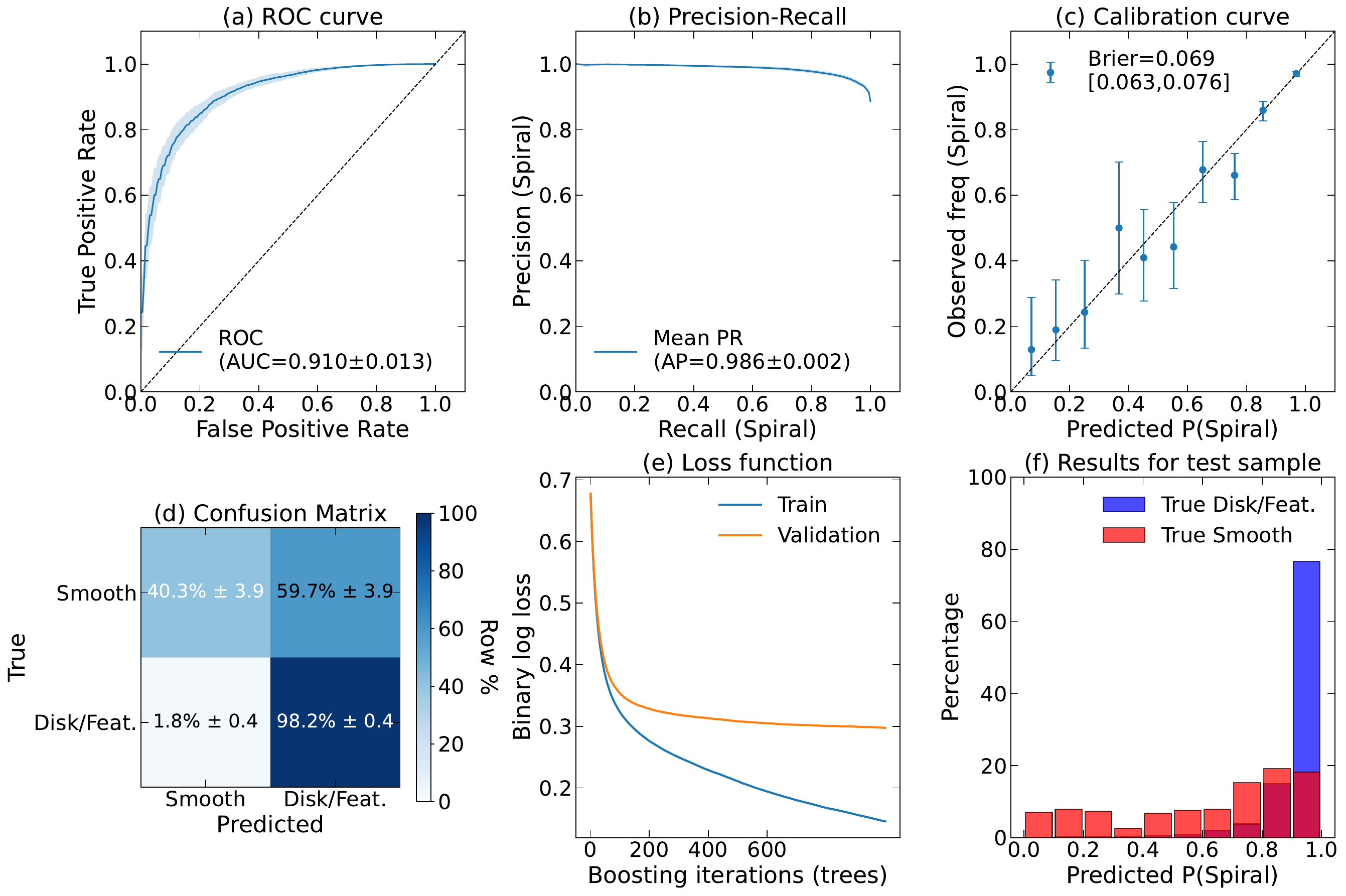}
    \caption{Similar to Fig.~\ref{fig:lightgbm_results}, but using the the smooth and disk/feature as training samples.}
    \label{fig:lightgbm_gz_comparison}
\end{figure*}

Finally, we show in Fig.~\ref{fig:lightgbm_gz_comparison} the lightGBM performance when using the "smooth" and "disk/feature" subsamples as training set. Notably, the performance is considerably worse than when we use the elliptical/spiral subsamples. While the accuracy for disk/feature is similar between Figs.\ref{fig:lightgbm_results} and \ref{fig:lightgbm_gz_comparison}, the major difference is found in the counterpart. Again, this reinforces our suggestion that the separation between "smooth" and "disk/feature" is considerably subjective, and does not link directly to the elliptical/spiral separation, especially in the case of ellipticals.

\section{Comparison between spectroscopic and photometric redshifts}
\label{ap:redshift}
\begin{figure}
    \centering
    \includegraphics[width=\linewidth]{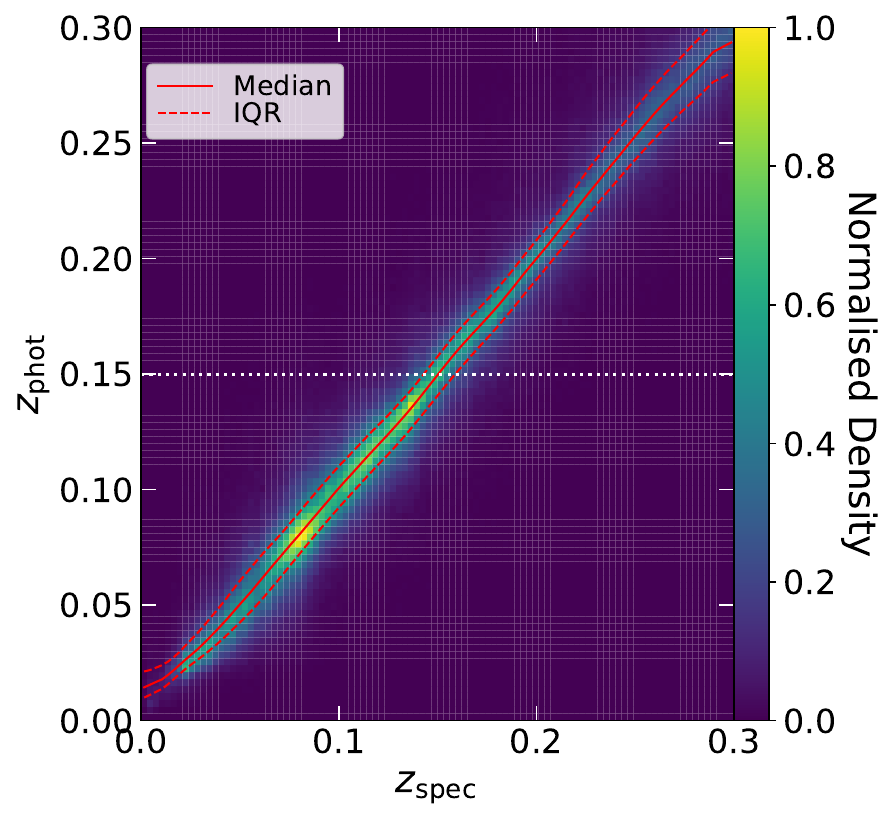}
    \caption{Number density of galaxies in the $z_{\rm spec}$ vs. $z_{\rm phot}$ diagram. The solid and dashed red lines denote the median and IQR of the distribution at a given $z_{\rm spec}$. The agreement between $z_{\rm spec}$ vs. $z_{\rm phot}$ ensures that we are not introducing bias in our morphological classification due to adopting a cut in photometric redshift.}
    \label{fig:redshift_comp}
\end{figure}

In this appendix, we present the comparison between spectroscopic and photometric redshift, which justifies our choice of applying a redshift threshold in our sample, even though using a photometric redshift. In Fig.~\ref{fig:redshift_comp}, we show the normalized density of galaxies in the $z_{\rm spec}$ vs. $z_{\rm phot}$ diagram. The plot encompasses a total of 819,043 galaxies. The red dashed lines denote the interquartile range (IQR), the solid red line shows the median at each $z_{\rm spec}$, and the white dotted line shows the threshold we adopt in this work. Notably, in the local universe ($z<0.3$) both quantities show excellent agreement, ensuring that we are not introducing bias in our morphological classifications due to uncertainties in the photometric redshift.

\section{Metrics definition}

The \texttt{galmex} package computes a comprehensive set of non-parametric morphological indices, each designed to capture different aspects of galaxy structure. Below we summarize their definitions:
\begin{itemize}
    \item Concentration (C):  
    Quantifies how centrally concentrated the light distribution is \citep{2000AJ....119.2645B}. It is defined as
    \begin{equation}
        C = 5 \, \log \left( \frac{r_{80}}{r_{20}} \right),
    \end{equation}
    where $r_{20}$ and $r_{80}$ are the radii enclosing 20\% and 80\% of the total flux, respectively. Larger values correspond to more bulge-dominated systems.
    \item Asymmetry (A):  
    Measures the degree of $180^\circ$ rotational symmetry \citep{2000ApJ...529..886C}. It is computed as
    \begin{equation}
        A = \min_{(x_c, y_c)} \left( \frac{\sum | I(i,j) - I_{180}(i,j) |}{\sum | I(i,j) |} \right) - 
            \left( \frac{\sum | B(i,j) - B_{180}(i,j) |}{\sum | I(i,j) |} \right),
    \end{equation}
    where $I(i,j)$ is the galaxy flux, $I_{180}(i,j)$ is the image rotated by $180^\circ$ about a center $(x_c,y_c)$, and the second term subtracts the contribution from background noise estimated in a representative segment ($B$) in the image containing only background pixels. The galaxy center is iteratively adjusted to minimize the galaxy term.
    \item Shape Asymmetry ($\rm A_S$):  
    Similar to $A$, but applied to the binary segmentation map instead of the flux image \citep{2016MNRAS.456.3032P}. Measures rotational asymmetry in the segmentation mask rather than in the flux distribution, thereby enhancing sensitivity to faint asymmetric structures such as tidal features \citep{2016MNRAS.456.3032P}. 
    The shape asymmetry is defined as
    \begin{equation}
    \label{eq:As}
        A_S = \frac{1}{2N_{\rm pix}} \sum_{i=1}^{N_{\rm pix}} \left| M(i,j) - M_{180}(i,j) \right|,
    \end{equation}
    where $M(i,j)$ is the binary segmentation map, $M_{180}(i,j)$ is its $180^\circ$ rotation about the galaxy center, and $N_{\rm pix}$ is the number of pixels in the mask. $A_S$ ranges from 0 for perfectly symmetric masks to 1 for completely asymmetric ones, and is particularly effective at identifying mergers and disturbed morphologies.
    \item Smoothness (S):  
    this measures the fraction of light in high-frequency structures \citep{2003ApJS..147....1C}. It is defined as
    \begin{equation}
        S = \frac{\sum | I(i,j) - I_S(i,j) |}{\sum | I(i,j) |},
    \end{equation}
    where $I_S(i,j)$ is a smoothed version of the image (in this case, convolved with a boxcar filter of width $0.25\,R_{\rm P}$). Therefore, high values of smoothness actually means a higher degree of clumpiness. Unlike the original definition, we omit the factor of 10 to ensure that it will be in range 0 to 1, similar to the other metrics.
    \item Second-order moment of light (M20):  
    Measures the spatial distribution of the brightest regions \citep{2004AJ....128..163L}. The total second-order moment is
    \begin{equation}
        M_{\rm tot} = \sum_i f_i \left[ (x_i - x_c)^2 + (y_i - y_c)^2 \right],
    \end{equation}
    where $f_i$ is the flux in pixel $i$, and $(x_c,y_c)$ is the galaxy center. $M_{20}$ is then
    \begin{equation}
        M_{20} = \log \left( \frac{\sum_i M_i}{M_{\rm tot}} \right), \quad \text{with } \sum f_i \leq 0.2 F_{\rm tot}.
    \end{equation}
    More negative values indicate compact, bulge-like structures, while higher values trace extended or clumpy star formation.
    \item Shannon Entropy (E):  
    Quantifies the uniformity in the flux distribution \citep{2015ApJ...814...55F}. Let $p_i = f_i / \sum_j f_j$ be the normalized flux distribution. Then
    \begin{equation}
        E = -\sum_{i=1}^{N_{\rm p}} p_i \log p_i,
    \end{equation}
    where $N_{\rm p}$ is the number of bins used in computation. Differently from previous works, instead of fixing the number of bins for all galaxies, we define the bin width for each galaxy using the relation $IQR / N^{1/3}$, where IQR is the inter-quartile range ($Q_{75} - Q_{25}$) and N is the number of pixels in the segmentation mask. Lower entropy values correspond to centrally concentrated systems, while higher values indicate more uniform, disk-like distributions.
    \item Gini Index (G):  
    Measures the inequality of the flux distribution across pixels \citep{2004AJ....128..163L}. For pixel fluxes $f_i$ sorted in ascending order,
    \begin{equation}
        G = \frac{1}{\bar{f} N_{\rm p}(N_{\rm p}-1)} \sum_{i=1}^{N_{\rm p}} (2i - N_{\rm p} - 1) f_i,
    \end{equation}
    where $\bar{f}$ is the mean pixel flux. $G$ ranges from 0 (uniform distribution) to 1 (all flux in one pixel). Bulge-dominated systems tend to have high $G$.
    \item Gradient Pattern Asymmetry (G2):  
    Based on Gradient Pattern Analysis, $G_2$ measures bilateral asymmetries in the image gradient field \citep{rosa2018gradient}. The gradient vector field is constructed across the image, pairing vectors equidistant from the galaxy center. Symmetric pairs are discarded, while asymmetric vectors are used to define the ``confluence'' parameter
    \begin{equation}
        cf = \frac{\left|\sum_i v_a^i \right|}{\sum_i |v_a^i|},
    \end{equation}
    where $v_a^i$ are the asymmetric vectors. The $G_2$ index is then given by
    \begin{equation}
        G_2 = \frac{V_{\rm A}}{V} \, (1 - cf),
    \end{equation}
    where $V_{\rm A}$ is the number of asymmetric vectors, $V$ the total number of pixels, and $cf$ the confluence correction. 
\end{itemize}

\section{Description of adopted image pre-processing}
\label{ap:image_preprocessing}

\begin{figure*}
    \centering
    \includegraphics[width=0.8\textwidth]{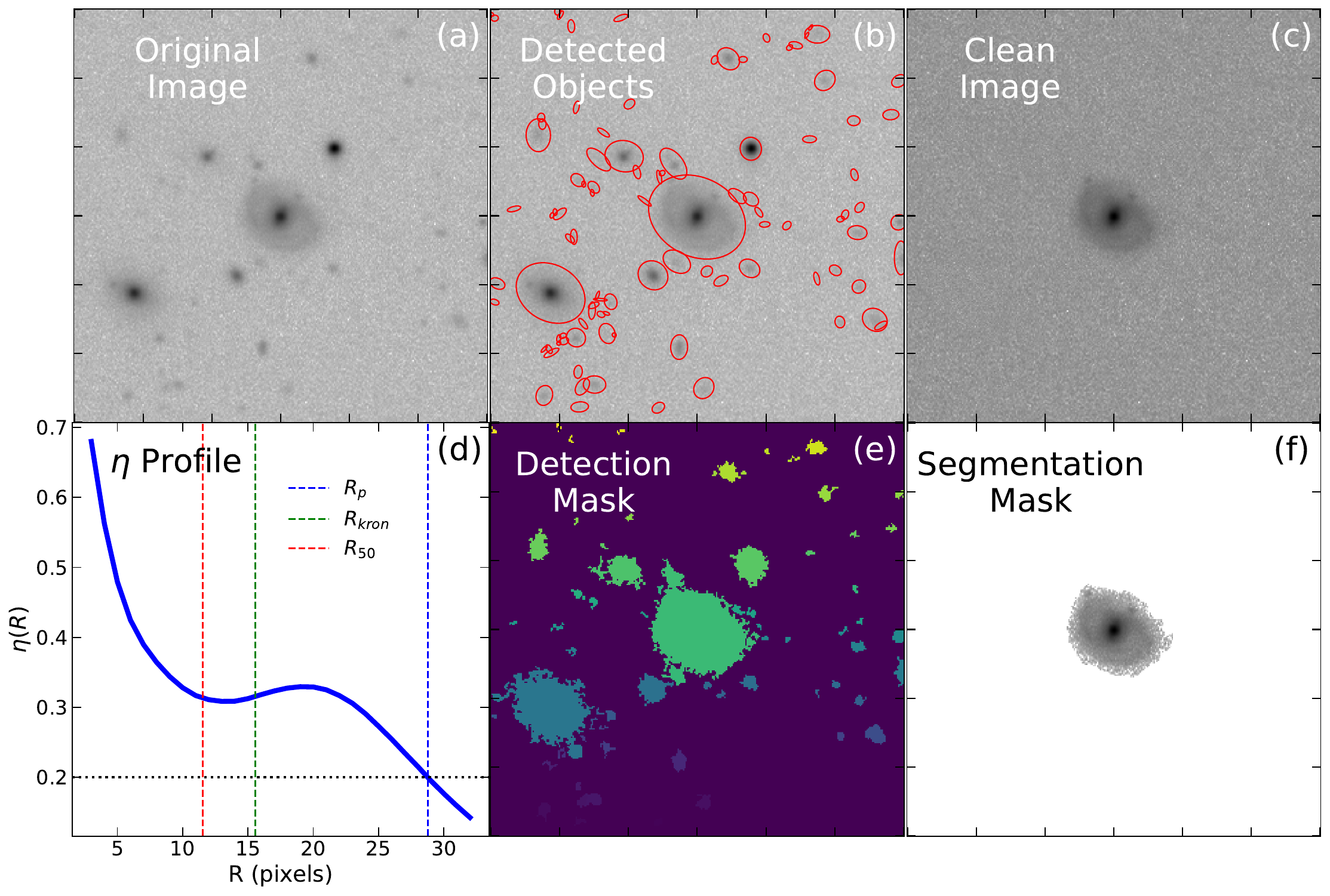}
    \caption{Illustration of the \texttt{galmex} pre-processing steps. 
    Panels show: (a) the original image; (b) object detection; (c) clean image; (d) Petrosian profile with key radii marked ($R_p$, $R_{\rm kron}$, $R_{50}$); (e) detection mask; and (f) the final segmentation mask used for morphological measurements.}
    \label{fig:galmex_preproc}
\end{figure*}

In this appendix, Fig.~\ref{fig:galmex_preproc} shows an example of the performance of the pre-processing pipeline adopted.

\section{The effect of aperture geometry in petrosian radius estimate}
\label{ap:rp}

The Petrosian radius is commonly defined as the radius $r_{\rm P}$ at which the ratio between the mean surface brightness in an annulus around $r$ and the mean surface brightness within $r$ reaches a fixed value $\eta_{\rm crit}$ (typically $\eta_{\rm crit}=0.2$):
\begin{equation}
\eta(r) \equiv 
\frac{\langle I \rangle_{[0.8\,r,\,1.25\,r]}}
     {\langle I \rangle_{< r}}
= \eta_{\rm crit}.
\end{equation}
In practice, most implementations use circular apertures, i.e. $r$ is the circular radius and the annulus is a circular ring.  However, galaxies are generally not circularly symmetric. For an intrinsically flattened galaxy with semi--major axis $a$ and semi--minor axis $b = q a$ ($q=b/a<1$), the isophotes are better described by ellipses.  In that case, the ``natural'' Petrosian radius is an elliptical semi--major axis $r_{\rm P,ell}$ measured in elliptical coordinates.

When circular apertures are used for an intrinsically elliptical system, the Petrosian annulus inevitably includes a substantial fraction of pixels that belong mostly to the sky background or to unrelated sources (``trash'' pixels).  A simple way to see this is to compare a circular annulus with radius $r=a$ to an elliptical annulus with the same semi--major axis $a$ and axis ratio $q$.  The area of the circular annulus scales as $A_{\rm circ} \propto \pi r^{2}$, whereas the area of the corresponding elliptical annulus scales as $A_{\rm ell} \propto \pi a b = \pi q a^{2}$.  For a given $a$, only a fraction $\simeq q$ of the circular annulus overlaps the galaxy isophotes, while a fraction $\simeq 1-q$ samples mainly background.  For a highly flattened system with $q=0.3$, this implies that roughly $70\,\%$ of the pixels in the circular annulus are effectively
``trash'' pixels, whereas for a round galaxy $q\approx 1$ this effect is negligible.

\begin{figure}
    \centering
    \includegraphics[width=0.8\linewidth]{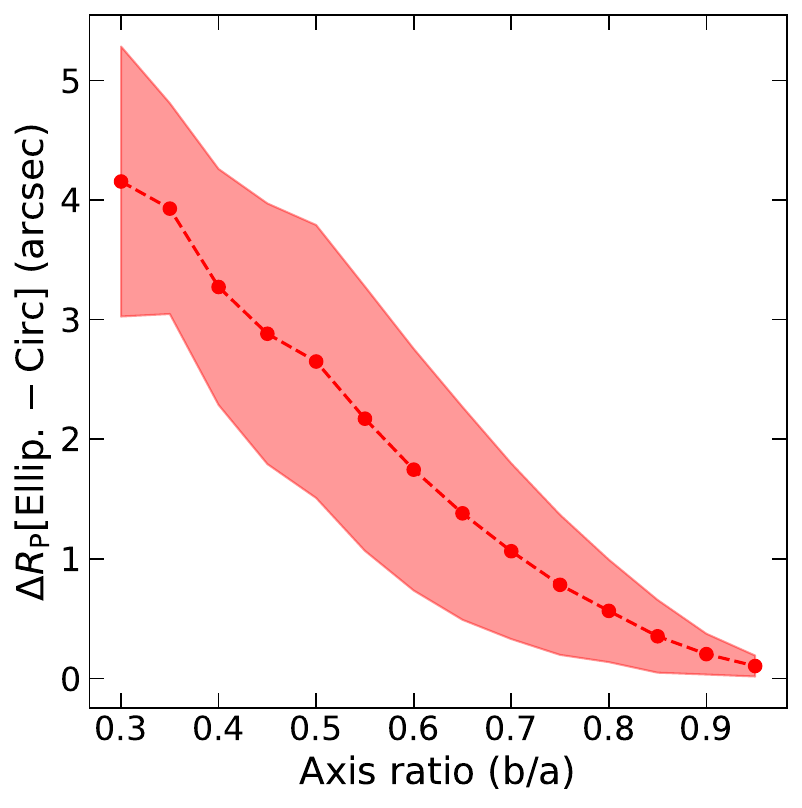}
    \caption{Difference between the Petrosian radius estimated using elliptical and circular apertures, as a function of the axis ratio of the galaxy. Notably, the difference increases for decreasing axis ratio, highlighting the effect of adopting mismatching geometry when calculating the characteristic radii.}
    \label{fig:rp_diff}
\end{figure}

Because the Petrosian ratio $\eta(r)$ is defined as a mean surface brightness in the annulus, the inclusion of a large and $q$-dependent fraction of background pixels systematically lowers $\langle I \rangle_{[0.8\,r,\,1.25\,r]}$ with respect to the elliptical case.  The mean surface brightness within $r$, $\langle I \rangle_{<r}$, is less affected because it is dominated by high S/N galaxy pixels.  As a consequence, $\eta(r)$ computed from circular apertures declines more rapidly with increasing $r$ than the corresponding elliptical $\eta_{\rm ell}(r)$, and the condition $\eta(r) = \eta_{\rm crit}$ is reached at a smaller radius:
\begin{equation}
R_{\rm P,circ} < R_{\rm P,ell},
\end{equation}
with the bias increasing as the axis ratio $q$ decreases.  This behavior is highlighted in Fig.~\ref{fig:rp_diff}, which shows the difference between the circular and elliptical Petrosian radius as a function of the axis ratio (b/a) of the object. This is calculated using our joint Sp and Ell samples. Notably, the use of circular apertures for objects with small axis ratio can introduce bias up to 5 arcseconds, which in the DECam resolution means $\sim20$ pixels, and can directly affect the creation of the segmentation mask, for instance. The bias is driven by geometry: for a flattened system, a circular annulus with radius equal to the semi-major axis inevitably includes a substantial number of pixels that lie beyond the galaxy isophotes, especially toward the galaxy’s outskirts. In contrast, an elliptical annulus with the same semi-major axis but matched axis ratio traces the isophotes and therefore better follows the true light distribution.

\section{Effect of segmentation in metrics estimation}
\label{ap:segmentation}

\begin{figure*}
    \centering
    \includegraphics[width=0.8\textwidth]{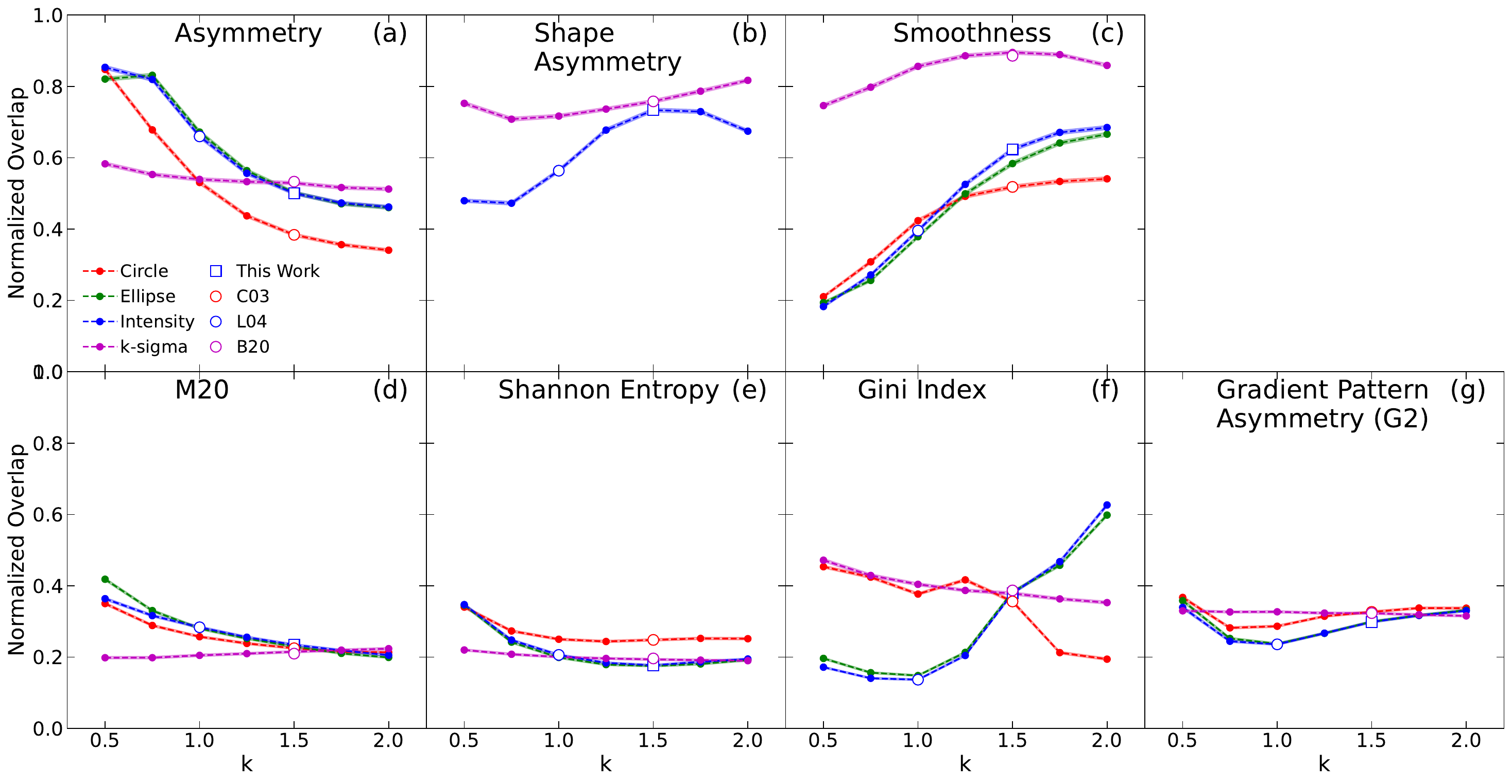}
    \caption{Variation of the overlap between Sp and Ell galaxies distributions for each segmentation-mask dependent non-parametric index (each panel). We present the results for 4 different types of segmentation (each colored curve), and highlight the choice of previous works in the literature in comparison to the adopted in this work (non-filled symbols). See the text for the definition of the meaning of $k$ for each segmentation method.}
    \label{fig:different_segm}
\end{figure*}

In this appendix we present how segmentation masks can affect the observed separation between Ell and Sp for each non-parametric index. Notably, the only index that does not rely on segmentation is C, thus it is not included in this analysis. In Fig.~\ref{fig:different_segm} we show how the overlap (calculated through Eq.~\ref{eq:OVL_1D}) between Sp and Ell distributions vary as a function of chosen segmentation. We considered four methods to define the segmentation mask, and parametrized the segmentation mask through the parameter $k$ as follows:
\begin{itemize}
    \item Circular aperture (red curve) -- circle with radius $k\times R_{\rm p}$;
    \item Elliptical aperture (green curve) -- elliptise with semi-major axis $k\times R_{\rm p}$;
    \item Intensity limited (blue curve) -- only pixels with intensity greater than $I(k \times R_{\rm P})$ are kept in the segmentation mask;
    \item k-sigma (magenta curve) -- the segmentation mask retrieved from the SExtractor detection when using a detection threshold equal to $k$ (note that in this case, in opposition to the others, larger $k$ means more restrict!);
\end{itemize}
In addition, we highlight using different symbols the segmentation masks used in previous works: 1) \citet{2003ApJS..147....1C} (C03, empty red circle) -- where the CAS parameters are defined; 2) \citet{2004AJ....128..163L} (L04, empty blue circle) -- inclusion of Gini index and M20; 3) \citet{2020A&C....3000334B} (B20, empty magenta circle) -- where they included Shannon Entropy, and G2; and 4) this work -- where we adopt the CA$\rm [A_S]$S + MEGG sytem as input for a machine learning method to estimate spiral probabilities (empty blue square). In particular for the $\rm A_S$ parameter (panel b), we show only the intensity limited and k-sigma results, as, by definition, the other two yields a shape asymmetry equals to 0, as can be seen from equation \ref{eq:As}.

Comparison between the different panels shows that A, $\rm A_S$, and S are the indices more sensitive to the choice of the segmentation mask. On the other hand, panels (d), (e), and (g) shows that M20, E, and G2 are the most stable with respect to the segmentation mask choice, which highlights these indexes as more robust with respect to variations in the pixel values distribution considered. Particularly for Gini index, shown in panel (f), overlap increases with $k$ for the ellipse and intensity limited cases, which can be explained by the inclusion of pixels ''close´´ to the background, thus setting smaller values for the "lowest income" of a pixel. In overall, our selection of intensity ensures that we are always sampling the same portion of the galaxy luminosity profile, and the choice of $k=1.0$ guarantees that we are not getting any artificial increased overlap between Sp and Ell galaxies. 

\section{Metrics comparison with \texttt{Statmorph}}

\begin{figure*}
    \centering
    \includegraphics[width=0.8\textwidth]{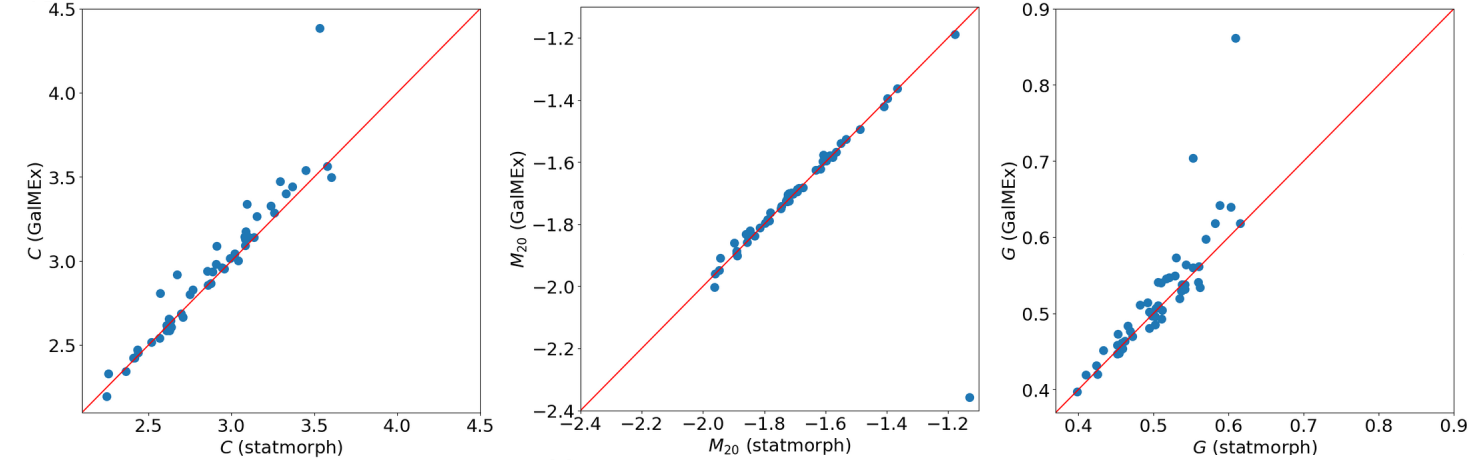}
    \caption{Comparison between the C (top panel), G (middle panel), and M20 (bottom panel) metrics estimated using the \texttt{galmex} (y-axis) and \texttt{statmorph} (x-axis) packages. Notably, both show good agreement with respect to the estimated values, irrespective of the panel considered.}
    \label{fig:galmex_comp}
\end{figure*}

In this Appendix, we show a simple comparison between the metrics C, G, and M20 estimated using \texttt{galmex} and \texttt{statmorph} for a sample of 50 randomly selected galaxies. The C index is select due to its independence of the segmentation mask, while G and M20 are selected due to \texttt{statmorph} measuring those indexes within a segmentation mask analogue to the one adopted in this work. In overall, the comparison, shown in Fig.\ref{fig:galmex_comp}, reveals agreement between the codes, and deviations from 1-to-1 line can be explained due to small deviations in the provided segmentation mask. Notably, the major advantage of \texttt{galmex} over \texttt{statmorph} is the flexibility of the first. For each metric, in \texttt{galmex}, the user can select the "rule" used to define the segmentation mask, whilst this is hard coded in \texttt{statmorph}. Additionally, all the pre-processing steps can be easily fine tuned within the \texttt{galmex} graphical interface. Therefore, this test shows that the flexibility of \texttt{galmex} comes with no cost with respect to reliability of the metrics.

\section{Example table}
\label{ap:example_table}

In this appendix we present the structure of the catalog containing galaxy properties and non-parametric indices measured with \texttt{galmex} for the $\sim1.7$ million galaxies described in Section \ref{sec:data}. Table~\ref{tab:galmex_columns} shows the description of each column of the catalog made public, whereas Tables~\ref{tab:galmex_warnmask} and \ref{tab:galmex_errcode} show the description of warnings and errors, respectively.

\begin{table*}
    \centering
    \caption{Description of the columns of the provided catalog.}
    \label{tab:galmex_columns}
    \begin{tabular}{l|p{0.68\textwidth}|l}
    \hline
    Name & Description & Units \\
    \hline
    ls\_id & Unique object identifier in the Legacy Surveys/DECaLS database. & N/A \\
    ra & Right ascension (ICRS/J2000 as provided by the survey catalog). & degrees \\
    dec & Declination (ICRS/J2000 as provided by the survey catalog). & degrees \\
    a & Object semi-major axis (from detection/segmentation ellipse). & Pixels \\
    b & Object semi-minor axis (from detection/segmentation ellipse). & Pixels \\
    theta & Object position angle (from +x to major axis). & Radians \\
    npix & Number of pixels in the detection segmentation mask. & N/A \\
    rp & Petrosian radius computed with an elliptical aperture. & arcsec \\
    r50 & Radius containing 50\% of the flux computed with an elliptical aperture. & arcsec \\
    rkron & Kron radius. & arcsec \\
    maingalaxy\_flag & 1 if no object is detected at the image center (e.g., center pixel label $\le 0$). & 0/1 \\
    rflag\_pixels & Radius of the circular aperture used in flagging/neighbor checks ($1.5\,R_{\rm p}$). & Pixels \\
    edge\_flag & 1 if there are zero/NaN-valued pixels within the flagging area. & 0/1 \\
    N\_rcheck & Number of secondary objects within the flagging area. & N/A \\
    Nsec\_flag & 1 if there are more than 4 objects within the flagging area. & 0/1 \\
    N\_deltaMAG & Number of secondary objects with magnitude difference smaller than 1. & N/A \\
    minMAG\_diff & Magnitude difference to the brightest secondary object. & N/A \\
    dist\_minMAG\_diff & Distance to the brightest secondary object. & Pixels \\
    BrightObj\_flag & 1 if there is a bright object within the flagging radius (as defined in the pipeline). & 0/1 \\
    normDist\_closest & Distance to the closest secondary object, normalized by $R_{\rm p}$. & N/A \\
    mup & Average pixel value at $1\,R_{\rm p}$ (surface-brightness proxy at the Petrosian scale). & Image units \\
    C & Concentration index (e.g., based on $r_{80}$ and $r_{20}$). & N/A \\
    r20 & Radius containing 20\% of the flux (elliptical aperture). & Pixels \\
    r80 & Radius containing 80\% of the flux (elliptical aperture). & Pixels \\
    A & Asymmetry index. & N/A \\
    Ashape & Shape asymmetry index (``$A3$'' in the warning bitmask). & N/A \\
    S & Smoothness index. & N/A \\
    M20 & $M_{20}$ index. & N/A \\
    xc\_M20 & Pixel $x$ coordinate that minimizes the total second-order moment. & Pixel \\
    yc\_M20 & Pixel $y$ coordinate that minimizes the total second-order moment. & Pixel \\
    E & Shannon entropy index. & N/A \\
    Gini & Gini index. & N/A \\
    G2 & Gradient pattern asymmetry index (GPA). & N/A \\
    pspiral & LightGBM probability that the galaxy is spiral (only available for the 1.7 million catalog). & N/A \\
    status & Processing status: 0 = full success; 1 = not success (failed); 2 = success with warnings. & 0/1/2 \\
    warn\_mask & Warning bitmask (decoded in Table~\ref{tab:galmex_warnmask}). & see Table~\ref{tab:galmex_warnmask} \\
    err\_code & Error code explaining why processing failed (decoded in Table~\ref{tab:galmex_errcode}). & see Table~\ref{tab:galmex_errcode} \\
    \hline
    \end{tabular}
\end{table*}

\begin{table*}
    \centering
    \caption{Decoding of the \texttt{warn\_mask} bitmask. A non-zero \texttt{warn\_mask} indicates one or more warnings were raised during processing. Example: \texttt{warn\_mask}=3 means \texttt{FLAG\_FAIL} (1) + \texttt{CUTOUT\_FAIL} (2).}
    \label{tab:galmex_warnmask}
    \begin{tabular}{r|r|l|p{0.62\textwidth}}
    \hline
    Bit $(1\ll n)$ & Decimal & Name & Meaning (when it gets set in the code) \\
    \hline
    -- & 0 & NONE & No warnings (special value). \\
    $1\ll 0$ & 1 & FLAG\_FAIL & Flagging step failed . \\
    $1\ll 1$ & 2 & CUTOUT\_FAIL & Cutout extraction failed. \\
    $1\ll 2$ & 4 & A1\_FAIL & Conselice asymmetry computation failed. \\
    $1\ll 3$ & 8 & A2\_FAIL & Ferrari asymmetry computation failed. \\
    $1\ll 4$ & 16 & A3\_FAIL & Shape asymmetry (Ashape / ``A3'') computation failed. \\
    $1\ll 5$ & 32 & S1\_FAIL & Conselice smoothness computation failed. \\
    $1\ll 6$ & 64 & S2\_FAIL & Ferrari smoothness computation failed. \\
    $1\ll 7$ & 128 & M20\_FAIL & $M_{20}$ computation failed. \\
    $1\ll 8$ & 256 & ENTROPY\_FAIL & Shannon entropy computation failed. \\
    $1\ll 9$ & 512 & GINI\_FAIL & Gini computation failed. \\
    $1\ll 10$ & 1024 & G2\_FAIL & G2 (GPA) computation failed. \\
    \hline
    \end{tabular}
\end{table*}

\begin{table*}
    \centering
    \caption{Meaning of the \texttt{err\_code} values.}
    \label{tab:galmex_errcode}
    \begin{tabular}{r|l|p{0.62\textwidth}}
    \hline
    Decimal value & Name & Meaning \\
    \hline
    0 & OK & Completed the pipeline (even if there were warnings). \\
    10 & FITS\_READ & Couldn’t open/read the FITS image. \\
    20 & SEP\_FAIL & SEP/detection failure. \\
    21 & NO\_CENTER\_OBJ & No galaxy detected at the image center. \\
    22 & SEG\_EMPTY & Empty/no pixels in segmentation mask. \\
    23 & SEP\_PIXSTACK & SEP “pixstack/internal pixel buffer full” type failure. \\
    24 & TOO\_MANY\_DETECTIONS & Too many detections ($\ge 1000$). \\
    30 & PETRO\_BAD & Petrosian radii invalid ($R_{\rm p}$ not finite or $\le 0$). \\
    31 & RECENTER\_FAIL & Recentering would exceed image bounds. \\
    32 & CUTOUT\_FAIL & Cutout failure. \\
    40 & ASYM\_FAIL & Asymmetry failure. \\
    41 & SMOOTH\_FAIL & Smoothness failure. \\
    50 & M20\_FAIL & $M_{20}$ failure. \\
    51 & ENTROPY\_FAIL & Entropy failure. \\
    52 & GINI\_FAIL & Gini failure. \\
    53 & G2\_FAIL & G2 failure. \\
    60 & FLAG\_FAIL & Flagging failure. \\
    99 & UNKNOWN & Anything not classified into one of the above. \\
    \hline
    \end{tabular}
\end{table*}

\end{appendix}

\end{document}